\documentclass[twocolumn]{aastex63}
\usepackage{natbib}
\usepackage{enumerate}
\usepackage{graphicx}
\usepackage{rotating}
\usepackage{epstopdf}
\usepackage{amsmath}
\usepackage{amssymb}
\usepackage{bm}
\usepackage{color}
\usepackage{multirow}
\usepackage{booktabs, makecell, multirow}
\usepackage{rotating}

\usepackage{tikz,xcolor,hyperref}

\newcommand{\Hb}{H$\beta$ }

\newcommand{\comment}[1]{}

\newcommand{\logfrmssigma}{$\log_{10}(f_{{\rm rms},{\sigma}})$}

\newcommand{\mrka}{Mrk 1392}

\newcommand{\pga}{PG 2209+184}
\newcommand{\rbsa}{RBS 1917}
\newcommand{\mrkb}{Mrk 110}
\newcommand{\rbsb}{RBS 1303}
\newcommand{\mcga}{MCG +04-22-042}

\newcommand{\mrkc}{Mrk 9}

\newcommand{\mrkd}{Mrk 1048}
\newcommand{\rxja}{RXJ 2044.0+2833}
\newcommand{\mrke}{Mrk 841}
\newcommand{\three}{3C 382}
\newcommand{\ark}{Ark 120}
\newcommand{\npmg}{NPM1G+27.0587}

\usepackage{float}
\usepackage{aliascnt}
\newaliascnt{eqfloat}{equation}
\newfloat{eqfloat}{h}{eqflts}
\floatname{eqfloat}{Equation}

\newcommand*{\ORGeqfloat}{}
\let\ORGeqfloat\eqfloat
\def\eqfloat{%
  \let\ORIGINALcaption\caption
  \def\caption{%
    \addtocounter{equation}{-1}%
    \ORIGINALcaption
  }%
  \ORGeqfloat
}

\defcitealias{pancoast14b}{P14}
\defcitealias{Grier++17}{G17}
\defcitealias{2018ApJ...866...75W}{W18}
\defcitealias{2020ApJ...902...74W}{W20}
\defcitealias{bentz2021detailed}{B21}
\defcitealias{pancoast11}{P11}

\newcommand{\pgalogmbh}{$7.53^{+0.19}_{-0.20}$}
\newcommand{\pgaellip}{$0.54^{+0.10}_{-0.15}$}
\newcommand{\pgaflow}{$0.24^{+0.17}_{-0.16}$}
\newcommand{\pgathetae}{$24^{+23}_{-16}$}
\newcommand{\pgasigmaturb}{$0.01^{+0.05}_{-0.01}$}
\newcommand{\pgarmean}{$ 16.27^{+1.00}_{-0.95}$}
\newcommand{\pgarmedian}{$15.2^{+1.1}_{-1.0}$}
\newcommand{\pgarmin}{$2.8^{+1.4}_{-1.6}$}
\newcommand{\pgasigmar}{$22^{+12}_{-8}$}
\newcommand{\pgataumean}{$15.75^{+0.74}_{-0.77}$}
\newcommand{\pgataumedian}{$12.95^{+0.87}_{-0.88}$}
\newcommand{\pgabeta}{$0.88^{+0.14}_{-0.15}$}
\newcommand{\pgathetao}{$29.1^{+11.0}_{-8.4}$}
\newcommand{\pgathetai}{$30.2^{+8.7}_{-6.9}$}
\newcommand{\pgakappa}{$-0.09^{+0.12}_{-0.15}$}
\newcommand{\pgagamma}{$1.40^{+0.38}_{-0.28}$}
\newcommand{\pgaxi}{$0.73^{+0.16}_{-0.18}$}
\newcommand{\pgainout}{$-0.40^{+0.09}_{-0.09}$}

\newcommand{\mcgalogmbh}{$7.59^{+0.42}_{-0.28}$}
\newcommand{\mcgaellip}{$0.39^{+0.21}_{-0.18}$}
\newcommand{\mcgaflow}{$0.27^{+0.18}_{-0.19}$}
\newcommand{\mcgathetae}{$19^{+20}_{-13}$}
\newcommand{\mcgasigmaturb}{$0.01^{+0.02}_{-0.00}$}
\newcommand{\mcgarmean}{$9.94^{+1.20}_{-0.98}$}
\newcommand{\mcgarmedian}{$6.24^{+1.01}_{-0.87}$}
\newcommand{\mcgarmin}{$1.05^{+0.65}_{-0.58}$}
\newcommand{\mcgasigmar}{$21.5^{+35.3}_{-9.7}$}
\newcommand{\mcgataumean}{$9.17^{+0.95}_{-0.81}$}
\newcommand{\mcgataumedian}{$5.58^{+0.81}_{-0.70}$}
\newcommand{\mcgabeta}{$1.40^{+0.20}_{-0.19}$}
\newcommand{\mcgathetao}{$13.6^{+6.9}_{-4.9}$}
\newcommand{\mcgathetai}{$11.3^{+5.8}_{-5.0}$}
\newcommand{\mcgakappa}{$-0.14^{+0.44}_{-0.27}$}
\newcommand{\mcgagamma}{$1.65^{+0.26}_{-0.36}$}
\newcommand{\mcgaxi}{$0.43^{+0.35}_{-0.26}$}
\newcommand{\mcgainout}{$-0.55^{+0.27}_{-0.17}$}
\newcommand{\mrkblogmbh}{$7.17^{+0.67}_{-0.26}$}
\newcommand{\mrkbellip}{$0.60^{+0.15}_{-0.20}$}
\newcommand{\mrkbflow}{$0.66^{+0.22}_{-0.39}$}
\newcommand{\mrkbthetae}{$13.7^{+15.5}_{-9.9}$}
\newcommand{\mrkbsigmaturb}{$0.01^{+0.04}_{-0.01}$}
\newcommand{\mrkbrmean}{$17.6^{+1.6}_{-1.5}$}
\newcommand{\mrkbrmedian}{$13.9^{+2.0}_{-1.8}$}
\newcommand{\mrkbrmin}{$ 1.22^{+0.40}_{-0.44}$}
\newcommand{\mrkbsigmar}{$47^{+13}_{-14}$}
\newcommand{\mrkbtaumean}{$18.8^{+2.0}_{-1.90}$}
\newcommand{\mrkbtaumedian}{$13.7^{+1.9}_{-1.8}$}
\newcommand{\mrkbbeta}{$1.20^{+0.09}_{-0.09}$}
\newcommand{\mrkbthetao}{$27^{+16}_{-13}$}
\newcommand{\mrkbthetai}{$19.9^{+9.6}_{-11}$}
\newcommand{\mrkbkappa}{$-0.41^{+0.42}_{-0.06}$}
\newcommand{\mrkbgamma}{$1.59^{+0.29}_{-0.36}$}
\newcommand{\mrkbxi}{$0.88^{+0.09}_{-0.19}$}
\newcommand{\mrkbinout}{$0.30^{+0.21}_{-0.63}$}

\newcommand{\mrkclogmbh}{$7.09^{+0.22}_{-0.23}$}
\newcommand{\mrkcellip}{$0.12^{+0.17}_{-0.08}$}
\newcommand{\mrkcflow}{$0.27^{+0.20}_{-0.18}$}
\newcommand{\mrkcthetae}{$45^{+15}_{-28}$}
\newcommand{\mrkcsigmaturb}{$0.01^{+0.04}_{-0.01}$}
\newcommand{\mrkcrmean}{$11.8^{+3.6}_{-2.7}$}
\newcommand{\mrkcrmedian}{$8.0^{+2.8}_{-1.9}$}
\newcommand{\mrkcrmin}{$ 2.21^{+0.81}_{-0.67}$}
\newcommand{\mrkcsigmar}{$36^{+152}_{-18}$}
\newcommand{\mrkctaumean}{$10.1^{+2.3}_{-2.2}$}
\newcommand{\mrkctaumedian}{$5.6^{+1.5}_{-1.5}$}
\newcommand{\mrkcbeta}{$1.44^{+0.12}_{-0.15}$}
\newcommand{\mrkcthetao}{$45^{+17}_{-17}$ }
\newcommand{\mrkcthetai}{$42^{+12}_{-15}$ }
\newcommand{\mrkckappa}{$0.02^{+0.11}_{-0.11}$}
\newcommand{\mrkcgamma}{$1.56^{+0.29}_{-0.33}$}
\newcommand{\mrkcxi}{$0.52^{+0.21}_{-0.20}$}
\newcommand{\mrkcinout}{$-0.59^{+0.23}_{-0.20}$}
\newcommand{\rxjalogmbh}{$7.09^{+0.17}_{-0.17}$}
\newcommand{\rxjarmean}{$35.7^{+7.8}_{-6.3}$}
\newcommand{\rxjarmedian}{$28.3^{+7.5}_{-5.4}$}
\newcommand{\rxjarmin}{$4.2^{+1.5}_{-1.1}$}
\newcommand{\rxjasigmar}{$55^{+45}_{-17}$ }
\newcommand{\rxjataumean}{$30.2^{+4.8}_{-5.3}$}
\newcommand{\rxjataumedian}{$18.8^{+3.5}_{-3.9}$}
\newcommand{\rxjabeta}{$1.12^{+0.08}_{-0.08}$}
\newcommand{\rxjathetao}{$51^{+15}_{-12}$}
\newcommand{\rxjathetai}{$42.5^{+9.6}_{-8.4}$}
\newcommand{\rxjakappa}{$-0.20^{+0.33}_{-0.19}$}
\newcommand{\rxjagamma}{$1.37^{+0.44}_{-0.29}$}
\newcommand{\rxjaxi}{$0.17^{+0.28}_{-0.12}$}
\newcommand{\rxjaellip}{$0.41^{+0.32}_{-0.29}$}
\newcommand{\rxjaflow}{$0.22^{+0.19}_{-0.15}$}
\newcommand{\rxjathetae}{$34^{+32}_{-21}$}
\newcommand{\rxjainout}{$-0.37^{+0.17}_{-0.23}$}
\newcommand{\rxjasigmaturb}{$0.01^{+0.03}_{-0.01}$} 

\newcommand{\rbsalogmbh}{$7.04^{+0.23}_{-0.35}$}
\newcommand{\rbsarmean}{$9.3^{+1.4}_{-1.6}$}
\newcommand{\rbsarmedian}{$5.0^{+1.3}_{-1.1}$}
\newcommand{\rbsarmin}{$1.42^{+0.39}_{-0.40}$}
\newcommand{\rbsataumean}{$9.0^{+1.9}_{-1.7}$}
\newcommand{\rbsataumedian}{$4.6^{+1.2}_{-1.2}$}
\newcommand{\rbsasigmar}{$27^{+25}_{-12}$}
\newcommand{\rbsabeta}{$1.63^{+0.13}_{-0.16}$}
\newcommand{\rbsathetao}{$25.1^{+9.2}_{-7.5}$}
\newcommand{\rbsathetai}{$20.2^{+9.9}_{-3.9}$}
\newcommand{\rbsakappa}{$-0.29^{+0.35}_{-0.14}$}
\newcommand{\rbsagamma}{$1.48^{+0.32}_{-0.32}$}
\newcommand{\rbsaxi}{$0.68^{+0.25}_{-0.35}$}
\newcommand{\rbsasigmaturb}{$0.01^{+0.04}_{-0.01}$}
\newcommand{\rbsathetae}{$20^{+21}_{-15}$}
\newcommand{\rbsaflow}{$0.59^{+0.28}_{-0.39}$}
\newcommand{\rbsaellip}{$0.59^{+0.14}_{-0.17}$}
\newcommand{\rbsainout}{$0.24^{+0.21}_{-0.63}$}
\newcommand{\mrkalogmbh}{$ 8.16^{+0.11}_{-0.13}$}
\newcommand{\mrkaellip}{$0.81^{+0.04}_{-0.06}$}
\newcommand{\mrkaflow}{$0.74^{+0.18}_{-0.18}$}
\newcommand{\mrkathetae}{$25^{+14}_{-15}$}
\newcommand{\mrkasigmaturb}{$0.01^{+0.04}_{-0.01}$}
\newcommand{\mrkarmean}{$56.0^{+12.0}_{-7.2}$}
\newcommand{\mrkarmedian}{$51.6^{+12.2}_{-8.6}$}
\newcommand{\mrkarmin}{$41^{+11}_{-12}$}
\newcommand{\mrkasigmar}{$31^{+54}_{-18}$}
\newcommand{\mrkataumean}{$38.7^{+4.6}_{-4.2}$}
\newcommand{\mrkataumedian}{$34.8^{+4.6}_{-4.6}$}
\newcommand{\mrkabeta}{$1.29^{+0.51}_{-0.74}$}
\newcommand{\mrkathetao}{$41.2^{+5.3}_{-4.8}$}
\newcommand{\mrkathetai}{$25.5^{+3.4}_{-2.8}$}
\newcommand{\mrkakappa}{$0.26^{+0.18}_{-0.25}$}
\newcommand{\mrkagamma}{$1.53^{+0.32}_{-0.33}$}
\newcommand{\mrkaxi}{$0.25^{+0.28}_{-0.18}$}
\newcommand{\mrkainout}{$ 0.16^{+0.05}_{-0.03}$}
\newcommand{\rbsblogmbh}{$ 6.79^{+0.19}_{-0.11}$}
\newcommand{\rbsbellip}{$0.18^{+0.17}_{-0.11}$}
\newcommand{\rbsbflow}{$0.75^{+0.17}_{-0.19}$}
\newcommand{\rbsbthetae}{$8.3^{+8.8}_{-5.8}$}
\newcommand{\rbsbsigmaturb}{$0.01^{+0.02}_{-0.00}$}
\newcommand{\rbsbrmean}{$12.9^{+1.3}_{-1.2}$}
\newcommand{\rbsbrmedian}{$10.1^{+1.3}_{-1.2}$}
\newcommand{\rbsbrmin}{$0.25^{+0.24}_{-0.18}$}
\newcommand{\rbsbsigmar}{$13.6^{+3.8}_{-2.3}$}
\newcommand{\rbsbtaumean}{$13.7^{+1.3}_{-1.3}$}
\newcommand{\rbsbtaumedian}{$9.8^{+1.2}_{-1.1}$}
\newcommand{\rbsbbeta}{$0.94^{+0.07}_{-0.07}$}
\newcommand{\rbsbthetao}{$34.0^{+8.9}_{-10}$}
\newcommand{\rbsbthetai}{$29.1^{+7.7}_{-9.0}$}
\newcommand{\rbsbkappa}{$-0.48^{+0.05}_{-0.01}$}
\newcommand{\rbsbgamma}{$1.85^{+0.11}_{-0.21}$}
\newcommand{\rbsbxi}{$0.60^{+0.22}_{-0.16}$}
\newcommand{\rbsbinout}{$0.80^{+0.11}_{-0.18}$}

\newcommand{\mrkdlogmbh}{$7.79^{+0.44}_{-0.48}$}
\newcommand{\mrkdellip}{$0.73^{+0.09}_{-0.13}$}
\newcommand{\mrkdflow}{$0.74^{+0.18}_{-0.19}$}
\newcommand{\mrkdthetae}{$15^{+15}_{-10}$}
\newcommand{\mrkdsigmaturb}{$0.01^{+0.04}_{-0.01}$}
\newcommand{\mrkdrmean}{$14.2^{+7.8}_{-7.6}$}
\newcommand{\mrkdrmedian}{$11.3^{+7.3}_{-6.2}$}
\newcommand{\mrkdrmin}{$3.3^{+3.0}_{-2.1}$}
\newcommand{\mrkdsigmar}{$14.1^{+22}_{-8.7}$}
\newcommand{\mrkdtaumean}{$11.5^{+6.6}_{-6.0}$}
\newcommand{\mrkdtaumedian}{$ 8.2^{+6.2}_{-4.5}$}
\newcommand{\mrkdbeta}{$1.12^{+0.28}_{-0.28}$}
\newcommand{\mrkdthetao}{$31^{+14}_{-10.0}$}
\newcommand{\mrkdthetai}{$21.5^{+9.4}_{-9.4}$}
\newcommand{\mrkdkappa}{$0.10^{+0.28}_{-0.38}$}
\newcommand{\mrkdgamma}{$1.47^{+0.33}_{-0.32}$}
\newcommand{\mrkdxi}{$0.30^{+0.42}_{-0.20}$}
\newcommand{\mrkdinout}{$0.24^{+0.11}_{-0.09}$}

\newcommand{\mrkelogmbh}{$7.62^{+0.50}_{-0.30}$}
\newcommand{\mrkeellip}{$0.33^{+0.24}_{-0.22}$}
\newcommand{\mrkeflow}{$0.45^{+0.36}_{-0.29}$}
\newcommand{\mrkethetae}{$51^{+20}_{-27}$}
\newcommand{\mrkesigmaturb}{$0.01^{+0.05}_{-0.01}$}
\newcommand{\mrkermean}{$14.1^{+7.1}_{-4.4}$}
\newcommand{\mrkermedian}{$10.6^{+5.6}_{-3.4}$}
\newcommand{\mrkermin}{$2.0^{+1.8}_{-1.1}$}
\newcommand{\mrkesigmar}{$14.7^{+15}_{-5.8}$}
\newcommand{\mrketaumean}{$13.5^{+4.6}_{-3.8}$}
\newcommand{\mrketaumedian}{$8.9^{+3.3}_{-2.5}$}
\newcommand{\mrkebeta}{$1.08^{+0.18}_{-0.16}$}
\newcommand{\mrkethetao}{$41^{+11}_{-11}$}
\newcommand{\mrkethetai}{$30^{+11}_{-15}$}
\newcommand{\mrkekappa}{$-0.23^{+0.43}_{-0.14}$}
\newcommand{\mrkegamma}{$1.41^{+0.42}_{-0.29}$}
\newcommand{\mrkexi}{$0.68^{+0.23}_{-0.41}$}
\newcommand{\mrkeinout}{$-0.33^{+0.55}_{-0.31}$}

\newcommand{\arklogmbh}{$8.26^{+0.12}_{-0.17}$}
\newcommand{\arkellip}{$0.14^{+0.02}_{-0.03}$}
\newcommand{\arkflow}{$0.25^{+0.17}_{-0.17}$}
\newcommand{\arkthetae}{$7.2^{+6.5}_{-4.8}$}
\newcommand{\arksigmaturb}{$0.01^{+0.02}_{-0.00}$}
\newcommand{\arkrmean}{$19.2^{+2.6}_{-2.2}$}
\newcommand{\arkrmedian}{$17.9^{+2.1}_{-2.4}$}
\newcommand{\arkrmin}{$1.16^{+1.4}_{-0.89}$}
\newcommand{\arksigmar}{$36^{+47}_{-21}$}
\newcommand{\arktaumean}{$12.8^{+1.4}_{-1.3}$}
\newcommand{\arktaumedian}{$11^{+1.5}_{-1.9}$}
\newcommand{\arkbeta}{$0.89^{+0.09}_{-0.10}$}
\newcommand{\arkthetao}{$32.0^{+7.1}_{-8.1}$}
\newcommand{\arkthetai}{$13.6^{+3.5}_{-3.2}$}
\newcommand{\arkkappa}{$0.26^{+0.18}_{-0.22}$}
\newcommand{\arkgamma}{$1.73^{+0.20}_{-0.55}$}
\newcommand{\arkxi}{$0.02^{+0.04}_{-0.01}$}
\newcommand{\arkinout}{$-0.85^{+0.02}_{-0.03}$}

\newcommand{\npmglogmbh}{$7.64^{+0.40}_{-0.36}$}  
\newcommand{\npmgellip}{$0.44^{+0.19}_{-0.18}$}
\newcommand{\npmgflow}{$0.26^{+0.18}_{-0.19}$}
\newcommand{\npmgthetae}{$36^{+35}_{-24}$}
\newcommand{\npmgsigmaturb}{$0.01^{+0.05}_{-0.01}$}
\newcommand{\npmgrmean}{$12.2^{+4.3}_{-3.5}$}
\newcommand{\npmgrmedian}{$7.2^{+2.8}_{-2.0}$}
\newcommand{\npmgrmin}{$ 2.69^{+1.6}_{-0.95}$}
\newcommand{\npmgsigmar}{$22.0^{+18}_{-9.2}$}
\newcommand{\npmgtaumean}{$10.7^{+3.1}_{-2.8}$}
\newcommand{\npmgtaumedian}{$6.0^{+2.0}_{-1.6}$}
\newcommand{\npmgbeta}{$1.54^{+0.13}_{-0.13}$}
\newcommand{\npmgthetao}{$18^{+11}_{-9.1}$}
\newcommand{\npmgthetai}{$19^{+11}_{-8.5}$}
\newcommand{\npmgkappa}{$-0.14^{+0.40}_{-0.25}$}
\newcommand{\npmggamma}{$1.39^{+0.38}_{-0.27}$}
\newcommand{\npmgxi}{$ 0.11^{+0.37}_{-0.09}$}
\newcommand{\npmginout}{$ -0.41^{+0.25}_{-0.18}$}

\shorttitle{LAMP 2016: Dynamical Modeling} 
\shortauthors{Villafa\~na et al.}
 
\begin{document}
\title{The Lick AGN Monitoring Project 2016: Dynamical Modeling of Velocity-Resolved H$\beta$ Lags in Luminous Seyfert Galaxies}
\author[0000-0002-1961-6361]{Lizvette Villafa\~na}
\affiliation{Department of Physics and Astronomy, University of California, Los Angeles, CA 90095-1547, USA}

\author[0000-0002-4645-6578]{Peter R. Williams}
\affiliation{Department of Physics and Astronomy, University of California, Los Angeles, CA 90095-1547, USA}

\author[0000-0002-8460-0390]{Tommaso Treu}
\affiliation{Department of Physics and Astronomy, University of California, Los Angeles, CA 90095-1547, USA}
%email{tt@astro.ucla.edu}

\author{Brendon J. Brewer}
\affiliation{Department of Statistics, The University of Auckland, Private Bag 92019, Auckland 1142, New Zealand}

\author[0000-0002-3026-0562]{Aaron J. Barth}
\affiliation{Department of Physics and Astronomy, 4129 Frederick Reines Hall, University of California, Irvine, CA 92697, USA}

\author[0000-0002-1912-0024]{Vivian U}
\affiliation{Department of Physics and Astronomy, 4129 Frederick Reines Hall, University of California, Irvine, CA 92697, USA}
\affiliation{Department of Physics and Astronomy, University of California, Riverside, CA 92521, USA}

\author[0000-0003-2064-0518]{Vardha N. Bennert}
\affiliation{Physics Department, California Polytechnic State University, San Luis Obispo CA 93407, USA}

\author{H. Alexander Vogler}
\affiliation{Department of Physics and Astronomy, 4129 Frederick Reines Hall, University of California, Irvine, CA 92697, USA}
\affiliation{Department of Physics and Astronomy, University of California, 1 Shields Avenue, Davis, CA 95616, USA}

\author[0000-0001-8416-7059]{Hengxiao Guo}
\affiliation{Department of Physics and Astronomy, 4129 Frederick Reines Hall, University of California, Irvine, CA 92697, USA}
%\email{barth@uci.edu}

\author[0000-0002-2816-5398]{Misty C. Bentz}
\affiliation{Department of Physics and Astronomy, Georgia State University, Atlanta, GA 30303, USA}

\author[0000-0003-4693-6157]{Gabriela Canalizo}
\affiliation{Department of Physics and Astronomy, University of California, Riverside, CA 92521, USA}

\author[0000-0003-3460-0103]{Alexei V. Filippenko}
\affiliation{Department of Astronomy, University of California, 501 Campbell Hall, Berkeley, CA 94720-3411, USA}
\affiliation{Miller Institute for Basic Research in Science, University of California, Berkeley, CA 94720, USA}

\author[0000-0002-3739-0423]{Elinor Gates}
\affiliation{Lick Observatory, P.O. Box 85, Mt. Hamilton, CA 95140, USA}

\author{Frederick Hamann}
\affiliation{Department of Physics and Astronomy, University of California, Riverside, CA 92521, USA}

\author[0000-0003-0634-8449]{Michael D. Joner}
\affiliation{Department of Physics and Astronomy, N283 ESC, Brigham Young University, Provo, UT 84602, USA}

\author{Matthew A. Malkan}
\affiliation{Department of Physics and Astronomy, University of California, Los Angeles, CA 90095-1547, USA}

\author[0000-0002-8055-5465]{Jong-Hak Woo}
\affil{Astronomy Program, Department of Physics and Astronomy, Seoul National University, 1 Gwanak-ro, Gwanak-gu, Seoul 08826, Korea}
\affil{SNU Astronomy Research Center, Seoul National University, 1 Gwanak-ro, Gwanak-gu, Seoul 08826, Republic of Korea}

\author{Bela Abolfathi}
\affiliation{Department of Physics and Astronomy, 4129 Frederick Reines Hall, University of California, Irvine, CA 92697, USA}

\author[0000-0002-8860-1032]{L. E.~Abramson}
\affiliation{Carnegie Observatories, 813 Santa Barbara Street, Pasadena, CA 91101, USA}

\author{Stephen F. Armen}
\affiliation{Department of Astronomy, San Diego State University, San Diego, CA 92182-1221, USA}
  
\author{Hyun-Jin Bae}
\affil{Astronomy Program, Department of Physics and Astronomy, Seoul National University, 1 Gwanak-ro, Gwanak-gu, Seoul 08826, Korea}

\author{Thomas Bohn}
\affiliation{Department of Physics and Astronomy, University of California, Riverside, CA 92521, USA}

\author[0000-0001-6301-570X]{Benjamin D. Boizelle}
\affiliation{Department of Physics and Astronomy, N283 ESC, Brigham Young University, Provo, UT 84602, USA}
\affiliation{Department of Physics and Astronomy, 4129 Frederick Reines Hall, University of California, Irvine, CA 92697, USA}
%\affiliation{Mitchell Institute for Fundamental Physics and Astronomy, Department of Physics and Astronomy, Texas A\&M University, 4242 TAMU, College Station, TX 77843, USA}

\author[0000-0002-4924-444X]{Azalee Bostroem}
\affiliation{Department of Physics and Astronomy, University of California, 1 Shields Avenue, Davis, CA 95616, USA}
\affiliation{DiRAC Institute, Department of Astronomy, University of Washington, 3910 15th Avenue, NE, Seattle, WA 98195, USA}

\author{Andrew Brandel}
\affiliation{Department of Physics and Astronomy, 4129 Frederick Reines Hall, University of California, Irvine, CA 92697, USA}

\author[0000-0001-5955-2502]{Thomas G. Brink}
\affiliation{Department of Astronomy, University of California, 501 Campbell Hall, Berkeley, CA 94720-3411, USA}
%\email{tgbrink@berkeley.edu}

%\author{Dan Carson}
%\affiliation{Department of Physics and Astronomy, 4129 Frederick Reines Hall, University of California, Irvine, CA 92697, USA}

\author{Sanyum Channa}
\affiliation{Department of Physics, University of California, Berkeley, CA 94720, USA}
\affiliation{Department of Physics, Stanford University, Stanford, CA 94305, USA}
% \email{sanyum_channa@berkeley.edu}

\author{M. C. Cooper}
\affiliation{Department of Physics and Astronomy, 4129 Frederick Reines Hall, University of California, Irvine, CA 92697, USA}

\author[0000-0002-2248-6107]{Maren Cosens}
%\affiliation{Physics Department, California Polytechnic State University, San Luis Obispo CA 93407, USA}
\affiliation{Physics Department, University of California, San Diego, 9500 Gilman Drive, La Jolla, CA 92093, USA}
\affiliation{Center for Astrophysics and Space Sciences, University of California, San Diego, 9500 Gilman Drive, La Jolla, CA 92093, USA}
%\email{mcosens@ucsd.edu}

\author{Edward Donohue}
\affiliation{Physics Department, California Polytechnic State University, San Luis Obispo CA 93407, USA}
\affiliation{Booz Allen, 1615 Murray Canyon Road, Suite 8000, San Diego, CA 92108, USA}
%\email{edwardd1@gmail.com}

\author[0000-0002-8425-0351]{Sean P. Fillingham}
\affiliation{Department of Physics and Astronomy, 4129 Frederick Reines Hall, University of California, Irvine, CA 92697, USA}
%\affiliation{Astronomy Department, Box 351580, University of Washington, Seattle, WA 98195, USA}

\author[0000-0002-9280-1184]{Diego Gonz\'{a}lez-Buitrago}
\affiliation{Department of Physics and Astronomy, 4129 Frederick Reines Hall, University of California, Irvine, CA 92697, USA}
\affiliation{Universidad Nacional Aut\'onoma de M\'exico, Instituto de Astronom\'ia, AP 106,  Ensenada 22860, BC, M\'exico}

\author[0000-0002-7232-101X]{Goni Halevi}
\affiliation{Department of Astronomy, University of California, 501 Campbell Hall, Berkeley, CA 94720-3411, USA}
\affiliation{Department of Astrophysical Sciences, Princeton University, 4 Ivy Lane, Princeton, NJ 08544, USA}

\author{Andrew Halle}
\affiliation{Department of Physics, University of California, Berkeley, CA 94720, USA}
%\email{ahalle@berkeley.edu}

\author[0000-0003-0034-5909]{Carol E. Hood}
\affiliation{Department of Physics, California State University, San Bernardino, 5500 University Parkway, San Bernardino, CA 92407, USA}

\author[0000-0003-1728-0304]{Keith Horne}
\affiliation{SUPA Physics and Astronomy, University of St~Andrews, North Haugh, St~Andrews, KY16 9SS, Scotland, UK}
%\collaboration{LCOGT AGN Key Project}

\author{J. Chuck Horst}
\affiliation{Department of Astronomy, San Diego State University, San Diego, CA 92182-1221, USA}

\author{Maxime de Kouchkovsky}
\affiliation{Department of Astronomy, University of California, 501 Campbell Hall, Berkeley, CA 94720-3411, USA}

\author{Benjamin Kuhn}
\affiliation{Space Telescope Science Institute, 3700 San Martin Drive, Baltimore, MD 21218, USA}
\affiliation{Department of Astronomy, San Diego State University, San Diego, CA 92182-1221, USA}

\author[0000-0001-8367-7591]{Sahana Kumar}
\affil{Department of Astronomy, University of California, 501 Campbell Hall, Berkeley, CA 94720-3411, USA}
\affil{Department of Physics, Florida State University, 77 Chieftan Way, Tallahassee, FL 32306, USA}

\author[0000-0001-7839-1986]{Douglas C. Leonard}
\affiliation{Department of Astronomy, San Diego State University, San Diego, CA 92182-1221, USA}

\author{Donald Loveland}
\affiliation{Physics Department, California Polytechnic State University, San Luis Obispo CA 93407, USA}
\affiliation{Lawrence Livermore National Laboratory, 7000 East Avenue, Livermore, CA 94550, USA}
%\email{donaldloveland1@yahoo.com}

\author{Christina Manzano-King}
\affiliation{Department of Physics and Astronomy, University of California, Riverside, CA 92521, USA}

\author{Ian McHardy}
\affiliation{University of Southampton, Highfield, Southampton, SO17 1BJ, UK} 

\author[0000-0003-1263-808X]{Ra\'ul Michel}
\affiliation{Instituto de Astronom\'ia, Universidad Nacional Aut\'onoma de M\'exico, AP 877, Ensenada, Baja California, C.P. 22830 M\'exico}

\author{Melanie Kae B. Olaes}
\affiliation{Department of Astronomy, San Diego State University, San Diego, CA 92182-1221, USA}
  
\author[0000-0001-9877-1732]{Daeseong Park}
\affiliation{Department of Astronomy and Atmospheric Sciences, Kyungpook National University, Daegu, 41566, Republic of Korea}
\affiliation{Korea Astronomy and Space Science Institute, Daejeon, 34055, Republic of Korea}

\author{Songyoun Park}
\affil{Astronomy Program, Department of Physics and Astronomy, Seoul National University, 1 Gwanak-ro, Gwanak-gu, Seoul 08826, Korea}

\author{Liuyi Pei}
\affiliation{Department of Physics and Astronomy, 4129 Frederick Reines Hall, University of California, Irvine, CA 92697, USA}

\author{Timothy W. Ross}
\affiliation{Department of Astronomy, University of California, 501 Campbell Hall, Berkeley, CA 94720-3411, USA}

\author[0000-0003-4852-8958]{Jordan N. Runco}
\affiliation{Department of Physics and Astronomy, University of California, Los Angeles, CA 90095-1547, USA}

\author[0000-0002-8429-4100]{Jenna Samuel}
\affiliation{Department of Astronomy, The University of Texas at Austin, 2515 Speedway, Stop C1400, Austin, TX 78712, USA}
\affiliation{Department of Physics and Astronomy, University of California, 1 Shields Avenue, Davis, CA 95616, USA}

\author[0000-0003-3136-9532]{Javier S\'{a}nchez}
\affiliation{Department of Physics and Astronomy, 4129 Frederick Reines Hall, University of California, Irvine, CA 92697, USA}
\affiliation{Fermi National Accelerator Laboratory, Kirk Rd. \& Pine St, Batavia, IL 60510, USA}
\affiliation{Kavli Institute for Cosmological Physics, 5640 South Ellis Avenue, Chicago, IL 60637, USA}

\author{Bryan Scott}
\affiliation{Department of Physics and Astronomy, University of California, Riverside, CA 92521, USA}

\author[0000-0003-3432-2094]{Remington O. Sexton}
\affiliation{Department of Physics and Astronomy, University of California, Riverside, CA 92521, USA}
\affiliation{U.S. Naval Observatory, 3450 Massachusetts Ave NW, Washington, DC 20392-5420, USA}
\affiliation{Department of Physics and Astronomy, George Mason University, 4400 University Dr, Fairfax, VA 22030-4444, USA}

\author{Jaejin Shin}
\affil{Astronomy Program, Department of Physics and Astronomy, Seoul National University, 1 Gwanak-ro, Gwanak-gu, Seoul 08826, Korea}

\author{Isaac Shivvers}
\affiliation{Department of Astronomy, University of California, 501 Campbell Hall, Berkeley, CA 94720-3411, USA}

\author[0000-0002-4202-4188]{Chance L. Spencer}
\affiliation{Physics Department, California Polytechnic State University, San Luis Obispo CA 93407, USA}
%\affiliation{Gemini Observatory, Hilo, Hawaii}
\affiliation{Department of Physics, California State University Fresno, Fresno, CA 93740-8031, USA}
%\email{chancesum@gmail.com}

\author[0000-0002-3169-3167]{Benjamin E. Stahl}
\affiliation{Department of Astronomy, University of California, 501 Campbell Hall, Berkeley, CA 94720-3411, USA}
\affiliation{Department of Physics, University of California, Berkeley, CA 94720, USA}
%\email{benjamin_stahl@berkeley.edu}

\author{Samantha Stegman}
\affiliation{Department of Astronomy, University of California, 501 Campbell Hall, Berkeley, CA 94720-3411, USA}
\affiliation{Department of Chemistry, University of Wisconsin, Madison, WI 53706, USA}
%\affiliation{Argonne National Laboratory, 9700 S. Cass Avenue, Lemont, IL 60439, USA}
%\email{sstegman@wisc.edu}

\author[0000-0001-9685-7049]{Isak Stomberg}
\affiliation{Physics Department, California Polytechnic State University, San Luis Obispo CA 93407, USA}
\affiliation{Deutsches Elektronen-Synchrotron DESY, 22607 Hamburg, Germany}
%\affiliation{KTH Royal Institute of Technology, Stockholm, Sweden}
%\email{isakst@kth.se}

\author{Stefano Valenti}
\affiliation{Department of Physics and Astronomy, University of California, 1 Shields Avenue, Davis, CA 95616, USA}

\author[0000-0002-1881-5908]{Jonelle L. Walsh}
\affiliation{George P. and Cynthia W. Mitchell Institute for Fundamental Physics and Astronomy, Department of Physics \& Astronomy, Texas A\&M University, 4242 TAMU, College Station, TX 77843, USA}

\author{Heechan Yuk}
\affiliation{Department of Astronomy, University of California, 501 Campbell Hall, Berkeley, CA 94720-3411, USA}
\affiliation{Department of Physics and Astronomy, University of Oklahoma, 440 W. Brooks St., Norman, OK 73019, USA} 

\author{WeiKang Zheng}
\affiliation{Department of Astronomy, University of California, 501 Campbell Hall, Berkeley, CA 94720-3411, USA}
%\email{vbennert@calpoly.edu}

\correspondingauthor{Lizvette Villafa\~na}
\email{lvillafana@astro.ucla.edu}

\begin{abstract}
We have modeled the velocity-resolved reverberation response of the \Hb broad emission line in nine Seyfert 1 galaxies from the Lick Active Galactic Nucleus (AGN) Monitioring Project 2016 sample, drawing inferences on the geometry and structure of the low-ionization broad-line region (BLR) and the mass of the central supermassive black hole. Overall, we find that the \Hb BLR is generally a thick disk viewed at low to moderate inclination angles. We combine our sample with prior studies and investigate line-profile shape dependence, such as $\log_{10}(\rm{FWHM}/\sigma)$, on BLR structure and kinematics and search for any BLR luminosity-dependent trends. We find marginal evidence for an anticorrelation between the profile shape of the broad \Hb emission line and the Eddington ratio, when using the root-mean-square spectrum. However, we do not find any luminosity-dependent trends, and conclude that AGNs have diverse BLR structure and kinematics, consistent with the hypothesis of transient AGN/BLR conditions rather than systematic trends. 
\end{abstract}
\section{Introduction}
\label{sect: intro}
Over the last few decades, reverberation mapping has enabled the black hole (BH) mass measurements of over 70 active galactic nuclei (AGNs) and facilitated the use of single-epoch BH mass measurements across cosmic time \citep{2015PASP..127...67B}. Despite the technique's ability to resolve the BH's sphere of influence in time, much remains unknown about the broad emission line region (BLR). And while the promise of velocity-resolved reverberation mapping has increased significantly over the last decade, analysis requires recovery of a nontrivial transfer function. 

In principle, velocity-resolved reverberation mapping \citep{blandford82} can provide insight into the BLR structure and kinematics by mapping the BLR response as a function of line-of-sight velocity. However, doing so requires a high signal-to-noise ratio (S/N), high cadence, and a lengthy observational campaign, and thus has only been applied to $\sim 30$ AGNs over roughly the last decade  \citep[e.g.,][]{bentz09,Denney++09,denney10, barth11,barth11b, grier13b, Du++16, Pei++17, 2018ApJ...866..133D, 2018ApJ...869..142D, 2021ApJ...920....9L, 2021ApJ...912...92F}. Nonetheless, information regarding the BLR collected from these campaigns is not straightforward, as the BLR structure and kinematics are embedded in the so-called transfer function. 

The transfer function describes the time-delay distribution across a broad emission line as a function of line-of-sight velocity \citep[]{horne94,skielboe15}. In other words, the transfer function can be thought of as a map from the AGN stochastic continuum variations to the emission-line response at some line-of-sight velocity $v_z$, after some time delay $\tau$, \citep{peterson93}, and is expressed as 
\begin{equation}
    L(v_z,t)=\int^\infty_{-\infty} \Psi(v_z,\tau)~C(t-\tau)~d\tau , \quad
\end{equation}
where $L(v_z,t)$ is the emission-line luminosity at line-of-sight velocity $v_z$ at observed time $t$, $C(t)$ is the AGN continuum light curve, and $\Psi(v_z,\tau)$ is the transfer function. Because the shape of the transfer function depends on the structure and kinematics of the BLR \citep{2004PASP..116..465H}, one can theoretically use the transfer function to constrain the BLR geometry. In practice, however, interpretation of a transfer function is nontrivial since different geometries can produce similar features.%, for example. 

As an alternative analysis, one can instead use the methods introduced by \citet[][hereafter \citetalias{pancoast11}]{pancoast11} to explore and constrain a phenomenological description of the BLR that is consistent with the reverberation mapping dataset. In this approach, using the Code for AGN Reverberation and Modeling of Emission Lines (\textsc{caramel}), the BLR emissivity is described in simple but flexible terms, allowing one to capture the key features expressed in the data in a statistically rigorous way. The posterior probability distribution function of parameters describing the geometry and kinematics of the line emissivity are derived through a diffusive nested sampling process. The parameter uncertainties account for the inevitable modeling approximation as described by \citeauthor{pancoast11} (\citeyear{pancoast11}, \citeyear{pancoast14b}) and briefly summarized in this paper.

This phenomenological model allows us to learn more about the BLR and has been applied to the low-ionization H$\beta$-emitting BLR of a total of 18 AGNs --- five from the Lick AGN Monitoring Project 2008 \citep[LAMP 2008;][hereafter \citetalias{pancoast14b}]{pancoast14b}, four from a 2010 AGN monitoring campaign at MDM Observatory \citep[AGN10;][hereafter \citetalias{Grier++17}]{Grier++17}, seven from the Lick AGN Monitoring Project 2011 \citep[LAMP 2011;][hereafter \citetalias{2018ApJ...866...75W}]{2018ApJ...866...75W}, one from the Space Telescope and Optical Reverberation Mapping Project \citep[AGNSTORM;][hereafter \citetalias{2020ApJ...902...74W}]{2020ApJ...902...74W}\footnote{NGC 5548 was previously modeled using data from the LAMP 2008 campaign. Modeling data from the  AGNSTORM campaign yields the same black hole mass but different geometry of the BLR. This is not surprising, as different aspects of the BLR  (luminosity and average size, for example) are known to vary over timescales of a few years \citep[see, e.g.,][]{2015ApJ...806..128D, 2018ApJ...856..108P, 2021ApJ...922..151K}. It is thus interesting to include the results of both campaigns in our analysis.}, and one from a monitoring campaign at Siding Spring Observatory \citep[SSO;][hereafter \citetalias{bentz2021detailed}]{bentz2021detailed}. These analyses found that the H$\beta$-emitting BLR is best described by a thick disk at a low to moderate inclination to our line of sight with near-circular Keplerian orbits and a contribution of inflow (with some outflow found by \citetalias{2018ApJ...866...75W}). 

In an attempt to gain further insight on the H$\beta$-emitting BLR structure and kinematics, we have expanded the sample of dynamically modeled AGNs from 18 to 27 by analyzing velocity-resolved reverberation mapping data for nine AGNs from the Lick AGN Monitoring Project 2016 campaign \citep[LAMP 2016;][]{u2021lick}. This paper is organized as follows. We describe our photometric and spectroscopic campaigns and briefly summarize the BLR model from~\citetalias{pancoast11} in Section~\ref{sect:data}.  Section~\ref{sec:results} presents the \textsc{caramel} BLR structure and kinematics of the nine LAMP 2016 sources modeled. With these results, we compare our model kinematics to those inferred by \citet{u2021lick} using traditional velocity-delay maps in Section~\ref{sec: discussion}. Finally, we combine our results with previous studies to create an extended sample that covers more than two decades in luminosity, and investigate luminosity-dependent trends and line-profile shape, e.g., $\log_{10}(\rm{FWHM}/\sigma)$, dependence on BLR structure and kinematics. We summarize our main conclusions in Section \ref{sec: summary}.

Throughout the paper, we have adopted H$_0 = 67.8 \rm{~km~s}^{-1}\rm{ ~Mpc}^{-1}$, 
$\Omega_{\rm m}$ = 0.308, and $\Omega_{\rm vac} = 0.692$ \citep{Planck16}.
\section{Data and Methods}
\label{sect:data}
\subsection{Photometric and Spectroscopic Data}
%\subsection{Data}
A detailed description of the photometric and spectroscopic monitoring data is provided by \citet{u2021lick}. In summary, \textit{V}-band photometric monitoring was carried out from February 2016 to May 2017 using a network of telescopes around the world, including the 0.76\,m Katzman Automatic Imaging Telescope \citep[KAIT;][]{filippenko01} and the Nickel telescope at Lick Observatory  on Mount Hamilton east of San Jose, California; the Las Cumbres Observatories Global Telescope (LCOGT) network \citep[]{Brown++13,2014SPIE.9149E..1EB}; the Liverpool Telescope at the Observatorio del Roque de Los Muchachos on the Canary island of La Palma, Spain \citep{2004SPIE.5489..679S}; the 1\,m Illinois Telescope at Mount Laguna Observatory (MLO) in the Laguna Mountains east of San Diego, California; the San Pedro M\'{a}rtir Observatory (SPM) 0.84\,m telescope at the Observatory Astron\'{o}mico Nacional located in Baja California, M\'{e}xico; the Fred Lawrence Whipple Observatory 1.2\,m telescope on Mount Hopkins, Arizona; and the 0.9\,m West Mountain Observatory (WMO) Telescope at the southern end of Utah Lake in Utah. Spectroscopic monitoring was carried out with the Kast double spectrograph on the 3\,m Shane telescope at Lick Observatory from 28 April 2016 to 6 May 2017; originally allocated 100 nights, a substantial fraction ($\sim 30 \%$) were unfortunately lost owing to poor weather. The total number of epochs for each object analyzed in this work can be found in Table \ref{table_agnproperties}. 

\begin{deluxetable*}{llcccccc}
\setlength{\tabcolsep}{7.5pt}
\tablecaption{AGN Observation Properties}
\tablehead{ 
\colhead{Galaxy} & 
\colhead{Alt. Name} &
\colhead{Redshift } &
\colhead{Monitoring Period} &
\colhead{$N_{\rm spec}$}  &
\colhead{S/N} &
\colhead{$N_{\rm phot}$} \\
\colhead{}&
\colhead{}&
\colhead{}&
\colhead{(UT)}&
\colhead{}&
\colhead{}}
\startdata
\pga & II Zw 171 & 0.07000 & 20160501$-$20161231 &40 & 32 & 9    \\ 
\rbsa & 2MASX J22563642+0525167 & 0.06600 & 20160601$-$20161231 & 32 & 39 & 9  \\
\mcga & & 0.03235 & 20160501$-$20170501 & 34 & 54 & 7   \\
NPM1G$+27.0587$ & 2MASX J18530389+2750275 &  0.06200 & 20160501$-$20161203 & 38 & 55 & 7 \\
%\mrkb & PG 0921+525 & 0.03529 & 20160501$-$20170501 & 41 & 74 & 6   \\ 
\mrka & $1505+0342$ & 0.03614 & 20160501$-$20170501 & 39 & 55 & 10 \\
\rbsb & CGS R14.01 & 0.04179 & 20160501$-$20170501 &22 & 67 & 5  \\
Mrk 1048 & NGC 985, VV 285 & 0.04314 & 20160808$-$20170216 & 27 & 88 & 5 \\
\rxja & & 0.05000 & 20160501$-$20161231 & 46 & 58 & 9 \\
\mrke & J15040+1026 & 0.03642 & 20160501$-$20170501 & 45 & 77 & 11  
%\ark & UGC 03271, Mrk 1095 & 0.3271 & 20160815$-$20170326 & 34 & 103 & 6 \\
%Mrk 9 & & 0.03987 & 20160501$-$20170501 & 33 & 78 & 5 \\ 
\enddata
\tablecomments{Observing information for the AGNs modeled in this work. The redshifts are from \citet{u2021lick}. $N_{\rm{spec}}$ represents the total number of spectroscopic observations for each source. S/N represents the median signal-to-noise ratio per pixel in the \Hb spectrum in the continuum at $(5100-5200)~(1+z)~\rm{\AA}$. $N_{\rm{phot}}$ represents the number of photometric nights for each source.} 
\label{table_agnproperties}
\end{deluxetable*} 

In total, 21 AGNs were observed during the campaign. Of those, nine had sufficient quality and continuum/\Hb variability for the analysis conducted in this paper. The reader is referred to \citet{u2021lick} for the full list of AGNs observed during this campaign. 

\begin{figure*}
    \centering
    \includegraphics[height= 12.5cm, keepaspectratio]{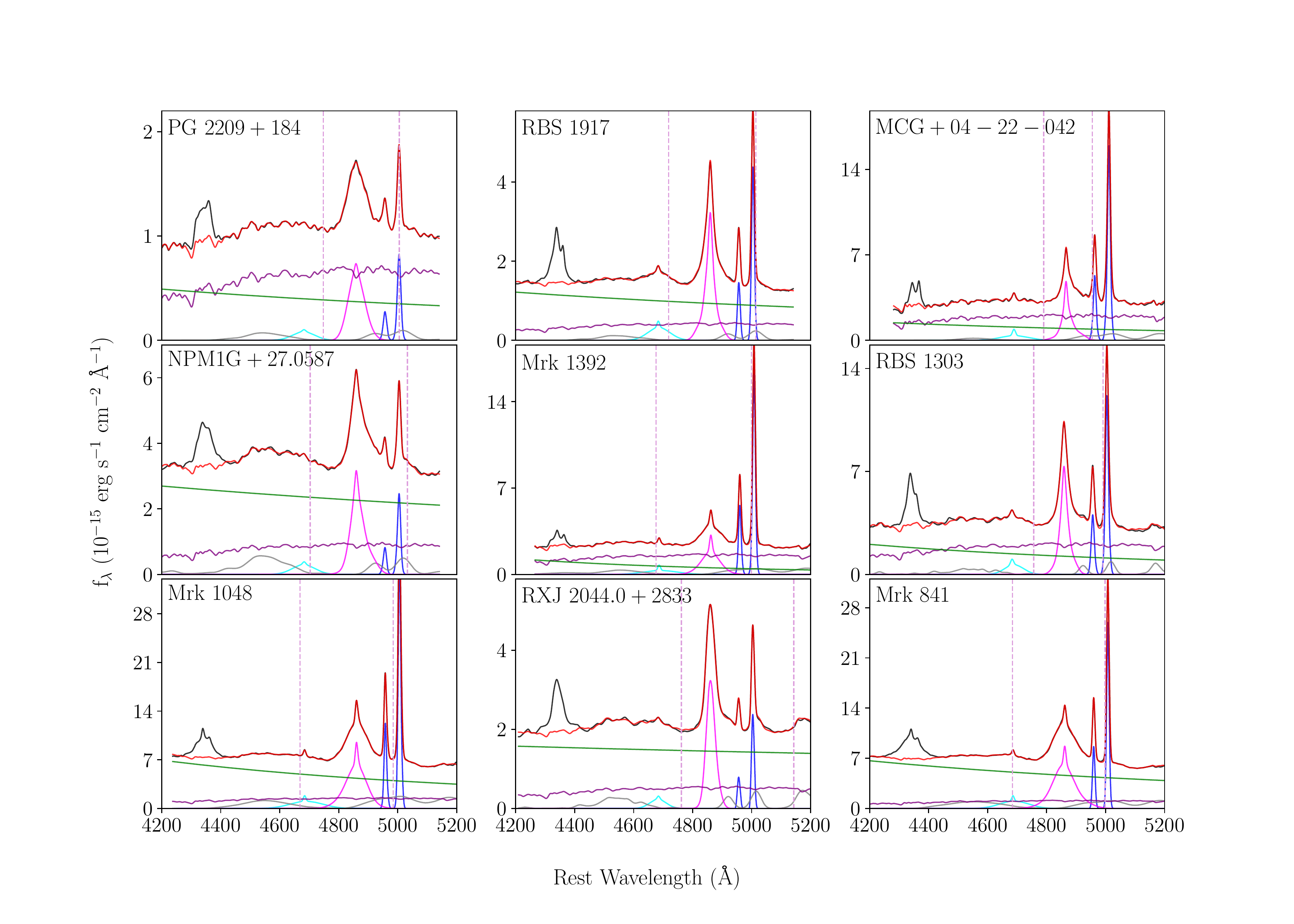}
    \caption{Spectral decomposition using the  K10 \ion{Fe}{2} template. The observed mean spectrum (black) for each galaxy is plotted alongside the decomposed model components: starlight (purple), AGN power-law continuum (green), \Hb $\lambda 4861$ (magenta), \ion{He}{2} $\lambda 4686$ (cyan), \ion{He}{1} $\lambda 5876$ (orange), \ion{Fe}{2} $\lambda(4500-5400)$ (grey), and [\ion{O}{3}] $\lambda5007$ (blue). The sum of the fit of these components is shown in red and the vertical dashed lines indicate the wavelength range used for fitting the H$\beta$-emitting BLR model.}
    \label{fig:spec_decomps}
\end{figure*}

To model the broad \Hb emission line, we must disentangle it from other features in the AGN spectrum, such as the  \ion{He}{1}, \ion{He}{2}, \ion{Fe}{2}, and [\ion{O}{3}] emission lines, the AGN continuum, and starlight. Our team isolates contributions of individual emission lines and continuum components within the vicinity of the \Hb emission line by fitting a multicomponent model to each night's spectrum (see Figure \ref{fig:spec_decomps}). A summary of the procedure, adopted from \citet{barth15} and used on the LAMP 2016 sample, is given by \citet{u2021lick}. 
\subsection{BLR Model}
\label{sec: model}
We model the H$\beta$-emitting BLR of each source using \textsc{caramel}, a phenomenological modeling code described in detail by \citetalias{pancoast11} and \citetalias{pancoast14b}. \textsc{caramel} models the BLR emission by sampling it with a distribution of test point particles surrounding the black hole located at the origin. Gravity is assumed to be the dominant force (i.e., radiation pressure is neglected). When ionizing light emitted from the central black hole reaches a particle, the particle instantaneously re-emits an emission line and the \textsc{caramel} model free parameters determine whether the re-emission is isotropic. The spatial distribution of the particles determines the associated time delay, while the line-of-sight velocity distribution determines the shape of the broad emission line profile. The spatial and velocity distributions of the point particles are constrained by a number of model parameters described by \citetalias{pancoast11} and \citetalias{pancoast14b}. Here we summarize some of the main parameters. 
\subsubsection{Geometry}
The spatial distribution of particles is described by angular and radial components. The radial distribution is drawn from a gamma distribution with shape parameter $\beta$ and mean $\mu$ that has been shifted from the origin by a minimum radius $r_{\rm{min}}$. Spherical symmetry is broken by introducing an opening-angle parameter, $\theta _o$, which can be interpreted as disk thickness with $\theta _o \rightarrow 0^{\circ}$ describing a razor-thin disk and $\theta_o \rightarrow 90^{\circ}$  describing a sphere. Inclination to the observer's line of sight is set by an inclination angle $\theta_i$, with $\theta_i=0^{\circ}$ representing a face-on view and $\theta_i=90^{\circ}$ representing an edge-on view. Three additional parameters ($\gamma$, $\xi$, and $\kappa$) allow for further asymmetry.

The extent to which the  emission is concentrated near the outer edges of the BLR is then determined by $\gamma$, which ranges in values from 1 to 2. A uniform distribution throughout the disk is described by $\gamma \rightarrow 1$ and a clustered distribution at the outer edges of the BLR disk is described by $\gamma \rightarrow 2$. The parameter $\xi$ permits obscuration along the midplane of the disk, with $\xi \rightarrow 0$ interpreted as a fully obscured (opaque) midplane and $\xi = 1$ as a fully transparent midplane (i.e., no obscuration). The parameter $\kappa$ is related to the relative brightness of each particle and controls how the continuum flux is radiated toward the observer as emission-line flux. While $\kappa = 0 $ represents isotropic emission,  $\kappa = -0.5$ represents preferential emission toward the origin (back toward the ionizing source) and $\kappa = 0.5$ represents preferential emission away from the origin (and away from the ionizing source). An observer viewing from $+ \infty$ along the \textit{x}-axis would view the first as preferential emission from the far side of the BLR, and the latter as preferential emission from the near side of the BLR. Preferential emission from the far side of the BLR can be interpreted as a result of self-shielding particles or obscuration (of the near-side BLR) by the torus, causing the BLR gas to appear to only re-emit back toward the ionizing source. Preferential emission from the near side might be due to an obstructed view of the far side of the BLR. 
\subsubsection{Dynamics}
Following the construction of the spatial distribution of particles, the BLR kinematics are then determined with a number of additional parameters. The fraction of particles with near-circular Keplerian orbits around the central black hole with mass $M_{\rm{BH}}$ is given by the $\mathit{f}_{\rm{ellip}}$ parameter. The remaining particles $(1 - f_{\rm{ellip}})$ are either inflowing ($\mathit{f}_{\rm{flow}} < 0.5$) or outflowing ($\mathit{f}_{\rm{flow}} > 0.5$) with velocities drawn from a Gaussian distribution centered on the radial escape velocity in the $v_r - v_{\phi}$ plane rotated by an angle, $\theta_e$, away from escape velocity toward circular velocity. Therefore $\theta_e \rightarrow 90^{\circ}$ indicates nearly circular orbits, $\theta \approx 45^{\circ}$ highly eccentric orbits, and $\theta_e \rightarrow 0^{\circ}$ a majority of particles are approaching escape velocity and are nearly unbound. 

Finally, we add the line-of-sight velocity component, $v_{\rm{turb}}$, of a randomly-orientated macroturbulent velocity vector to the particle's line-of-sight velocity. This macroturbulent contribution is drawn from a Gaussian distribution, $\mathcal{N}$, centered on 0 with standard deviation $\sigma_{\rm{turb}}$ and is dependent on the particle's circular velocity $v_{\rm{circ}}$:

\begin{equation}
    v_{\rm{turb}} = \mathcal{N}(0,\sigma_{\rm{turb}})|v_{\rm{circ}}| ,
\end{equation}

where $\sigma_{\rm{turb}}$ is the free parameter that determines the amount of contribution from macroturbulent velocities and can range from 0.001 to 0.1.

For simplicity, we summarize the BLR dynamics by an ``In.$-$Out." parameter as defined by \citetalias{2018ApJ...866...75W}, where values of $1$ indicate pure radial outflow and $-1$ indicate pure radial inflow:

\begin{equation}
    \rm{In.-Out.}= \rm{sgn}(\mathit{f}_{\rm{flow}}-0.5)\times (1-\mathit{f}_{\rm{ellip}})\times \cos({\theta_e}) .
\end{equation}

\subsubsection{Continuum Model and Implementation}
In order to use the parameterized spatial and velocity distributions described above to calculate the resulting broad emission-line profile at arbitrary times, we need an input continuum light curve that can \textit{also} be sampled at arbitrary times. To do this, we model the AGN continuum using Gaussian processes. This allows us to both interpolate between photometric measurements and extrapolate beyond the monitoring campaign, as well as propagate the associated uncertainties into the determination of the BLR model parameters. By combining the modelled continuum light curve with the BLR model parameters, a broad emission line profile can be produced for each spectroscopic epoch observed during the monitoring campaign. 

This last step requires the application of a smoothing parameter to account for minor differences in spectral resolution throughout the observational campaign, due to variable seeing conditions, for example. We assume that the narrow [\ion{O}{3}] emission line remains intrinsically constant throughout our monitoring campaign and use it to calibrate the smoothing parameter by comparing its measured width to its width taken from \citet{whittle92}. We then use this smoothing parameter to blur the modeled spectrum and combine with the modeled particle dynamics to produce the \Hb emission-line profile. 

Once the model emission-line profile is produced, we use a Gaussian likelihood function to compare the resulting spectra with the observed spectra, and adjust the BLR model parameters accordingly. We explore the model parameter space using \textsc{dnest4} \citep{brewer11b}, a diffusive nested sampling code that allows one to apply a likelihood softening parameter post-analysis. This parameter is a statistical ``temperature,'' $T$, which allows us to account for systematic uncertainty by increasing measurement uncertainty, as well as account for our simple model's inability to capture all the real details.  We select a value for $T$ that avoids overfitting while still achieving the highest levels of likelihood. 

\subsubsection{Model Limitations} 
\label{sec: model_limitations}
Before proceeding onto the discussion of our results, we would like to reiterate that \textsc{caramel} models the \textit{BLR emission}, rather than the underlying BLR gas distribution producing the emission lines. Our model does not include photoionization processes. Doing so would require additional assumptions about the gas density, temperature, metallicity distribution, and the relation between the observed \textit{V}-band continuum and the ionizing spectrum. Therefore, the interpretation of the model parameters discussed below in Section \ref{sec:results} is a reflection of the \Hb BLR \textit{emission}, rather than the underlying gas producing the emission lines. 

Additionally, our model is currently set up to only account for gravitational effects from the central BH and does not take into account the effects of radiation pressure. This is important to keep in mind when interpreting model results for high Eddington ratio AGNs. We note that the sources modeled in this work have moderate luminosities, with extinction-corrected $\log[\lambda L_{\lambda}(5100 \rm{\AA})/L_{\odot}]\approx 43.5$--43.9 \citep{u2021lick}. The precise Eddington ratio is difficult to determine, however, since bolometric correction factors may depend on the true Eddington ratio and other parameters. For the purposes of this work, we remain consistent in our calculations and apply the same bolometric correction factor as our prior studies \citepalias{pancoast14b,Grier++17,2018ApJ...866...75W}. Within our extended sample, we find that the Eddington ratio of the LAMP 2016 sources can also be considered moderate when compared to prior studies (e.g., \citetalias{Grier++17}), but find this model limitation worth noting as neglecting radiation pressure can potentially lead to biased results for sources with high Eddington ratio.

For further discussion of model limitations and the model improvements that are currently underway, the reader is referred to \citet{2020MNRAS.493.1227R}. 
\subsection{Searching for Trends with BLR Structure and Kinematics} \label{sec:trends}
In addition to learning more about the H$\beta$-emitting BLR, a primary goal of this program is to investigate the existence of any systematic trends in AGN BLR structure and kinematics. This is part our team's long-standing goal to gain insight on the nature of the BLR through our dynamic modeling approach, and ultimately improve BH mass estimators \citep[see][for  the latter]{v2022}. In this work, we specifically search for luminosity-dependent and line-profile shape dependency on BLR structure/dynamics.

We use the \textsc{idl} routine \texttt{linmix\_err} \citep{Kelly07} to perform a Bayesian linear regression in order to account for correlated measurement uncertainties. Doing so allows us to analyze the actual intrinsic correlation with any two parameters without worrying about a false increase due to correlated measurement uncertainties. This is especially important for our search for correlations with scale factor since individual scale factors are determined using our model black hole mass estimate, and therefore its uncertainties are connected to uncertainties in other model parameters.

To quantify the strength of any correlation, we compare the median fit slope to the $1\sigma$ uncertainty in the slope and determine our level of confidence using the following intervals we have defined previously (\citetalias{2018ApJ...866...75W}). We classify 0--2$\sigma$ as no evidence, 2--3$\sigma$ as marginal evidence, 3--5$\sigma$ as evidence, and $>5\sigma$ as conclusive evidence.

\movetabledown=20mm
\addtolength{\tabcolsep}{+4.5pt}
\begin{longrotatetable}
\begin{deluxetable*}{lccccccccccccc}
\tablecaption{BLR Model Parameter Values}
\tablehead{ 
\colhead{Parameter} &
\colhead{${\rm PG ~ 2209+184}$} & %pga 
\colhead{${\rm RBS ~ 1917}$} & %rbsa
\colhead{${\rm MCG+04-22-042}$} & %mcga
\colhead{${\rm NPM1G+27.0587}$} & %npmg
\colhead{${\rm Mrk ~ 1392}$} & %mrka 
\colhead{${\rm RBS ~ 1303}$} & %rbsb
\colhead{${\rm Mrk ~ 1048}$} & %mrkd
\colhead{${\rm RXJ ~ 2044.0+2833}$} & %rxja
\colhead{${\rm Mrk ~ 841}$} %mrke
}
\startdata
$\log_{10}(M_{\rm bh}/M_{\odot})$  & \pgalogmbh & \rbsalogmbh & \mcgalogmbh & \npmglogmbh & \mrkalogmbh & \rbsblogmbh & \mrkdlogmbh & \rxjalogmbh & \mrkelogmbh \\
%$7.53^{+0.19}_{-0.20}$ & $7.04^{+0.23}_{-0.35}$ & $7.59^{+0.42}_{-0.28}$ & $7.64^{+0.40}_{-0.36}$ & $8.16^{+0.11}_{-0.13}$ & $6.79^{+0.19}_{-0.11}$ & $7.79^{+0.44}_{-0.48}$ & $7.09^{+0.17}_{-0.17}$ & $7.62^{+0.50}_{-0.30}$\\
$r_{\rm mean}$ (light-days) & \pgarmean & \rbsarmean & \mcgarmean & \npmgrmean & \mrkarmean & \rbsbrmean & \mrkdrmean & \rxjarmean & \mrkermean \\
%$16.27^{+1.0}_{-0.95}$ & $9.3^{+1.4}_{-1.1}$ & $9.9^{+1.2}_{-0.98}$ & $12.2^{+4.3}_{-3.5}$ & $56.0^{+12}_{-7.2}$ & $12.9^{+1.3}_{-1.2}$ & $14.2^{+7.8}_{-7.6}$ & $35.7^{+7.8}_{-6.3}$ & $14.1^{+7.1}_{-4.4}$\\
$r_{\rm median}$ (light-days)  & \pgarmedian & \rbsarmedian & \mcgarmedian & \npmgrmedian & \mrkarmedian & \rbsbrmedian & \mrkdrmedian & \rxjarmedian & \mrkermedian \\ 
%$15.2^{+1.1}_{-1.0}$ & $5.0^{+1.3}_{-1.1}$ & $6.24^{+1.01}_{-0.87}$ & $7.2^{+2.8}_{-2.0}$ & $51.6^{+12}_{-8.6}$ & $10.1^{+1.3}_{-1.2}$ & $11.3^{+7.3}_{-6.2}$ & $28.3^{+7.5}_{-5.4}$ & $10.6^{+5.6}_{-3.4}$\\
$r_{\rm min}$ (light-days)  & \pgarmin & \rbsarmin & \mcgarmin & \npmgrmin & \mrkarmin & \rbsbrmin & \mrkdrmin & \rxjarmin & \mrkermin \\
%& $2.8^{+1.4}_{-1.6}$ & $1.42^{+0.39}_{-0.40}$ & $1.05^{+0.65}_{-0.58}$ & $2.69^{+1.6}_{-0.95}$ & $41^{+11}_{-12}$ & $0.25^{+0.24}_{-0.18}$ & $3.3^{+3.0}_{-2.1}$ & $4.2^{+1.5}_{-1.1}$ & $2.0^{+1.8}_{-1.1}$\\
$\sigma_r$ (light-days)  & \pgasigmar & \rbsasigmar & \mcgasigmar & \npmgsigmar & \mrkasigmar & \rbsbsigmar & \mrkdsigmar & \rxjasigmar & \mrkesigmar \\
%$22^{+12}_{-8.3}$ & $27^{+25}_{-12}$ & $21.5^{+35}_{-9.7}$ & $22.0^{+18}_{-9.2}$ & $31^{+54}_{-18}$ & $13.6^{+3.8}_{-2.3}$ & $14.1^{+22}_{-8.7}$ & $55^{+45}_{-17}$ & $14.7^{+15}_{-5.8}$\\
$\tau_{\rm mean}$ (days)  & \pgataumean & \rbsataumean & \mcgataumean & \npmgtaumean & \mrkataumean & \rbsbtaumean & \mrkdtaumean & \rxjataumean & \mrketaumean \\ 
%$15.75^{+0.74}_{-0.77}$ & $9.0^{+1.9}_{-1.7}$ & $9.17^{+0.95}_{-0.81}$ & $10.7^{+3.1}_{-2.8}$ & $38.7^{+4.6}_{-4.2}$ & $13.7^{+1.3}_{-1.3}$ & $11.5^{+6.6}_{-6.0}$ & $30.2^{+4.8}_{-5.3}$ & $13.5^{+4.6}_{-3.8}$\\
$\tau_{\rm median}$ (days)  & \pgataumedian & \rbsataumedian & \mcgataumedian & \npmgtaumedian & \mrkataumedian & \rbsbtaumedian & \mrkdtaumedian & \rxjataumedian & \mrketaumedian \\ 
%& $12.95^{+0.87}_{-0.88}$ & $4.6^{+1.2}_{-1.2}$ & $5.58^{+0.81}_{-0.70}$ & $6.0^{+2.0}_{-1.6}$ & $34.8^{+4.7}_{-4.6}$ & $9.8^{+1.2}_{-1.1}$ & $8.2^{+6.2}_{-4.5}$ & $18.8^{+3.5}_{-3.9}$ & $8.9^{+3.3}_{-2.5}$\\
$\beta$  & \pgabeta & \rbsabeta & \mcgabeta & \npmgbeta & \mrkabeta & \rbsbbeta & \mrkdbeta & \rxjabeta & \mrkebeta \\ 
%$0.88^{+0.14}_{-0.15}$ & $1.63^{+0.13}_{-0.16}$ & $1.40^{+0.20}_{-0.19}$ & $1.54^{+0.13}_{-0.13}$ & $1.29^{+0.51}_{-0.74}$ & $0.94^{+0.07}_{-0.07}$ & $1.12^{+0.28}_{-0.28}$ & $1.12^{+0.08}_{-0.08}$ & $1.08^{+0.18}_{-0.16}$\\
$\theta_o$ (degrees)  & \pgathetao & \rbsathetao & \mcgathetao & \npmgthetao &  \mrkathetao & \rbsbthetao & \mrkdthetao & \rxjathetao & \mrkethetao \\
%$29.1^{+11}_{-8.4}$ & $25.1^{+9.2}_{-7.4}$ & $13.6^{+6.9}_{-4.9}$ & $17.7^{+11.2}_{-9.0}$ & $41.2^{+5.2}_{-4.8}$ & $34.0^{+8.9}_{-10}$ & $31^{+14}_{-9.9}$ & $51^{+15}_{-12}$ & $41^{+11}_{-11}$\\
$\theta_i$ (degrees)  & \pgathetai & \rbsathetai & \mcgathetai & \npmgthetai &  \mrkathetai & \rbsbthetai & \mrkdthetai & \rxjathetai & \mrkethetai \\
%$30.2^{+8.7}_{-6.9}$ & $20.2^{+9.9}_{-3.9}$ & $11.3^{+5.8}_{-5.0}$ & $19.1^{+11}_{-8.5}$ & $25.5^{+3.4}_{-2.8}$ & $29.1^{+7.7}_{-9.0}$ & $21.5^{+9.4}_{-9.4}$ & $43^{+9.6}_{-8.4}$ & $30^{+11}_{-15}$\\
$\kappa$  & \pgakappa & \rbsakappa & \mcgakappa & \npmgkappa & \mrkakappa & \rbsbkappa & \mrkdkappa & \rxjakappa & \mrkekappa \\
%$-0.09^{+0.12}_{-0.15}$ & $-0.29^{+0.35}_{-0.14}$ & $-0.14^{+0.44}_{-0.27}$ & $-0.14^{+0.40}_{-0.25}$ & $0.26^{+0.18}_{-0.25}$ & $-0.48^{+0.05}_{-0.01}$ & $0.10^{+0.28}_{-0.38}$ & $-0.20^{+0.33}_{-0.19}$ & $-0.23^{+0.43}_{-0.14}$\\
$\gamma$  & \pgagamma & \rbsagamma & \mcgagamma & \npmggamma & \mrkagamma & \rbsbgamma & \mrkdgamma & \rxjagamma & \mrkegamma \\
%$1.4^{+0.4}_{-0.3}$ & $1.5^{+0.3}_{-0.3}$ & $1.6^{+0.3}_{-0.4}$ & $1.4^{+0.4}_{-0.3}$ & $1.5^{+0.3}_{-0.3}$ & $1.8^{+0.1}_{-0.2}$ & $1.5^{+0.3}_{-0.3}$ & $1.4^{+0.4}_{-0.3}$ & $1.4^{+0.4}_{-0.3}$\\
$\xi$  & \pgaxi & \rbsaxi & \mcgaxi & \npmgxi & \mrkaxi & \rbsbxi & \mrkdxi & \rxjaxi & \mrkexi \\
%$0.73^{+0.16}_{-0.18}$ & $0.68^{+0.25}_{-0.35}$ & $0.43^{+0.35}_{-0.26}$ & $0.11^{+0.37}_{-0.09}$ & $0.25^{+0.28}_{-0.18}$ & $0.60^{+0.22}_{-0.16}$ & $0.30^{+0.42}_{-0.20}$ & $0.17^{+0.28}_{-0.12}$ & $0.68^{+0.23}_{-0.41}$\\
$f_{\rm ellip}$  & \pgaellip & \rbsaellip & \mcgaellip & \npmgellip & \mrkaellip & \rbsbellip & \mrkdellip & \rxjaellip & \mrkeellip \\
%$0.54^{+0.10}_{-0.15}$ & $0.59^{+0.14}_{-0.17}$ & $0.39^{+0.21}_{-0.18}$ & $0.44^{+0.19}_{-0.18}$ & $0.81^{+0.04}_{-0.06}$ & $0.18^{+0.17}_{-0.11}$ & $0.73^{+0.09}_{-0.13}$ & $0.41^{+0.32}_{-0.29}$ & $0.33^{+0.24}_{-0.22}$\\
$f_{\rm flow}$  & \pgaflow & \rbsaflow & \mcgaflow & \npmgflow & \mrkaflow & \rbsbflow & \mrkdflow & \rxjaflow & \mrkeflow \\ 
%$0.24^{+0.17}_{-0.16}$ & $0.59^{+0.28}_{-0.39}$ & $0.27^{+0.18}_{-0.19}$ & $0.26^{+0.18}_{-0.19}$ & $0.74^{+0.18}_{-0.18}$ & $0.75^{+0.17}_{-0.19}$ & $0.74^{+0.18}_{-0.19}$ & $0.22^{+0.19}_{-0.15}$ & $0.45^{+0.36}_{-0.29}$\\
$\theta_e$ (degrees)  & \pgathetae & \rbsathetae & \mcgathetae & \npmgthetae & \mrkathetae & \rbsbthetae & \mrkdthetae & \rxjathetae & \mrkethetae \\
%$24^{+23}_{-16}$ & $20^{+21}_{-15}$ & $19^{+20}_{-13}$ & $36^{+35}_{-24}$ & $25^{+14}_{-15}$ & $8.3^{+8.8}_{-5.8}$ & $15^{+15}_{-10}$ & $34^{+32}_{-21}$ & $51^{+20}_{-27}$\\
${\rm In. - Out.}$  & \pgainout & \rbsainout & \mcgainout & \npmginout & \mrkainout & \rbsbinout & \mrkdinout & \rxjainout & \mrkeinout \\ 
%$-0.40^{+0.09}_{-0.09}$ & $0.24^{+0.21}_{-0.63}$ & $-0.55^{+0.27}_{-0.17}$ & $-0.41^{+0.25}_{-0.18}$ & $0.16^{+0.05}_{-0.03}$ & $0.80^{+0.11}_{-0.18}$ & $0.24^{+0.11}_{-0.09}$ & $-0.37^{+0.17}_{-0.23}$ & $-0.33^{+0.55}_{-0.31}$\\
$\sigma_{\rm turb}$  & \pgasigmaturb & \rbsasigmaturb & \mcgasigmaturb & \npmgsigmaturb & \mrkasigmaturb & \rbsbsigmaturb & \mrkdsigmaturb & \rxjasigmaturb & \mrkesigmaturb \\  %$0.014^{+0.046}_{-0.012}$ & $0.011^{+0.043}_{-0.009}$ & $0.006^{+0.019}_{-0.004}$ & $0.012^{+0.046}_{-0.009}$ & $0.010^{+0.042}_{-0.008}$ & $0.006^{+0.016}_{-0.004}$ & $0.011^{+0.038}_{-0.009}$ & $0.008^{+0.033}_{-0.006}$ & $0.011^{+0.049}_{-0.009}$\\
%$r_{\rm out}$ (light-days) & $48$ & $50$ & $176$ & $114$ & $155$ & $70$ & $76$ & $131$ & $88$ %\\
%$T$  & $25$ & $50$ & $175$ & $75$ & $150$ & $100$ & $500$ & $100$ & $500$\\
\enddata
\tablecomments{Median values and 68\% confidence intervals for BLR model parameters. Note that $r_{\rm{out}}$ is a fixed parameter, so we do not include uncertainties.}
\label{table_results}
\end{deluxetable*}
\end{longrotatetable}

\section{Results}
\label{sec: results }
\label{sec:results}
Of the 21 sources from our full sample, \citet{u2021lick} determine 16 sources to have reliable time lags. Of those 16, nine have sufficient data quality/variability to model using \textsc{caramel}. To verify that our model fits the data, we compare our continuum light curve, the \Hb line profile shape from a randomly selected night, and the resultant modeled integrated \Hb emission line to those observed.

We exclude results for three additional sources whose models were determined to only fit the data with moderate quality. We note, however, that although we chose not to include these results in our extended sample with prior studies, including these sources does not significantly change any findings presented in this work or that of \citet{v2022}. We include the \textsc{caramel} results in Appendix \ref{appendix} for readers who may still be interested in our model description of these three sources. 

\begin{figure*}
    \centering
    \includegraphics[height=11cm, keepaspectratio]{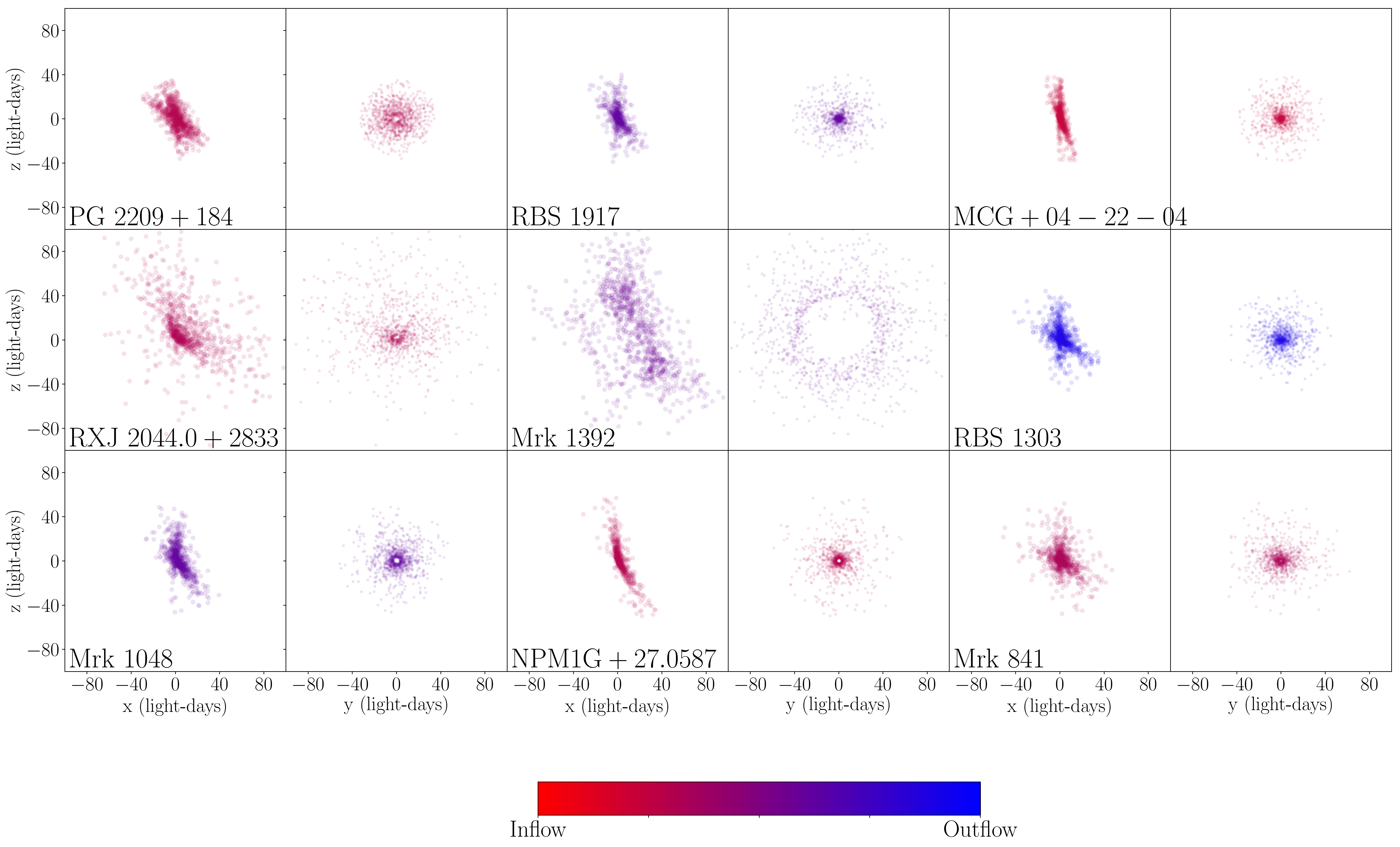}
    \caption{Geometric interpretation of BLR emission for the nine LAMP 2016 modelled sources using median parameter estimates. For each source, the left panel shows an edge-on view while the right panel shows a face-on view. Each circle corresponds to one point particle in the model. The geometries are color-coded to indicate whether the BLR dynamics exhibit inflow (red) or outflow (blue).}
    \label{fig:geo_all}
\end{figure*}

\begin{figure}
    \centering
    \includegraphics[width=\columnwidth, keepaspectratio]{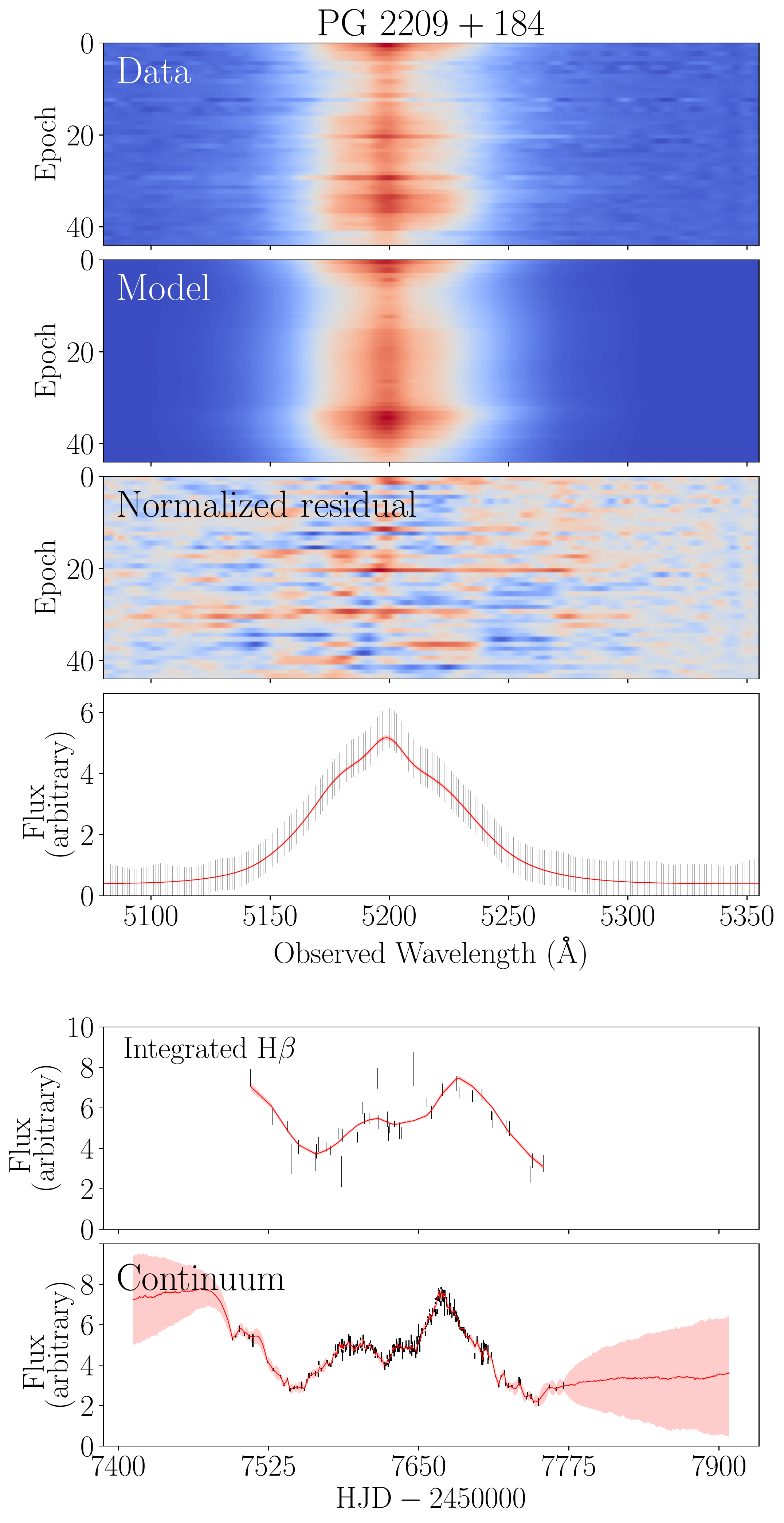}
    \caption{Model fits to the \Hb line profile, integrated \Hb flux, and AGN continuum flux for \pga. Labeling panels 1--6 from top to bottom, panels 1 and 2 show the observed intensity of the \Hb emission-line profile by observation epoch and the profile produced by one sample of the \textsc{caramel} BLR and continuum model. Panel 3 displays the resulting normalized residual. Panel 4 shows the observed \Hb profile of one randomly chosen epoch in black and the corresponding profile produced by the model in panel 2, in red. The corresponding error bars of the observed epoch have been multiplied by $\sqrt{T}$, where $T$ is the \textsc{dnest4} statistical ``temperature" that is used as a likelihood softening parameter post analysis.  Panels 5 and 6 illustrate the time series of the observed integrated \Hb and continuum flux in black and the corresponding model fits (of the model shown in panel 2) of the light curves in red.}
    \label{fig:model_fits_pg2209}
\end{figure}

\begin{figure}
    \centering
    \includegraphics[width=\columnwidth, keepaspectratio]{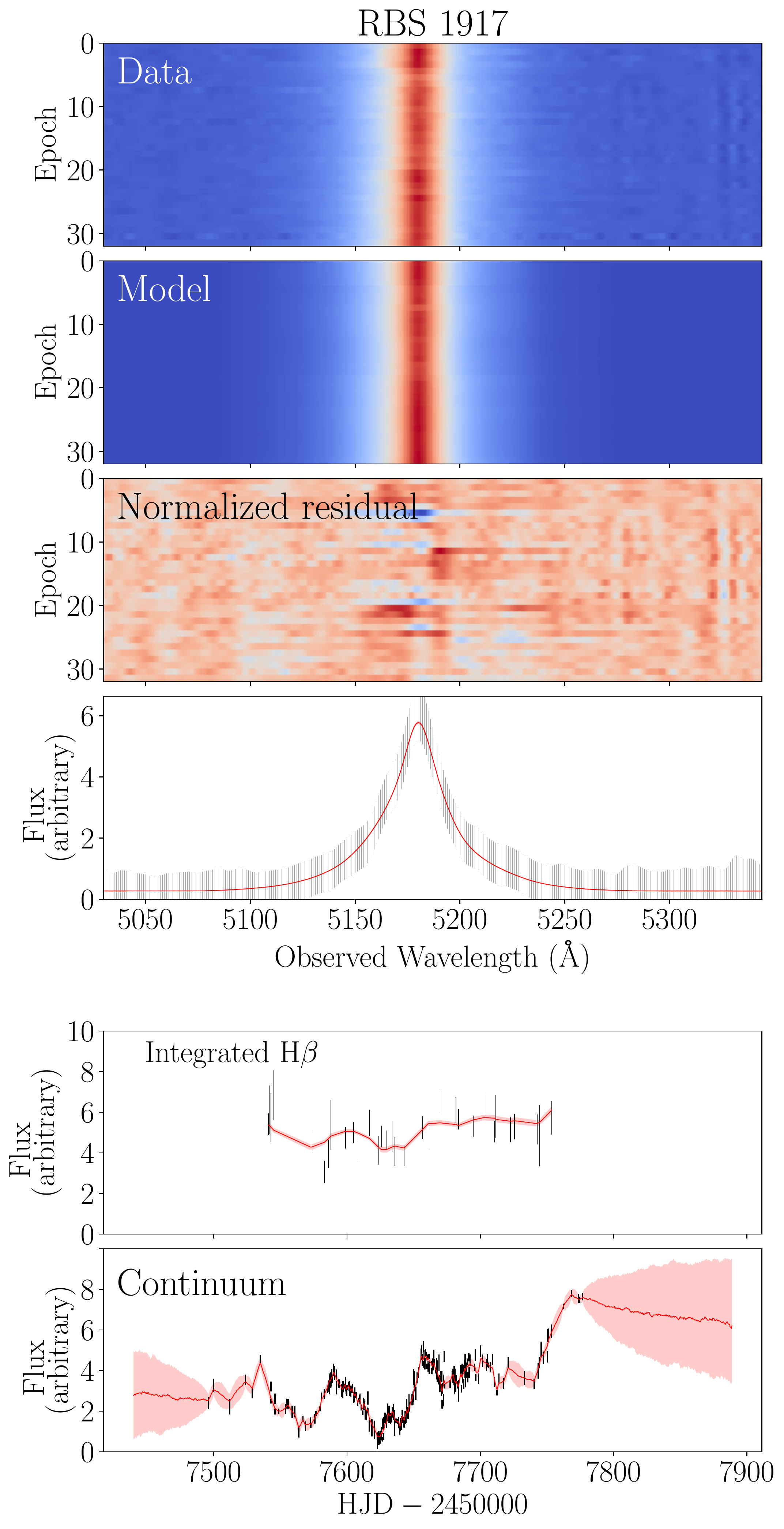}
    \caption{Model fits to the \Hb line profile, integrated \Hb flux, and AGN continuum flux for \rbsa. See Figure \ref{fig:model_fits_pg2209} caption for panel descriptions.}
    \label{fig:model_fits_rbs1917}
\end{figure}

\begin{figure}
    \centering
    \includegraphics[width=\columnwidth, keepaspectratio]{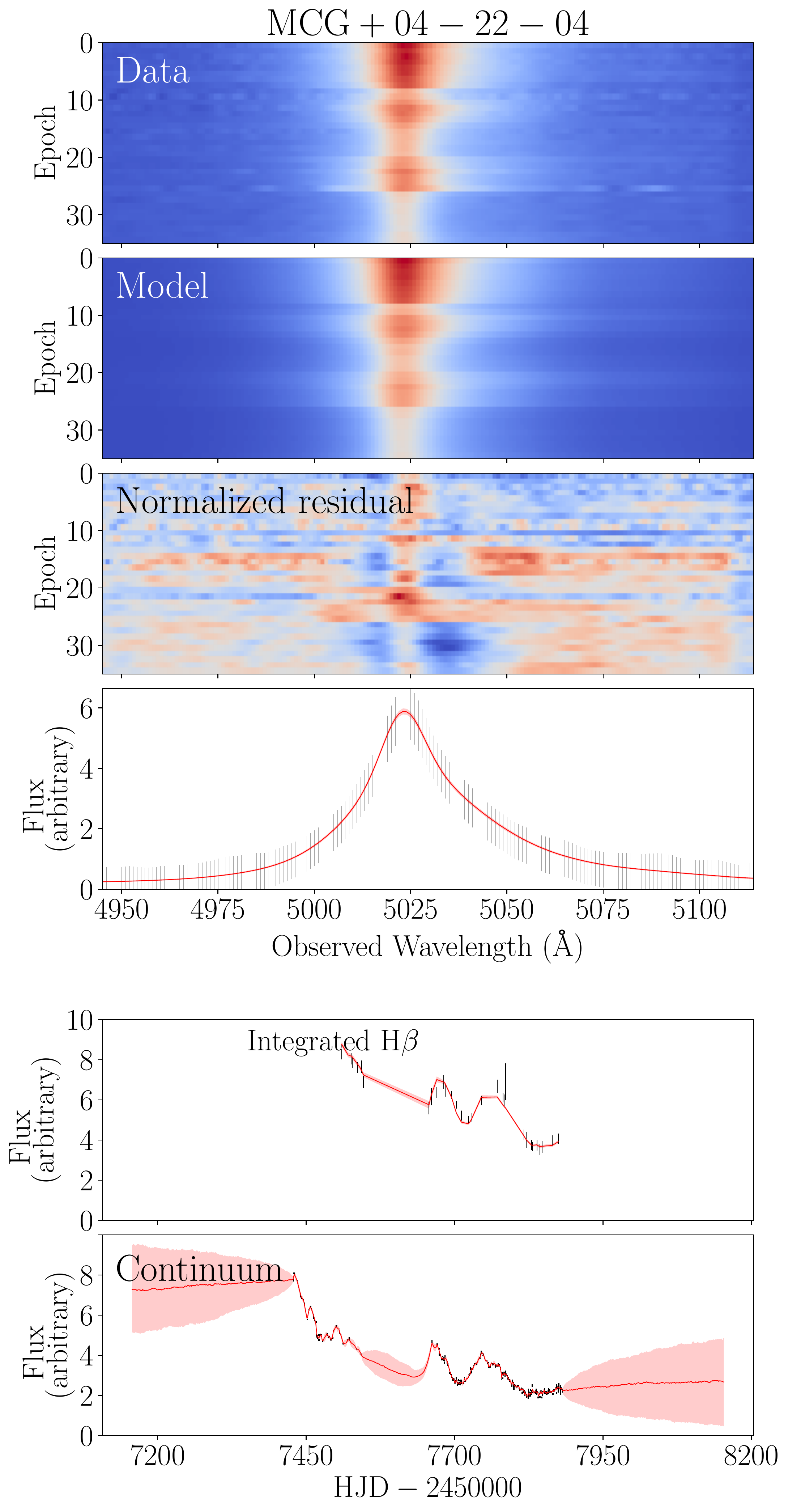}
    \caption{Model fits to the \Hb line profile, integrated \Hb flux, and AGN continuum flux for \mcga. See the Figure \ref{fig:model_fits_pg2209} caption for panel descriptions.}
    \label{fig:model_fits_mcg04}
\end{figure}

\begin{figure}
    \centering
    \includegraphics[width=\columnwidth, keepaspectratio]{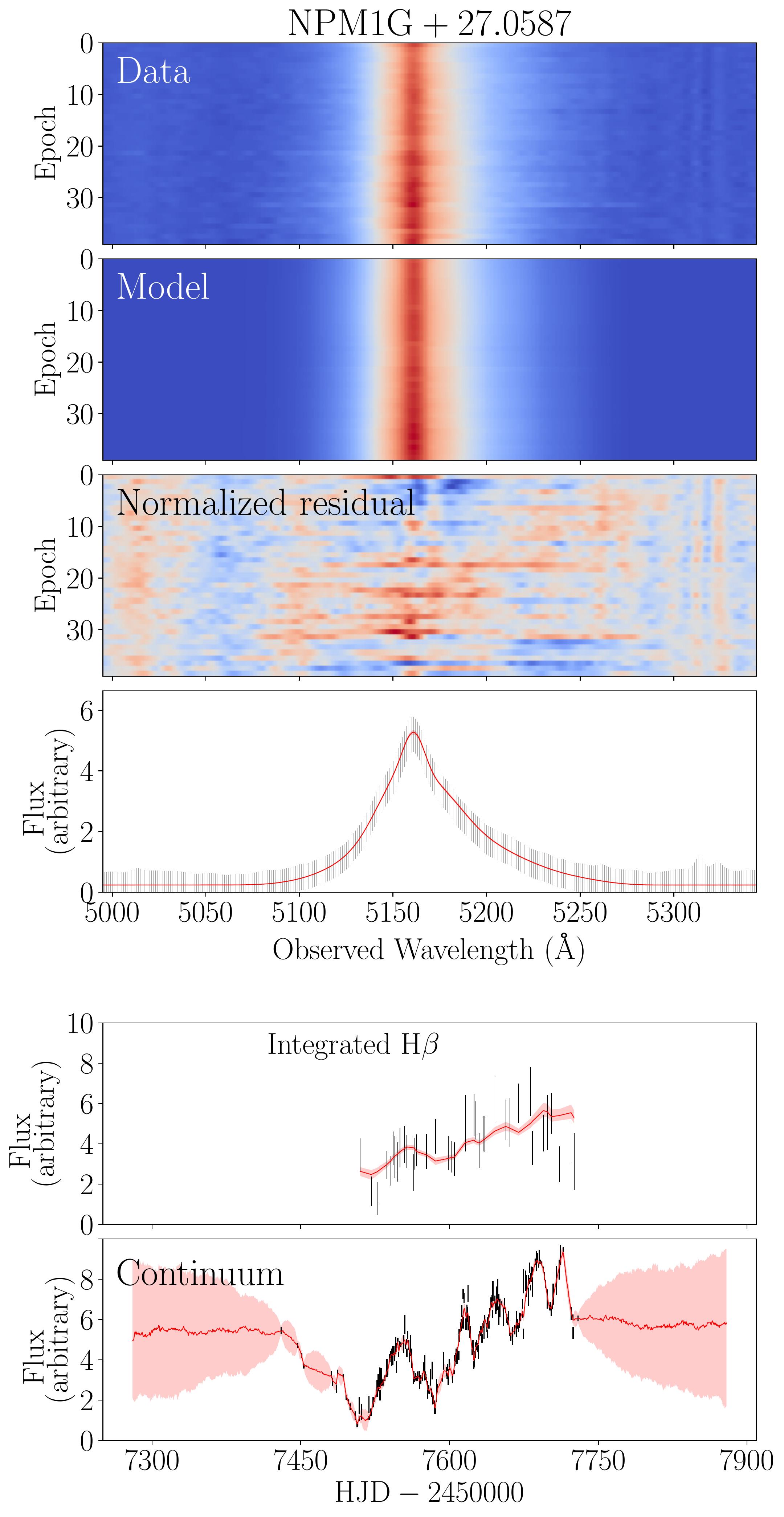}
    \caption{Model fits to the \Hb line profile, integrated \Hb flux, and AGN continuum flux for NPM1G+27.0597. See the Figure \ref{fig:model_fits_pg2209} caption for panel descriptions.}
    \label{fig:model_fits_npm1g}
\end{figure}

\begin{figure}
    \centering
    \includegraphics[width=\columnwidth, keepaspectratio]{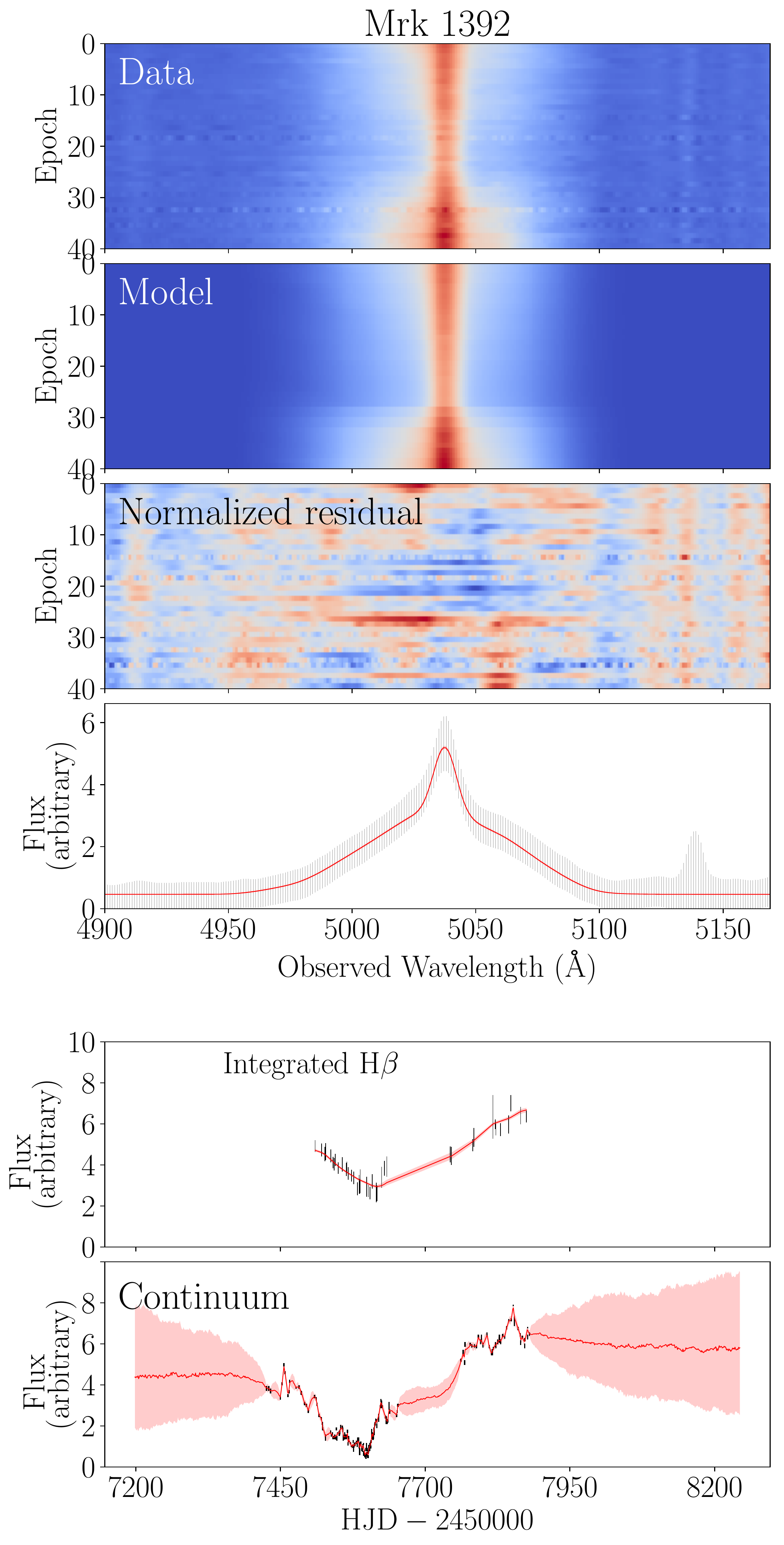}
    \caption{Model fits to the \Hb line profile, integrated \Hb flux, and AGN continuum flux for \mrka. See the Figure \ref{fig:model_fits_pg2209} caption for panel descriptions.}
    \label{fig:model_fits_mrk1392}
\end{figure}

\begin{figure}
    \centering
    \includegraphics[width=\columnwidth, keepaspectratio]{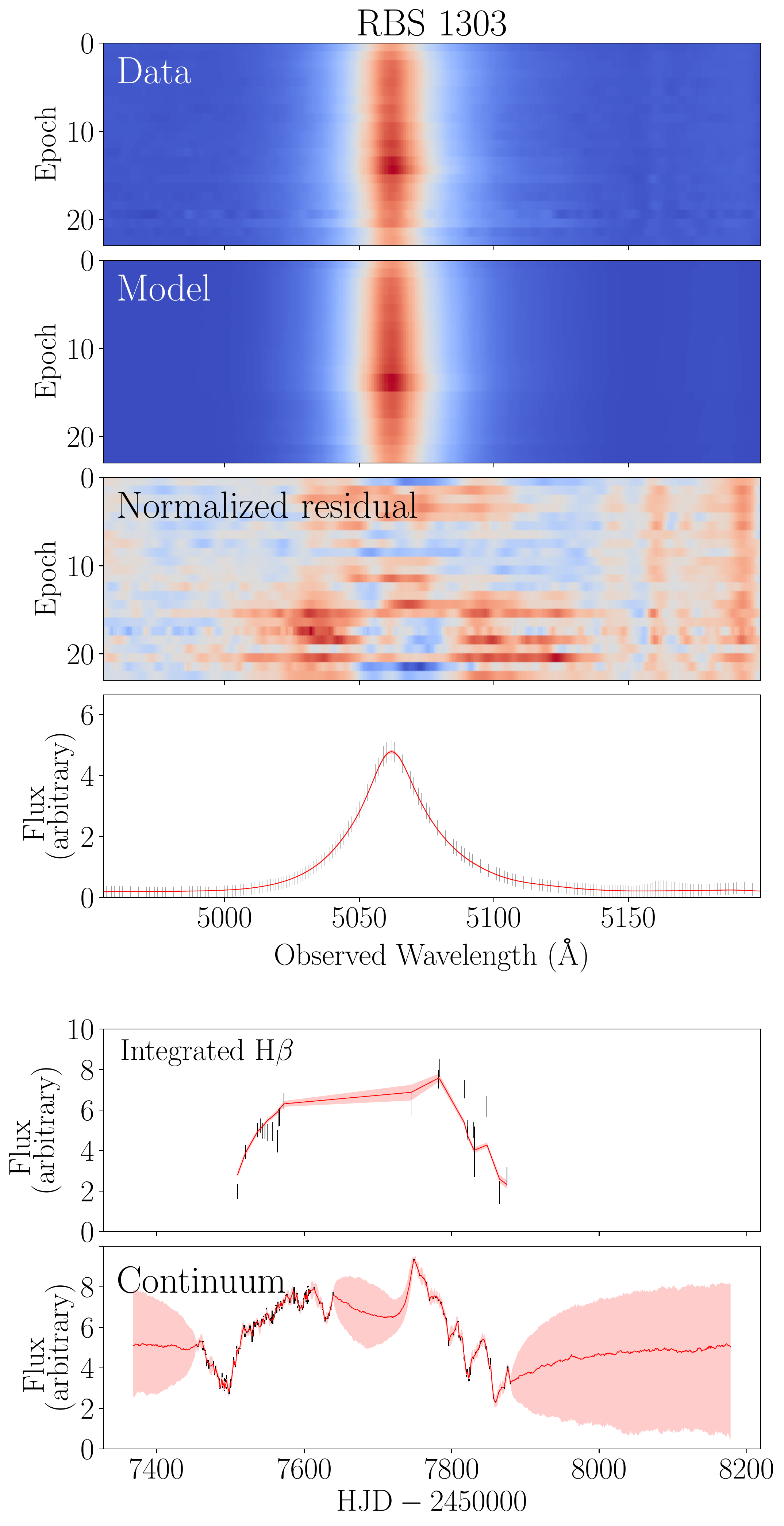}
    \caption{Model fits to the \Hb line profile, integrated \Hb flux, and AGN continuum flux for \rbsb. See the Figure \ref{fig:model_fits_pg2209} caption for panel descriptions.}
    \label{fig:model_fits_rbs1303}
\end{figure}

\begin{figure}
    \centering
    \includegraphics[width=\columnwidth, keepaspectratio]{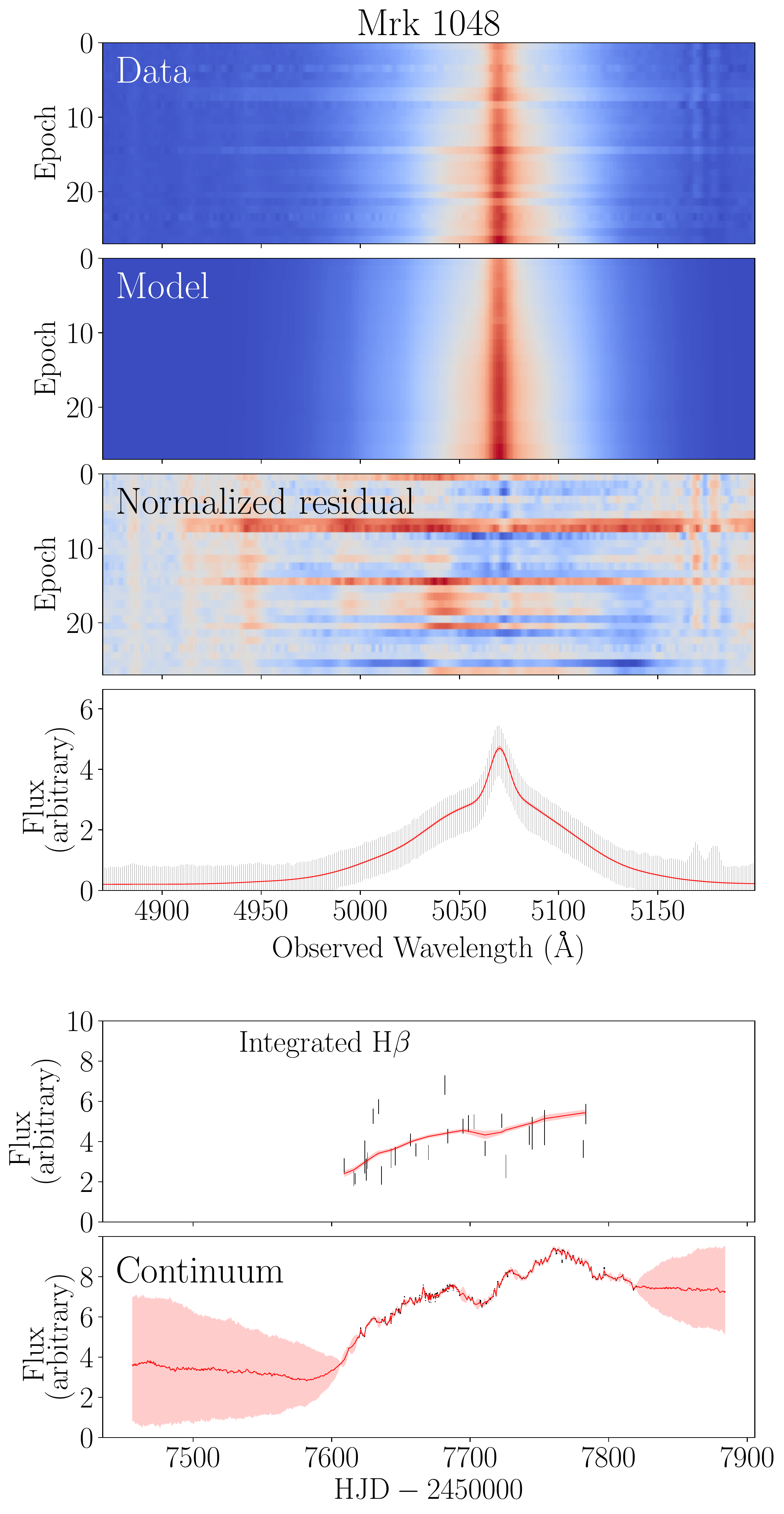}
    \caption{Model fits to the \Hb line profile, integrated \Hb flux, and AGN continuum flux for \mrkd. See the Figure \ref{fig:model_fits_pg2209} caption for panel descriptions.}
    \label{fig:model_fits_mrk1048}
\end{figure}

\begin{figure}
    \centering
    \includegraphics[width=\columnwidth, keepaspectratio]{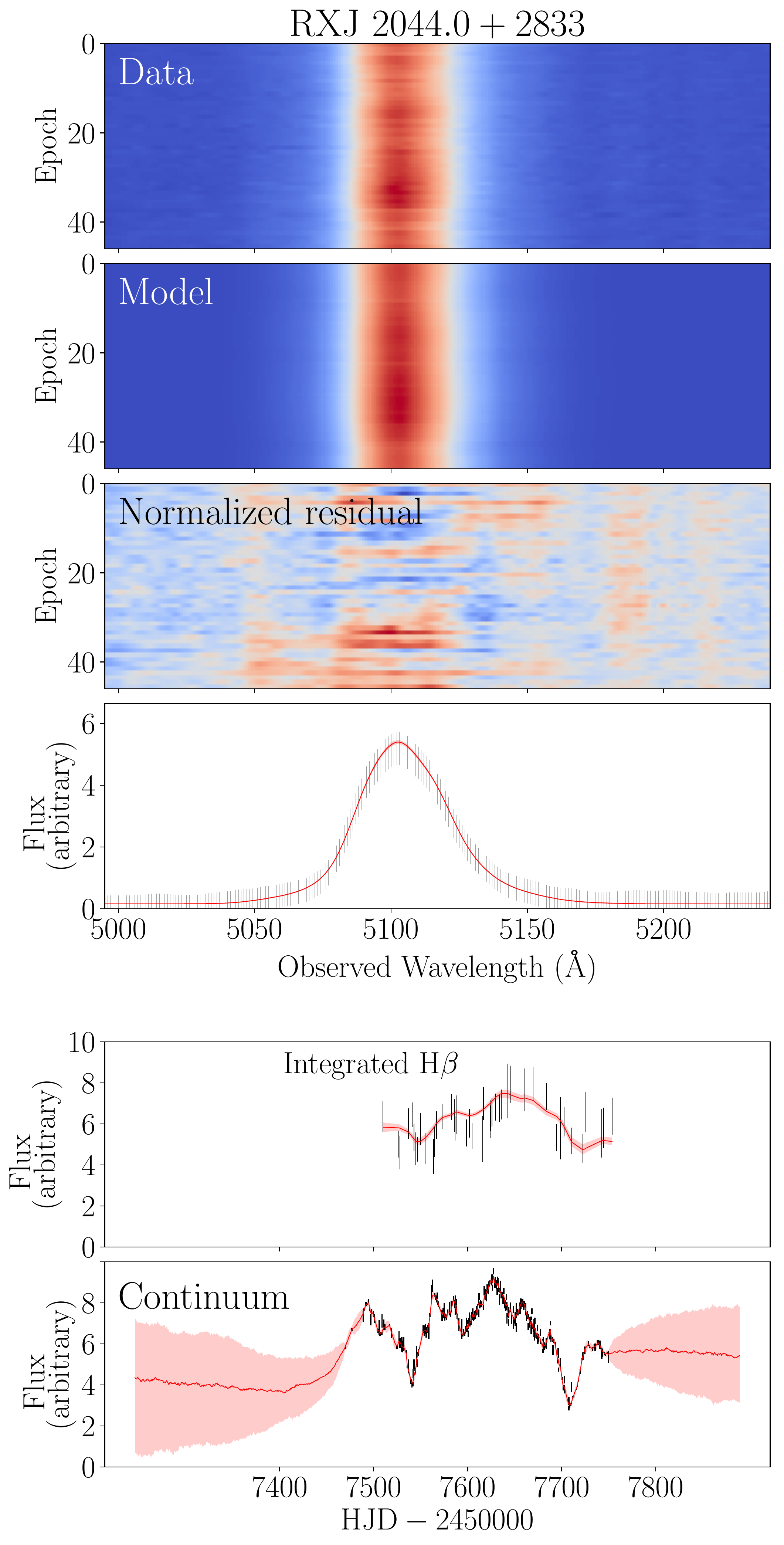}
    \caption{Model fits to the \Hb line profile, integrated \Hb flux, and AGN continuum flux for \rxja. See the Figure \ref{fig:model_fits_pg2209} caption for panel descriptions.}
    \label{fig:model_fits_rxj2044}
\end{figure}

\begin{figure}
    \centering
    \includegraphics[width=\columnwidth, keepaspectratio]{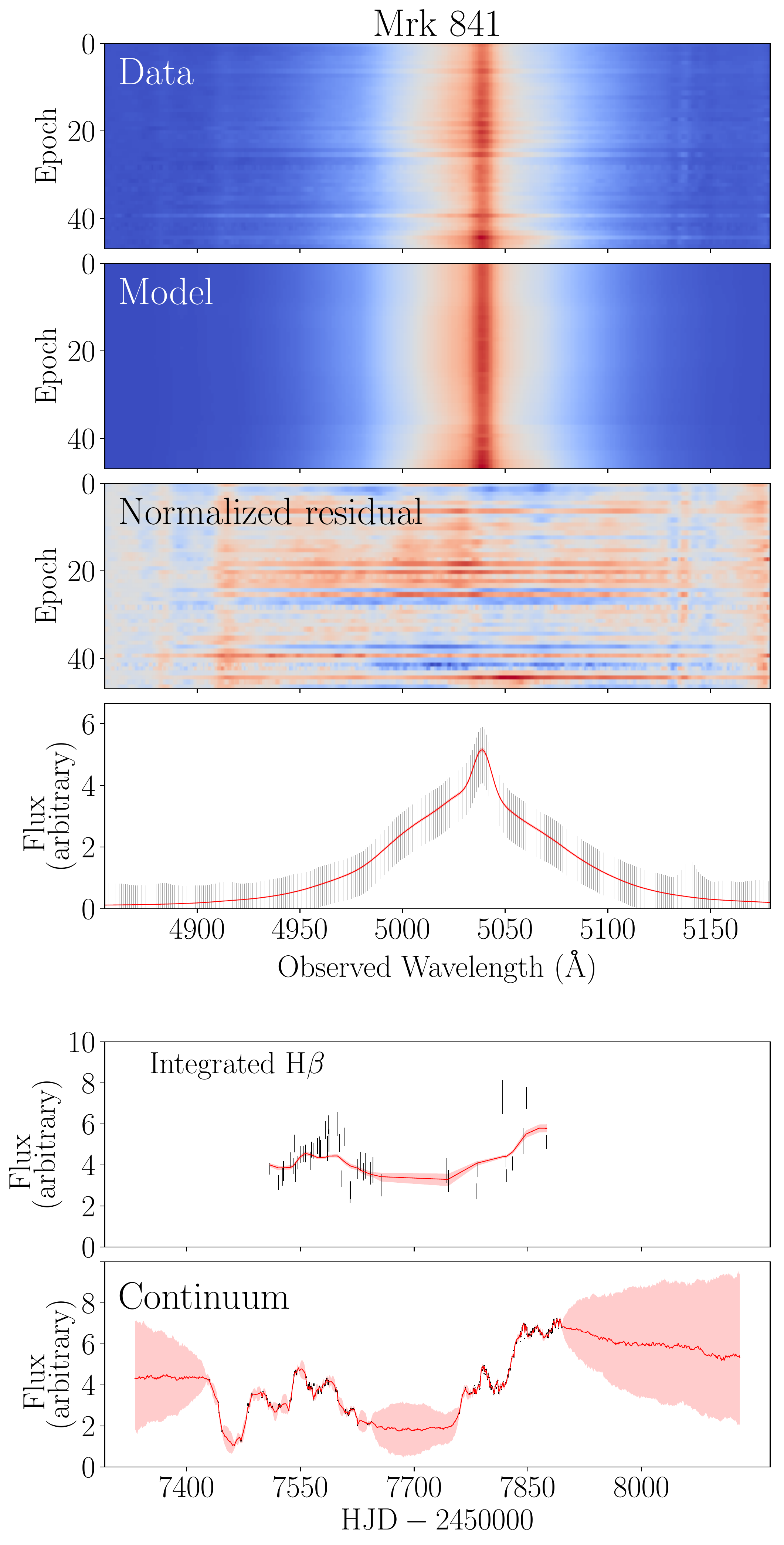}
    \caption{Model fits to the \Hb line profile, integrated \Hb flux, and AGN continuum flux for \mrke. See the Figure \ref{fig:model_fits_mcg04} caption for panel descriptions.}
    \label{fig:model_fits_mrk841}
\end{figure}

Here we present the details for the nine sources determined to have good model fits (Figures \ref{fig:model_fits_pg2209}--\ref{fig:model_fits_mrk841}). An overview of model parameter estimates is provided in Table \ref{table_results}. Overall, the H$\beta$-emitting BLR is best described as a thick disk observed at low to moderate inclination angles with diverse kinematics, as depicted in Figure \ref{fig:geo_all}. We find black hole mass estimates that are consistent (within at least $\sim 3 \sigma$) with velocity-resolved reverberation mapping estimates determined by \citet{u2021lick}, using a value of \logfrmssigma$=0.65$ for the virial coefficient.

\subsection{\pga}
Our model finds a BLR mean radius of $r_{\rm median} = $ \pgarmedian\ light-days and corresponding mean lag of $\tau_{\rm{median}} =$ \pgataumedian\ light-days. The opening and inclination angles are $\theta_o =$ \pgathetao\ degrees and $\theta_i = $ \pgathetai\ degrees, respectively, indicating a thick-disk structure slightly inclined toward the observer. Our model finds a strong preference for a transparent midplane ($\xi =$ \pgaxi), but is unable to constrain whether \Hb emission is isotropic/concentrated at the edges ($\gamma =$ \pgagamma) nor whether emission from the far/near sides of the BLR is preferred ($\kappa =$ \pgakappa). Dynamically, 54\% of particles have nearly circular orbits ($f_{\rm ellip} =$ \pgaellip) while the rest are on inflowing ($f_{\rm flow}= $ \pgaflow) orbits with velocities drawn from a distribution with center rotated $\theta_{\rm{e}} =$ \pgathetae\ degrees from escape velocity toward the circular velocity. Macroturbulent velocities are found to be insignificant with $\sigma_{\rm turb} = $ \pgasigmaturb. Finally, we find a black hole mass of $\log_{10}(M_{\rm bh}/M_{\odot}) =$ \pgalogmbh, which is consistent within our uncertainties with the velocity-resolved reverberation mapping estimate of $\log_{10}(M_{\rm bh}/M_{\odot}) = 7.46^{+0.1}_{-0.12}$ found by \citet{u2021lick}. 

\subsection{\rbsa}
Geometrically, our model predicts a BLR that is a relatively thick disk ($\theta_o$ = \rbsathetao\ degrees) inclined $\theta_i =$ \rbsathetai\  degrees toward the observer, with a median radius of $r_{\rm median} = $ \rbsarmedian\  light-days corresponding to an average time delay of $\tau_{\rm median} = $ \rbsataumedian\ light-days.  There is a slight preference for preferential \Hb emission from the far side of the BLR (with $\kappa = $ \rbsakappa) and a transparent BLR midplane ($\xi= $ \rbsaxi). Our model is unable to constrain, however, whether emission is uniformly emitted or concentrated at the edges ($\gamma = $\rbsagamma). Dynamically, 59\% of particles remain on circular bounded orbits ($f_{\rm ellip} = $\rbsaellip), and the remaining $\sim 40\%$ of particles exhibit outflowing ($f_{\rm flow} =$ \rbsaflow) behavior with velocities rotated $\theta_e =$ \rbsathetae\ degrees from the radial outflowing escape velocity toward a circular velocity. Additionally, the contribution from macroturbulent velocities is small with $\sigma_{\rm turb} = $  \rbsasigmaturb. Finally, our model estimates a black hole mass of $\log_{10}(M_{\rm bh}/M_{\odot}) =$ \rbsalogmbh, which is consistent within our uncertainties with the velocity-resolved reverberation mapping estimate of $\log_{10}(M_{\rm bh}/M_{\odot}) = 7.15^{+0.15}_{-0.25}$, found by \citet{u2021lick}.

\subsection{\mcga}
We find the H$\beta$-emitting BLR of \mcga\ to be best described by a slightly thick disk ($\theta_o = $ \mcgathetao\ degrees) inclined $\theta_i = $ \mcgathetai\ degrees toward the observer and median BLR radius of $r_{\text{median}} = $ \mcgarmedian\  light-days. Our model finds a preference for concentrated emission at the edges of the BLR ($\gamma =$ \mcgagamma) but is unable to constrain the transparency of the BLR midplane ($\xi =$ \mcgaxi) or whether emission from the far/near side of the BLR is preferred ($\kappa =$ \mcgakappa). Dynamically, our model finds a preference for 40\% of particles having nearly circular orbits ($f_{\rm ellip} = $ \mcgaellip) with the remaining particles having velocities drawn from a distribution with center rotated $\theta_{\rm{e}} =$ \mcgathetae\ degrees from inflowing ($f_{\rm flow} =$ \mcgaflow) escape velocity toward the circular velocity. The contribution from macroturbulent velocities is small, with $\sigma_{\rm turb} = $ \mcgasigmaturb. Finally, we find a black hole mass of $\log_{10}(M_{\rm bh}/M_{\odot}) =$ \mcgalogmbh, which is consistent within our $1.5\sigma$ uncertainty with the \cite{u2021lick} value of $\log_{10}(M_{\rm bh}/M_{\odot}) = 7.18^{+0.10}_{-0.10}$.
\subsection{\npmg}
The data best fit a moderately thick disk ($\theta_o =$ \npmgthetao) H$\beta$-emitting BLR, viewed at an inclination of $\theta_i =$ \npmgthetai\ degrees with a median radius of $r_{\rm{median}} =$ \npmgrmedian\ light-days. Our model finds a preference for an opaque BLR midplane with $\xi =$ \npmgxi\ but is unable to constrain whether the BLR prefers emission to the far/near side of the BLR ($\kappa =$ \npmgkappa) or is uniformly emitted/concentrated at the edges ($\gamma =$ \npmggamma). Dynamically, our model finds that a little under half of the particles have circular orbits ($f_{\rm ellip} =$ \npmgellip). The remaining particles having velocities drawn from a Gaussian $v_r - v_{\phi}$ distribution rotated $\theta_e =$ \npmgthetae\ degrees from radially inflowing ($f_{\rm flow}$ = \npmgflow) escape velocity to circular velocity. The contribution from macroturbulent velocities is small, with $\sigma_{\rm turb} = $ \npmgsigmaturb. Finally, we estimate a black hole mass of $\log_{10}(M_{\rm bh}/M_{\odot}) =$ \npmglogmbh\ that is consistent within $2\sigma$ uncertainties of the $\log_{10}(M_{\rm bh}/M_{\odot}) = 7.28^{+0.23}_{-0.43}$ estimate found by \citet{u2021lick} using a traditional reverberation mapping analysis.
\subsection{\mrka}
The H$\beta$-emitting BLR of this source is modeled as a thick disk ($\theta_o =$ \mrkathetao\ degrees) inclined $\theta_i = $ \mrkathetai\ degrees toward an observer with a median BLR radius of $r_{\rm{median}} =$ \mrkarmedian\ light-days. The data best fit a mostly opaque BLR midplane with $\xi =$ \mrkaxi\ with slight preferrential emission from the near side of the BLR ($\kappa =$ \mrkakappa) and mostly isotropic emission ($\gamma =$ \mrkagamma). Dynamically, our model suggests that $\sim 80\%$ of particles have nearly circular orbits with ($f_{\rm{ellip}} =$ \mrkaellip), with the remaining particles having velocities drawn from a Gaussian $v_r - v_{\phi}$ distribution rotated $\theta_e =$ \mrkathetae\ degrees from radially outflowing ($f_{\rm flow}$ = \mrkaflow) escape velocity to circular velocity. The contribution from macroturbulent velocities is small, with $\sigma_{\rm turb} = $ \mrkasigmaturb. Finally, we estimate a black hole mass of $\log_{10}(M_{\rm bh}/M_{\odot}) =$ \mrkalogmbh\ that is consistent within $< 3\sigma$ with the estimate found by \citet{u2021lick} ($\log_{10}(M_{\rm bh}/M_{\odot}) = 7.80^{+0.06}_{-0.07}$). 
\subsection{\rbsb}
The data is in best agreement with a thick disk BLR ($\theta_o =$ \rbsbthetao degrees) inclined $\theta_i =$ \rbsbthetai\ degrees toward an observer with a median radius of $r_{\rm{median}} =$ \rbsbrmedian\ light-days. The model finds a slight preference for a transparent BLR midplane ($\xi=$ \rbsbxi) and a strong preference for preferential emission from the far side of the BLR ($\kappa=$ \rbsbkappa) and concentrated emission toward the edges of the disk ($\gamma =$ \rbsbgamma). Dynamically, the model suggests $\sim 18\%$ of particles have nearly circular orbits ($f_{\rm ellip} =$ \rbsbellip), with the remaining particles having velocities drawn from a Gaussian $v_r - v_{\phi}$ distribution rotated $\theta_e =$ \rbsbthetae\ degrees from radially outflowing ($f_{\rm flow}$ = \rbsbflow) escape velocity to circular velocity. The contribution from macroturbulent velocities is small, with $\sigma_{\rm turb} = $ \rbsbsigmaturb. Finally, we find a black hole mass of $\log_{10}(M_{\rm bh}/M_{\odot}) =$ \rbsblogmbh\ that is 
consistent within $3.2 \sigma$ of the estimate given by \citet{u2021lick} ($\log_{10}(M_{\rm bh}/M_{\odot}) = 7.40^{+0.08}_{-0.14}$). 
\subsection{\mrkd}
The \Hb BLR emission for this source is best described by a thick disk ($\theta_o =$ \mrkdthetao) inclined $\theta_i =$ \mrkdthetai\ degrees toward an observer with a median BLR radius of $r_{\rm{median}} =$ \mrkdrmedian\ light-days. We find a slight preference for an opaque midplane ($\xi =$ \mrkdxi) but are unable to constrain whether \Hb emission is isotropic/concentrated at the edges ($\gamma =$ \mrkdgamma) or whether emission from the far/near side of the BLR is preferred ($\kappa =$ \mrkdkappa). Dynamically, our model suggests $\sim 73\%$ of the particles are on circular orbits ($f_{\rm ellip} =$ \mrkdellip), with the remaining particles having velocities drawn from a Gaussian $v_r - v_{\phi}$ distribution rotated $\theta_e =$ \mrkdthetae\ degrees from radially outflowing ($f_{\rm flow}$ = \mrkdflow) escape velocity toward circular velocity. The contribution from macroturbulent velocities is small, with $\sigma_{\rm turb} = $ \mrkdsigmaturb. Finally, we estimate a black hole mass of $\log_{10}(M_{\rm bh}/M_{\odot}) =$ \mrkdlogmbh that is consistent within $1\sigma$ uncertainties of the estimate $\log_{10}(M_{\rm bh}/M_{\odot}) = 7.38^{+0.34}_{-0.60}$, found by \citet{u2021lick}.
\subsection{\rxja}
Geometrically, the BLR is modeled as a thick disk ($\theta_o =$ \rxjathetao\ degrees) inclined $\theta_i =$ \rxjathetai\ degrees toward an observer with a mean BLR radius of $r_{\rm{median}} =$ \rxjarmedian\  light-days. The model finds slight preferences for an opaque BLR midplane ($\xi =$ \rxjaxi) and preferential emission from the far side of the BLR ($\kappa =$ \rxjakappa) but is unable to constrain whether \Hb emission is isotropic/concentrated at the edges ($\gamma =$ \rxjagamma). Dynamically, the model suggests that a little under half ($f_{\rm ellip} =$ \rxjaellip) of particles have nearly circular orbits, with the remaining particles having velocities drawn from a Gaussian $v_r - v_{\phi}$ distribution rotated $\theta_e =$ \rxjathetae\ degrees from radially inflowing ($f_{\rm flow}$ = \rxjaflow) escape velocity to circular velocity. The contribution from macroturbulent velocities is small, with $\sigma_{\rm turb} = $ \rxjasigmaturb. Finally, we find a black hole mass of $\log_{10}(M_{\rm bh}/M_{\odot}) =$ \rxjalogmbh\ that is consistent with the estimate of $\log_{10}(M_{\rm bh}/M_{\odot}) = 7.08^{+0.07}_{-0.08}$, found by \citet{u2021lick}.
\subsection{\mrke}
Our model indicates that the \Hb BLR emission is best described by a very thick disk ($\theta_o =$ \mrkethetao\ degrees) inclined $\theta_i =$ \mrkethetai\ degrees toward an observer with a median BLR radius of $r_{\rm{median}} =$ \mrkermedian\  light-days. The data prefer preferential emission from the far side of the BLR ($\kappa =$ \mrkekappa) and slightly prefer a mostly transparent midplane ($\xi =$ \mrkexi). Our model is unable to constrain, however, whether emission isotropic/concentrated at the edges ($\gamma =$ \mrkegamma). Dynamically, our model suggests that $\sim 33\%$ of particles are on circular orbits ($f_{\rm ellip} =$ \mrkeellip), with the remaining particles having velocities drawn from a Gaussian $v_r - v_{\phi}$ distribution rotated $\theta_e =$ \mrkethetae\ degrees from radially inflowing ($f_{\rm flow}$ = \mrkeflow) escape velocity to circular velocity. The contribution from macroturbulent velocities is small, with $\sigma_{\rm turb} = $ \mrkesigmaturb. Finally, we estimate a black hole mass of $\log_{10}(M_{\rm bh}/M_{\odot}) =$ \mrkelogmbh\ that is consistent with the estimate $\log_{10}(M_{\rm bh}/M_{\odot}) = 7.66^{+0.20}_{-0.21}$ found by \citet{u2021lick}.
\section{Discussion}
\label{sec: discussion}
Here we highlight our phenomenological model's capability to directly constrain the BLR kinematics that best fit the data. We compare our model's interpretations with those found by \citet{u2021lick} using traditional qualitative velocity-delay map interpretations. We then combine our results with those from previous studies and search for any luminosity-dependent trends or a line profile shape dependence on BLR structure and kinematics, to try to gain a better understanding of the H$\beta$-emitting BLR. 
%\subsection{Inferred \textsc{caramel} Time Lags Compared to Cross Correlation Results}
%\begin{figure}
    %\centering
    %\includegraphics[width=\columnwidth,keepaspectratio]{figures/time_lag_comparison.pdf}
    %\caption{Caption}
   % \label{fig:timelag_comparison}
%\end{figure}
%Before we highlight the straightforward analysis our modeling approach provides, we would like to first emphasize the utility of a traditional reverberation mapping analysis. 
%\textcolor{red}{[currently working on this/unsure if we will end up including]}

%\input{time_lag_table}
\subsection{Inferred \textsc{caramel} Kinematics Compared to Velocity-Delay Map Results}
\begin{figure}
    \centering
    \includegraphics[width=\columnwidth,keepaspectratio]{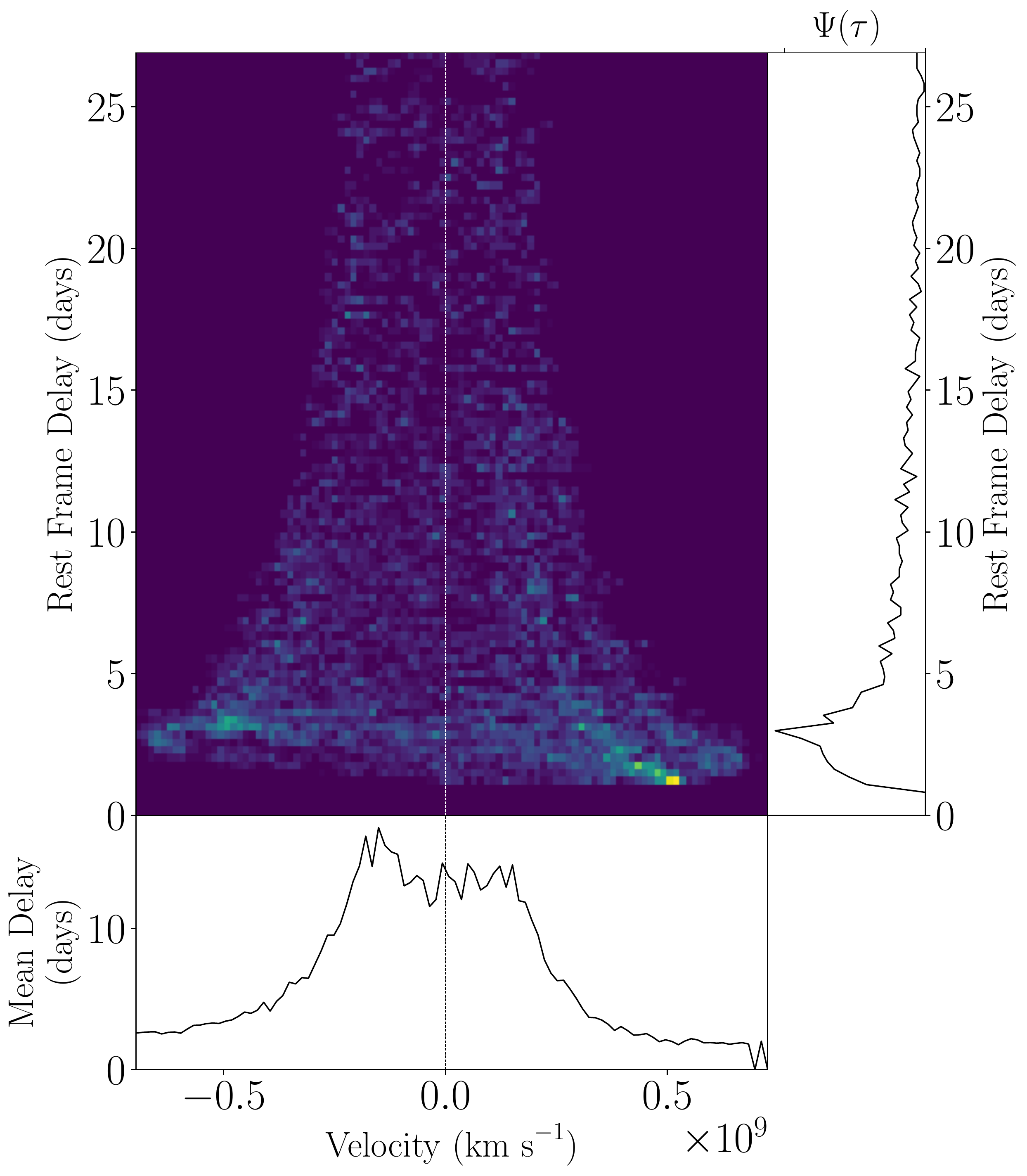}
    \caption{\npmg\ transfer function produced using median model parameter estimates. The right-hand panel shows the velocity-integrated transfer function and the bottom panel shows the average time lag for each velocity pixel.}
    \label{fig:npmg_tf}
\end{figure}

\begin{figure}
    \centering
    \includegraphics[width=\columnwidth,keepaspectratio]{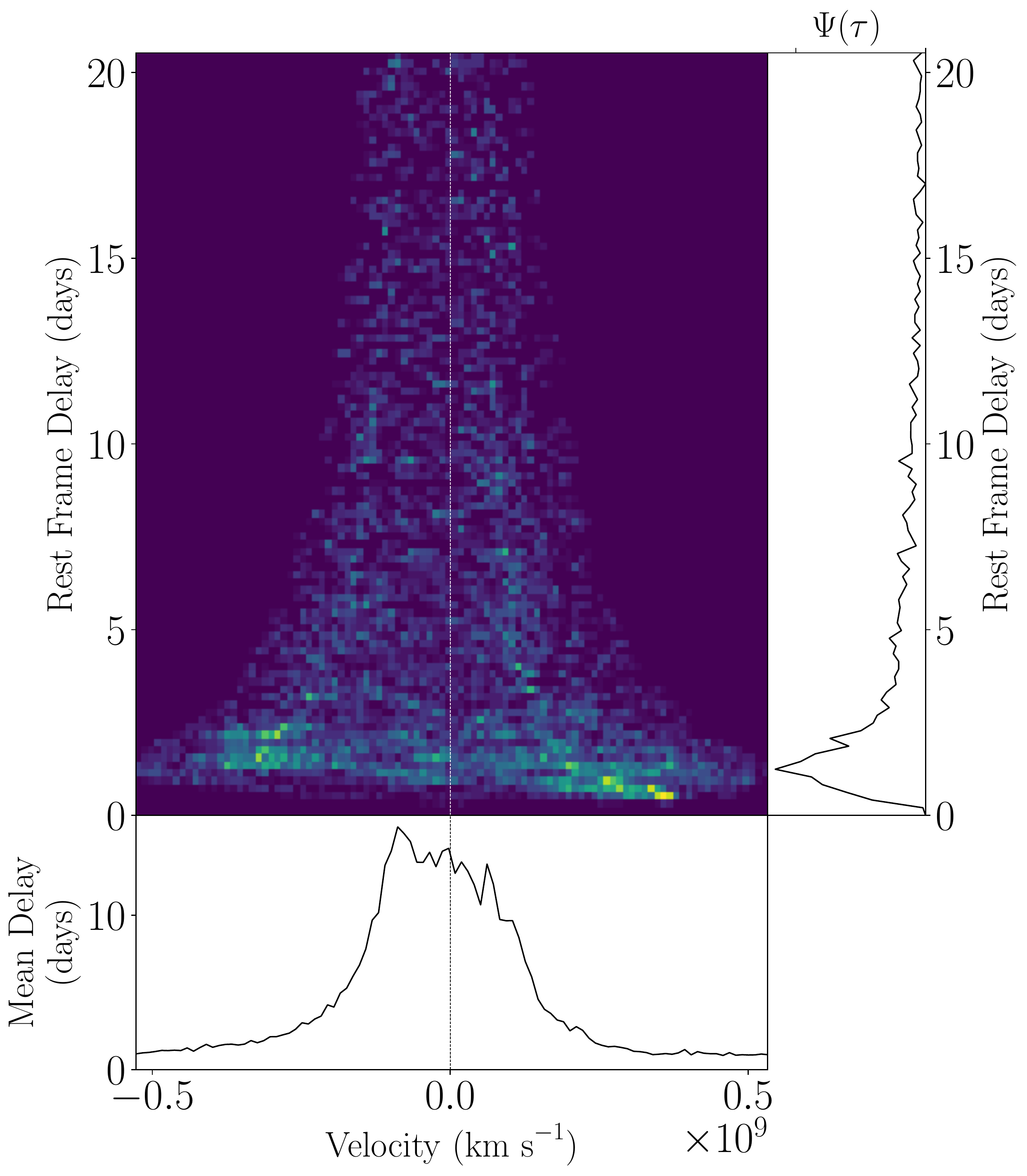}
    \caption{RBS 1917 transfer function produced using median model parameter estimates. See Figure \ref{fig:npmg_tf} caption for panel descriptions.}
    \label{fig:rbs1917_tf}
\end{figure}

\begin{figure}
    \centering
    \includegraphics[width=\columnwidth,keepaspectratio]{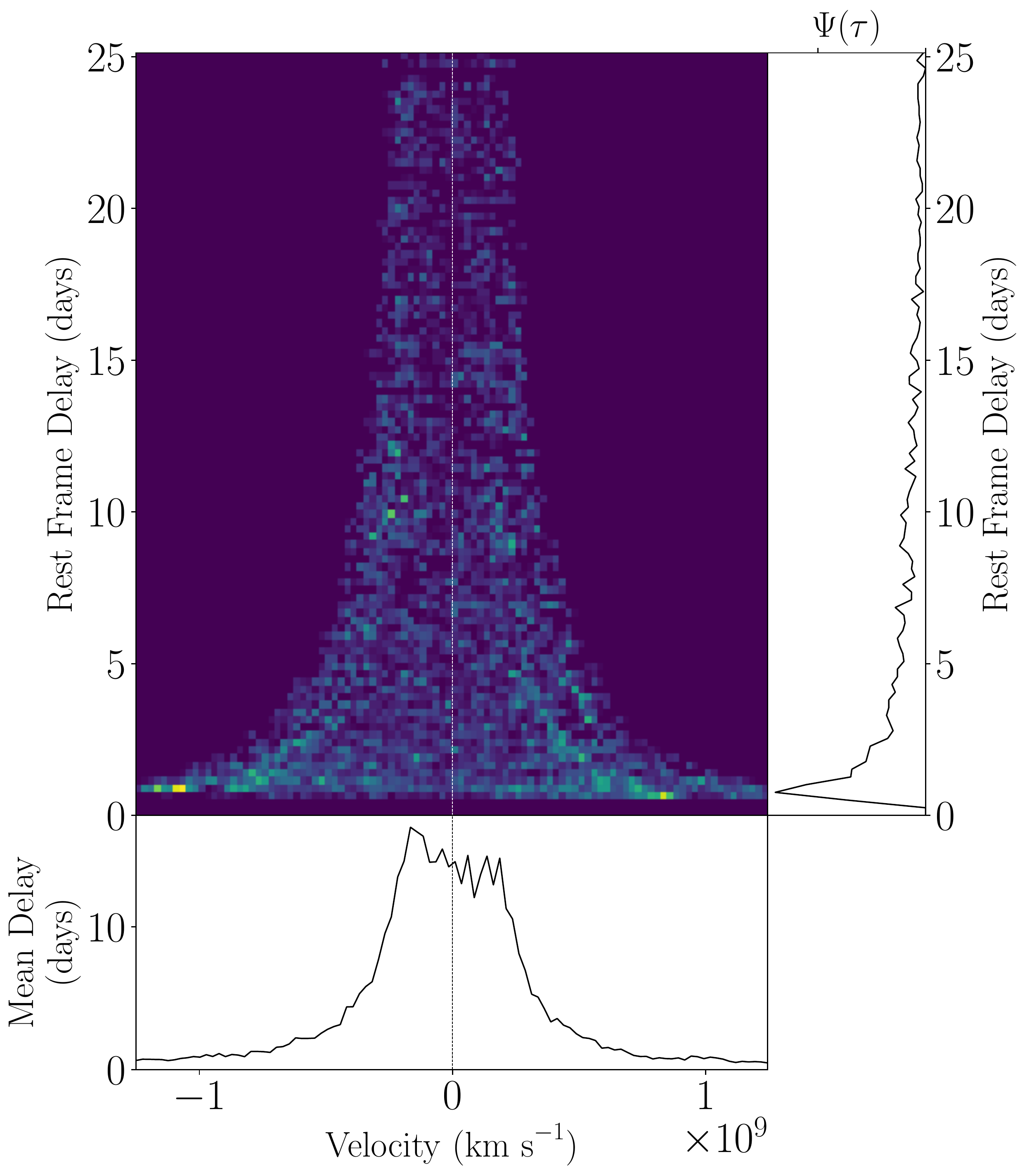}
    \caption{\mcga\ transfer function produced using median model parameter estimates. See Figure \ref{fig:npmg_tf} caption for panel descriptions.}
    \label{fig:mcg04_tf}
\end{figure}

\begin{figure}
    \centering
    \includegraphics[width=\columnwidth,keepaspectratio]{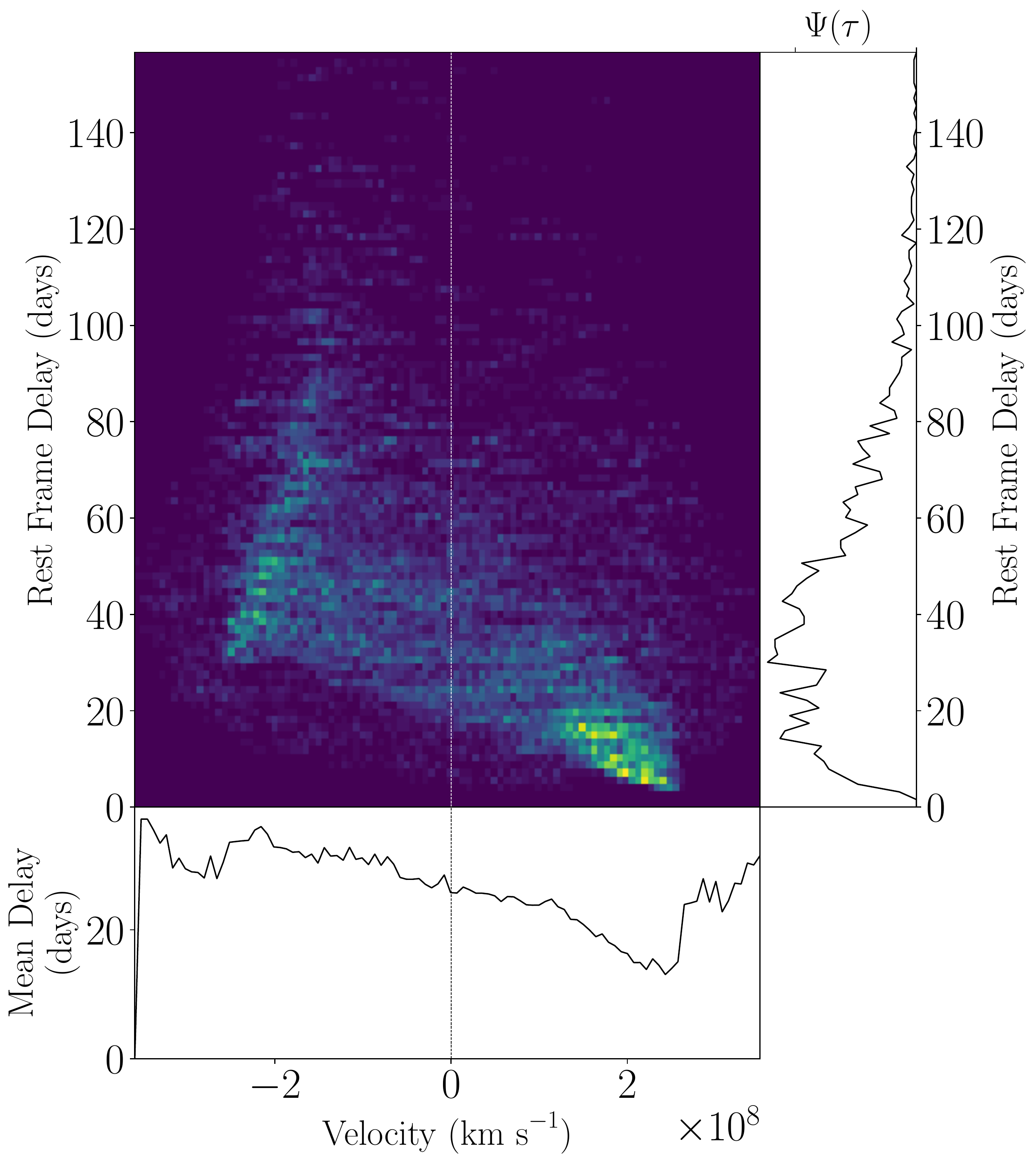}
    \caption{Mrk 1392 transfer function produced using median model parameter estimates. See Figure \ref{fig:npmg_tf} caption for panel descriptions.}
    \label{fig:mrk1392_tf}
\end{figure}

\begin{figure}
    \centering
    \includegraphics[width=\columnwidth,keepaspectratio]{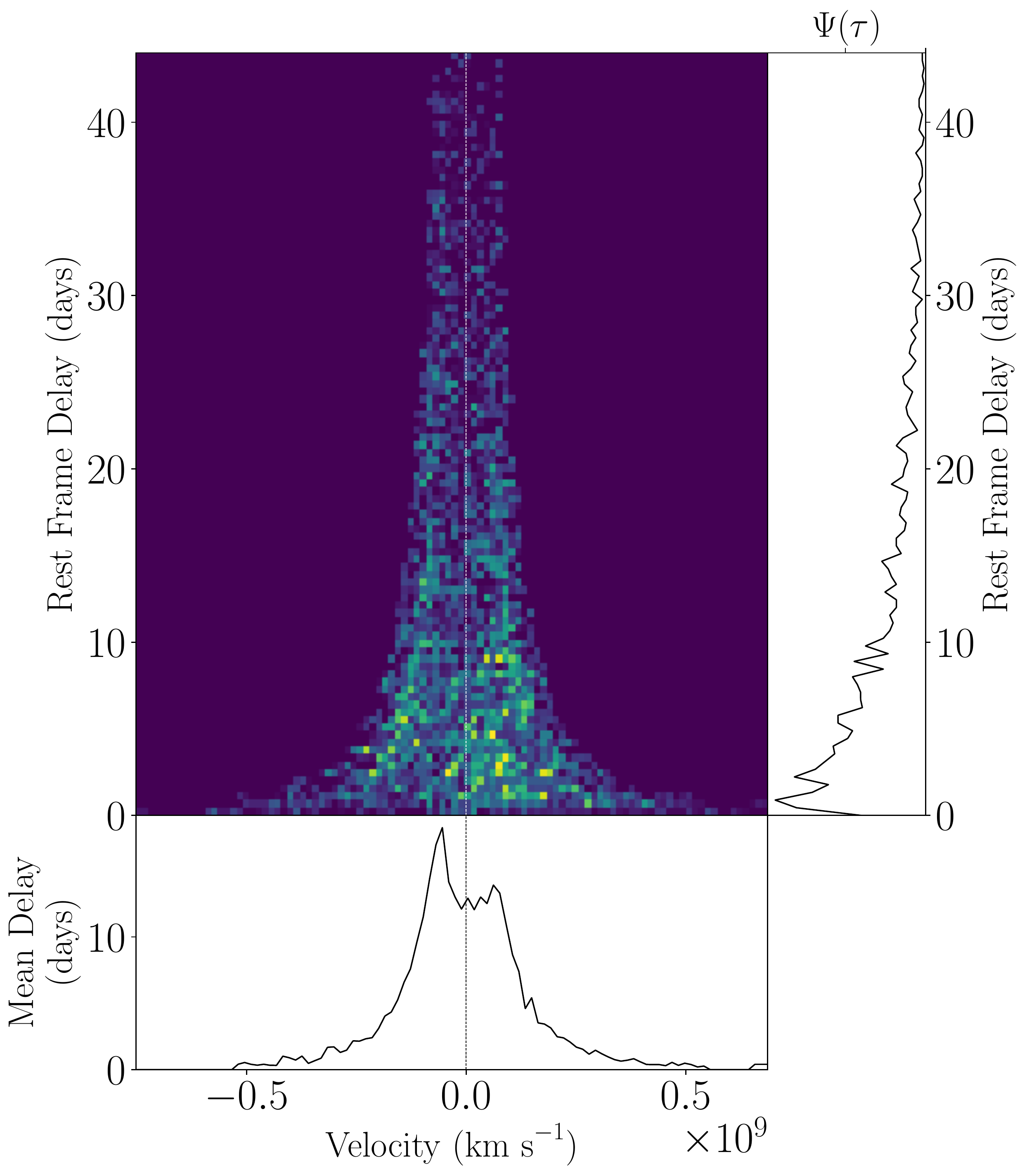}
    \caption{\rbsb\ transfer function produced using median model parameter estimates. See Figure \ref{fig:npmg_tf} caption for panel descriptions.}
    \label{fig:rbs1303_tf}
\end{figure}

\begin{figure}
    \centering
    \includegraphics[width=\columnwidth,keepaspectratio]{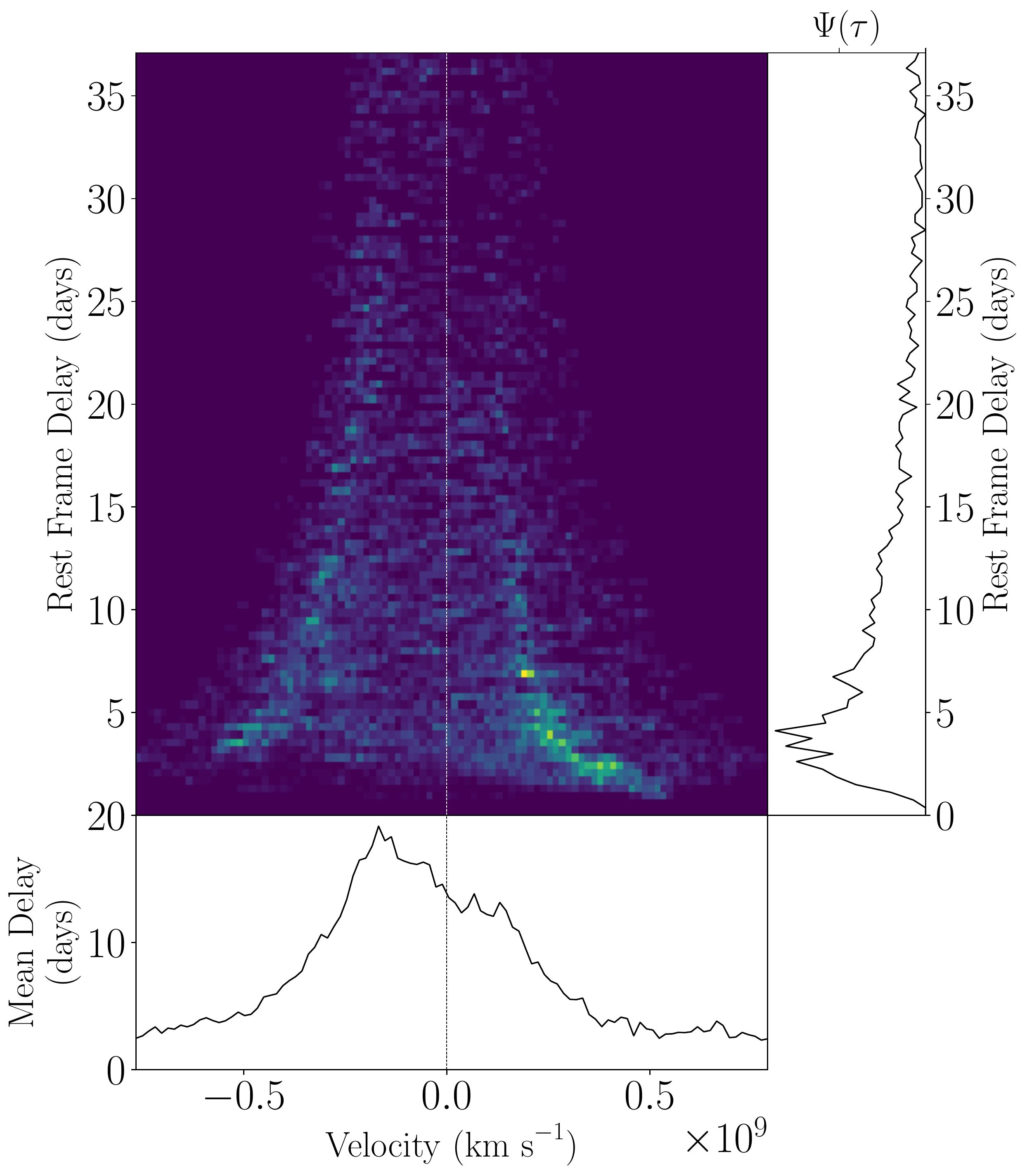}
    \caption{Mrk 1048 transfer function produced using median model parameter estimates. See Figure \ref{fig:npmg_tf} caption for panel descriptions.}
    \label{fig:mrk1048_tf}
\end{figure}

\begin{figure}
    \centering
    \includegraphics[width=\columnwidth,keepaspectratio]{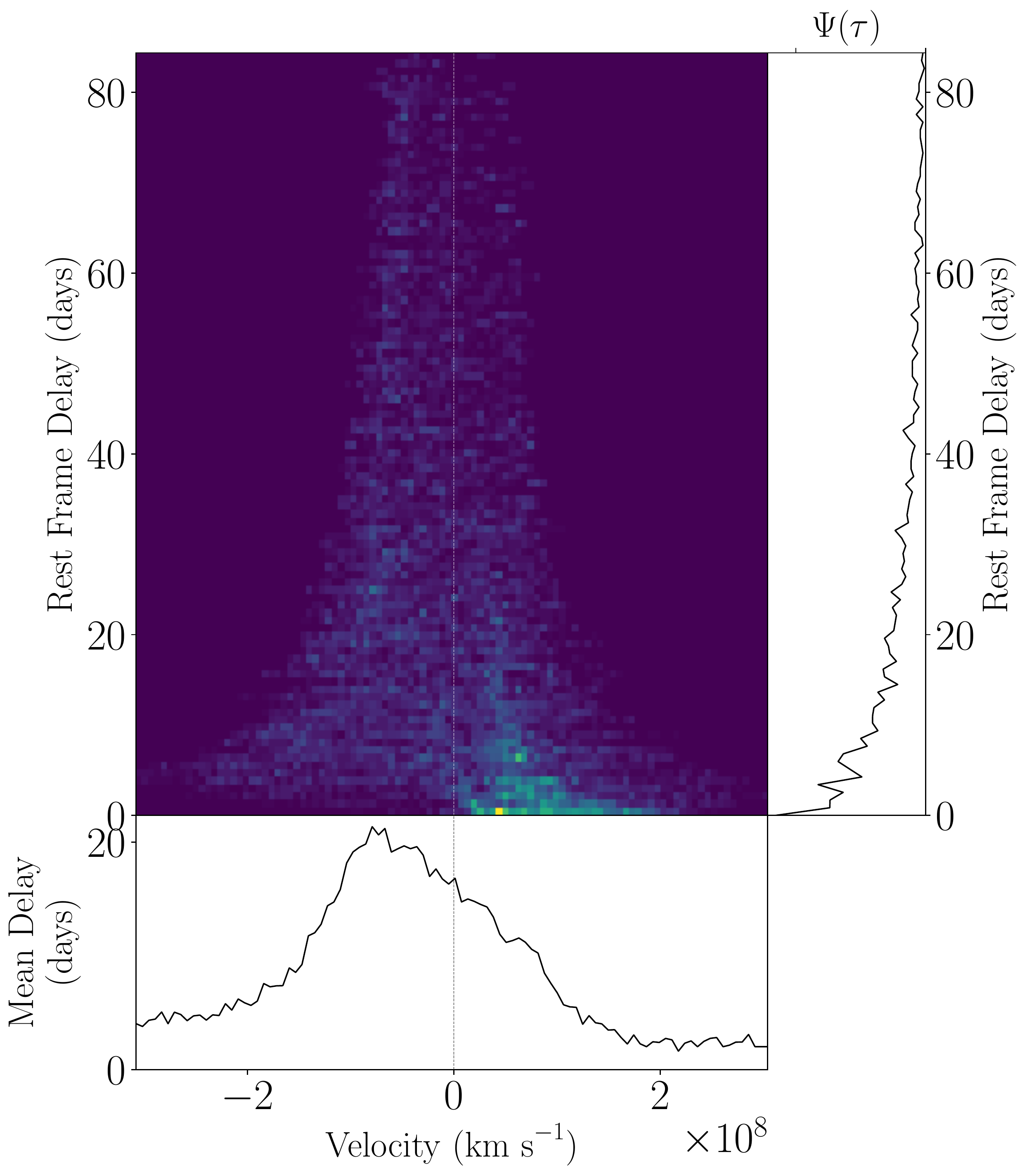}
    \caption{\rxja\ transfer function produced using median model parameter estimates. See Figure \ref{fig:npmg_tf} caption for panel descriptions.}
    \label{fig:rxj2044_tf}
\end{figure}

\begin{figure}
    \centering
    \includegraphics[width=\columnwidth,keepaspectratio]{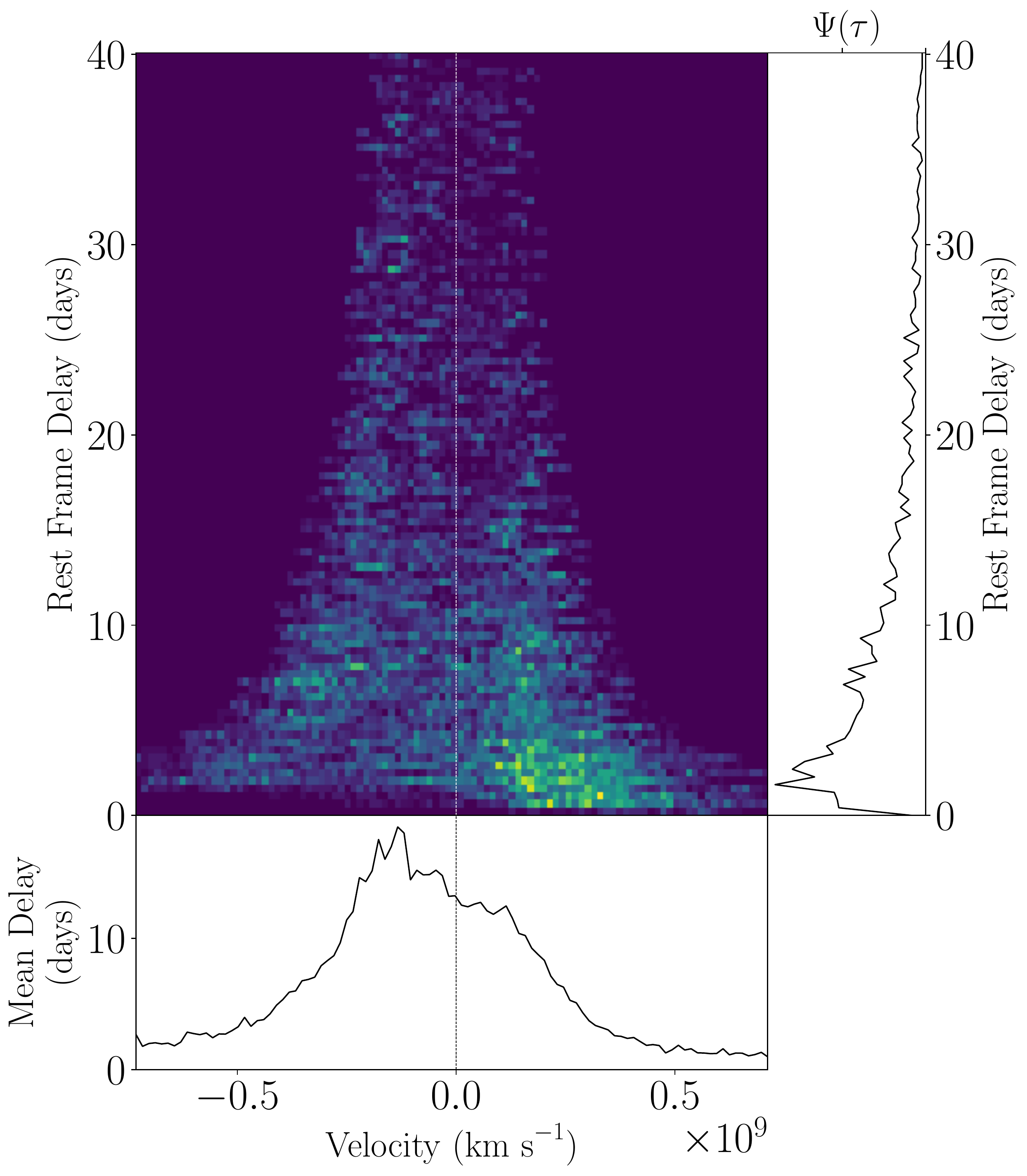}
    \caption{Mrk 841 transfer function produced using median model parameter estimates. See Figure \ref{fig:npmg_tf} caption for panel descriptions.}
    \label{fig:mrk841_tf}
\end{figure}

Overall, we find that roughly half of the sources in this work have interpretations consistent with those suggested by \citet{u2021lick}. In agreement with \citet{u2021lick}, we find infalling behavior ($\mathit{f}_{\rm{flow}} < 0.5$) in Mrk 841, \rxja, \npmg, and Mrk 1048, and outflowing ($\mathit{f}_{\rm{flow}} > 0.5$) behavior in \rbsb\ and RBS 1917. For the two sources which were interpreted to exhibit symmetric behavior (\mcga\ and Mrk 1392), our model allows for a more detailed analysis and finds a small fraction of particles exhibit outflowing behavior in %\mrkb and 
Mrk 1392 and a small fraction of particles in \mcga\ exhibiting inflowing behavior.  We now focus on \pga, whose flat velocity-resolved structures were difficult to describe with simple models. This in turn made it difficult to constrain the \Hb BLR kinematics \citep{u2021lick} using traditional reverberation mapping techniques. We present the recovered transfer functions for the remaining eight sources in Figures \ref{fig:npmg_tf}--\ref{fig:mrk841_tf}. 

\begin{figure}
    \centering
    \includegraphics[width=\columnwidth,keepaspectratio]{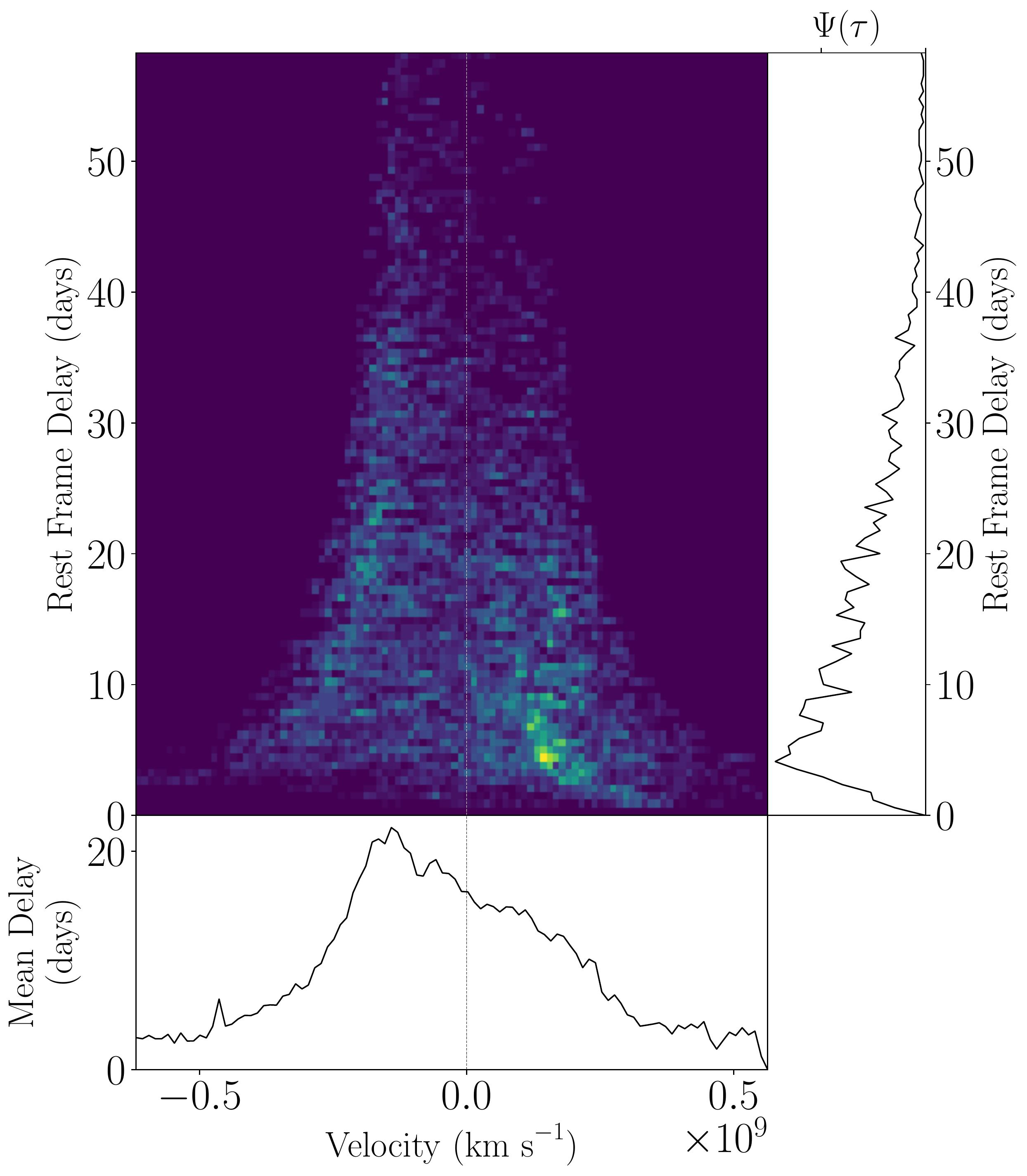}
    \caption{\pga\ transfer function produced using median model parameter estimates. See Figure \ref{fig:npmg_tf} caption for panel descriptions.}
    \label{fig:pg2209_tf}
\end{figure}

The transfer function constructed for \pga\ using \textsc{caramel} median value model parameters that best fit the data is found in Figure \ref{fig:pg2209_tf}. Our model suggests that $\sim 54\%$ of particles have nearly circular orbits ($f_{\rm{ellip}} =$ \pgaellip), with the remaining particles having velocities drawn from a Gaussian $v_r - v_{\phi}$ distribution rotated $\theta_e=$ \pgathetae\ degrees from radially inflowing ($f_{\rm flow}$=\pgaflow) escape velocity toward circular velocity. This can be summarized with the In.$-$Out. parameter defined by \citetalias{2018ApJ...866...75W}, with a value of \pgainout, suggesting that a majority of the remaining  ($1-f_{\rm{ellip}}$) $\approx 46\%$ particles exhibit radial inflow behavior. 

Our result emphasizes the qualitative interpretation of transfer functions, as the transfer function depicted in Figure \ref{fig:npmg_tf} could easily be interpreted as symmetric, which is consistent with Keplerian, disk-like rotation or random motion without any net radial inflow or outflow across the extended BLR. It appears that the asymmetric pattern associated with radial infalling gas is much more subtle and the slightly longer lags on the blue wing near zero velocity may not immediately be interpreted as radial infalling gas, since the asymmetry is not seen in the high-velocity component of the blue wing. (i.e., the slight top-hat profile shape is only slightly asymmetric from the center on the blue side). This example again emphasizes the difficulty in interpreting qualitatively the information embedded in velocity-resolved velocity delay maps and highlights a benefit of our quantitative forward modelling approach.
\subsection{Luminosity-Dependent Trends}
\label{sec: lum_dep_trends}
 Prior reverberation mapping studies searched for potential patterns in the velocity fields of the ionized H$\beta$-emitting regions. Using inferred kinematics from velocity-delay maps, \citet{Du++16} investigated whether any trends existed for super-Eddington accreting AGNs, since their stronger radiation pressure could induce pressure-driven winds %provide more energy
 and BLR outflow. With a small sample size, \citet{Du++16} concluded that BLR kinematics were diverse for super-Eddington accretion rate AGNs. 
 
 Although a similar trend seemed apparent with the modeled sources of the LAMP 2016 sample (as seen in the diversity of kinematics in Figure~\ref{fig:geo_all}), we increase the statistical power of our investigation by combining our sample with those from \citetalias{pancoast14b}, \citetalias{Grier++17}, \citetalias{2018ApJ...866...75W}, \citetalias{2020ApJ...902...74W}, and \citetalias{bentz2021detailed}. In particular, we search for correlations between BLR inclination, opening angle (disk thickness), and kinematics (outflow/inflow/symmetric behavior) with luminosity. We expect such trends to arise for example as a result of radiation pressure driven winds or by variation due to overall accretion rate.

\begin{figure*}
\centering
\includegraphics[height=14cm, keepaspectratio]{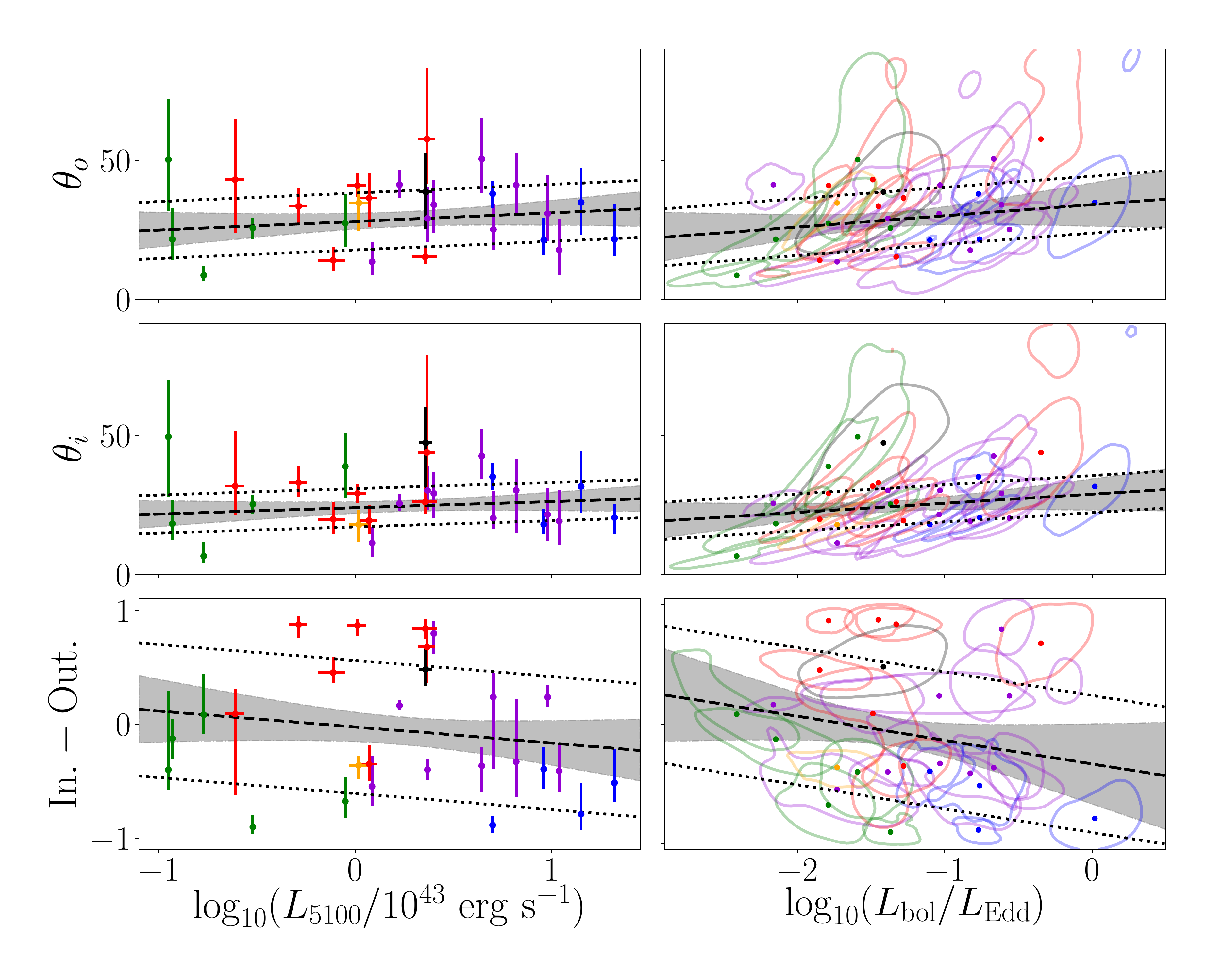}
\caption{Left panels: Correlations between $L_{5100}$ luminosity and select BLR model parameters. Right panels: Correlations between Eddington ratio and select BLR model parameters. The colored dots and contours show the median and 68\% confidence regions of the two-dimensional posterior probability distribution functions for each AGN. When the abscissa uncertainty is unavailable, the one-dimensional 68\% confidence interval is shown. The dashed black lines and gray shaded regions give the median and 68\% confidence intervals of the linear regression. Dotted lines are offset above and below the dashed line by the median value of the intrinsic scatter. Purple points are for the AGNs in this paper, red points are from \citetalias{2018ApJ...866...75W}, green points are from \citetalias{pancoast14b}, blue points are from \citetalias{Grier++17}, the black point is from \citetalias{2020ApJ...902...74W}, and the orange point is from \citetalias{bentz2021detailed}.}
\label{luminosity_dependent_trends}
\end{figure*}

We use both optical luminosity at 5100 \AA\ and the Eddington ratio.\footnote{We use a bolometric correction factor of nine, but would like to note that this only serves as a rough approximation and the actual bolometric correction factor may depend on Eddington ratio or other parameters.} The linear regression results are plotted in Figure~\ref{luminosity_dependent_trends} and the regression fit values are found in Table~\ref{tab:luminosity_regression}. With our combined sample, we do not find any significant luminosity-dependent trends; we do not find higher accretion rates to correlate with BLR outflow behavior and come to the same conclusion as \citet{Du++15}, that AGNs have diverse BLR geometry and kinematics. A possible interpretation of this diversity is that BLR geometry and kinematics experience ``weather-like" changes and cycle through a range of states on timescales of order a year or less \citep[see, e.g.,][]{2015ApJ...806..128D, 2018ApJ...856..108P, 2021ApJ...922..151K}.

\begin{deluxetable*}{c l c cccccc}
\tablecaption{Linear regression results for luminosity dependent trends}
\tablewidth{0pt}
\setlength{\tabcolsep}{11pt}
\tablehead{
\colhead{Luminosity} & 
\colhead{} &
\colhead{$\theta_o$ (${\rm deg.}$)} &
\colhead{$\theta_i$ (${\rm deg.}$)} &
\colhead{${\rm In.-Out.}$}
}
\startdata
\multirow{ 3}{*}{\begin{tabular}{l} $\log_{10}(L_{5100}/10^{43}{\rm ~erg~s}^{-1})$  \\ \end{tabular}}  & $\alpha$ &  28.1 $\pm$ $2.9$ & 23.9 $\pm$ $2.2$ & $-0.02$ $\pm$ $0.13$\\
 & $\beta$ &  3.04 $\pm$ $4.7$ &    2.3 $\pm$ $3.5$  &   $-0.15$ $\pm$ $0.2$\\
 & $\sigma_{\rm int}$ &  10 $\pm$ $56$ &  6.9 $\pm$ $34$ &  0.59 $\pm$ $0.13$ \\ \hline 
 \multirow{ 3}{*}{\begin{tabular}{l} $\log_{10}(L_{\rm bol}/L_{\rm Edd})$ \\ \end{tabular}}  & $\alpha$ & 34.0 $\pm$ $7.8$  & 28.7 $\pm$ $5.9$   & $-0.33$ $\pm$ $0.34$\\
 & $\beta$ & 4.0 $\pm$ $5.4$ & 3.2 $\pm$ $4.1$ &  $-0.2$ $\pm$ $0.24$\\
 & $\sigma_{\rm int}$ & 10 $\pm$ $53$    &  7 $\pm$ $30$&  0.57 $\pm$ $0.12$  \\ 
\enddata
\tablecomments{Linear regression results for optical $L_{5100}$ luminosity and Eddington ratio vs. BLR parameters using both the mean and rms spectrum. The parameter $\alpha$ represents the constant in the regression and $\beta$ represents the slope of the regression, while $\sigma_{\rm{int}}$ represents the standard deviation of the intrinsic scatter. The corresponding relationship is therefore given by $\rm{parameter}= \alpha + \beta\times\log_{10}(\rm{luminosity}) +\mathcal{N}(0,\sigma_{\rm int})$.}
\label{tab:luminosity_regression}
\end{deluxetable*}

\subsection{Line Profile Shape Dependence on BLR Structure and Kinematics}
\label{sec: lineprofileshape}
As suggested by \citet{collin06}, the ratio of the full width at half-maximum intensity (FWHM) of the line to the dispersion $\sigma_{\rm line}$ (i.e., the second moment of the line) may serve as a tracer for the physical parameters of the inner regions of an AGN. Since we expect BLR structure and dynamics to play a role in determining the line-profile shape, we might also expect to find correlations with AGN/BLR parameters. 

\begin{figure*}[]
\centering
\includegraphics[height=9cm,keepaspectratio]{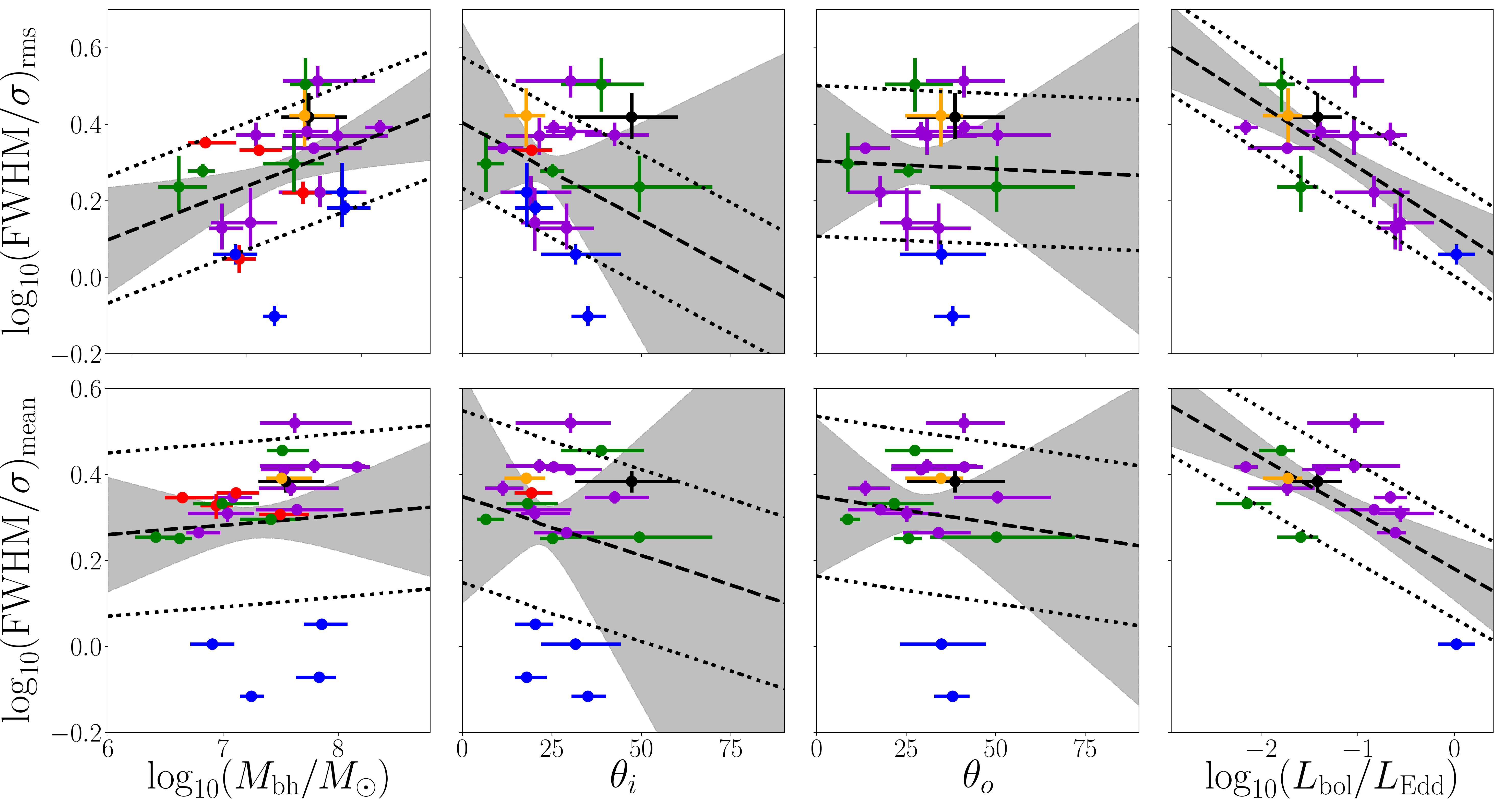}
\caption{Correlations between line-profile shape and black hole mass, BLR inclination angle, opening angle (disk thickness), and Eddington ratio. Top and bottom panels show line-profile shape determined using the mean and rms spectrum, respectively. The dashed black lines and gray shaded regions give the median and 68\% confidence intervals of the linear regression. Dotted lines are offset above and below the dashed line by the median value of the intrinsic scatter. Purple points are for the AGNs in this paper, red points are from \citetalias{2018ApJ...866...75W}, green points are from \citetalias{pancoast14b}, blue points are from \citetalias{Grier++17}, the black point is from \citetalias{2020ApJ...902...74W}, and the orange point is from \citetalias{bentz2021detailed}.}
\label{mean_line_profile_shape_blr}
\end{figure*}

For reference, $\log_{10}({\rm FWHM}/\sigma) \approx 0.37$ corresponds to a Gaussian-shaped line profile. Greater values correspond to a flat-topped shape while values less than 0.37 correspond to a narrower line profile shape with extended wings similar to a Lorentzian profile. We search for correlations between the line-profile shape with the following parameters: black hole mass $\log_{10}(M_{\rm bh}/M_{\odot}$), BLR inclination angle $\theta_i$, BLR opening angle, i.e. disk thickness, $\theta_o$, and Eddington ratio $L_{\rm{bol}}/L_{\rm{Edd}}$, using both the mean and root-mean-square (rms) spectrum. The linear regression results are shown in Figure \ref{mean_line_profile_shape_blr} with corresponding values found in Table \ref{tab:lineprofile_params_linear_regression}.

Although there appears to be an apparent correlation with black hole mass, using the levels of confidence we have defined in Section \ref{sec:trends}, we do not find it to be significant. A correlation with black hole mass would be expected if the size of the black hole somehow plays a role in the BLR structure and kinematics. Given the apparent correlation (see left most panels in Figure \ref{mean_line_profile_shape_blr}), a larger sample size with future studies may help further investigate the existence of such a correlation.

A correlation betweeen line-profile shape given by $\log(\rm{FWHM}/\sigma)$ and BLR inclination has been suggested in the past by \citet{collin06} and \citet{goad12}, but has been difficult to confirm since BLR inclination is generally not a direct observable. It is worth noting prior BLR-radio jet inclination studies in which orientation of the radio jet has been shown to be linked to the BLR \citep[e.g.,][]{2005AJ....130.1418J, Agudo_2012}, however, these measurements are limited and do not exist for the entire sample of reverberation mapped sources. We take advantage of the inclination estimates provided by our method to revisit the issue and do not find significant evidence of a correlation.\footnote{For readers who may recall that \citetalias{2018ApJ...866...75W} found marginal evidence for a correlation between scale factor and BLR inclination, we would like to reiterate that the lack of correlation found here is between BLR inclination and line-profile shape. For followup work regarding correlations between scale factor and AGN/BLR parameters using our newly extended sample, the reader is referred to \citet{v2022}.} We also do not find any correlation with disk thickness, $\theta_o$. 

We do, however, find marginal ($2.7\sigma$)  evidence for an anticorrelation between line-profile shape and Eddington ratio (when using the rms spectrum), which has also previously been suggested. \citet{collin06} found a similar correlation but cautioned that the Eddington rates were overestimated since the optical luminosity had not been corrected for host-galaxy starlight. Using host-galaxy starlight corrected optical luminosities from \citet{u2021lick}, we find the observed anticorrelation to be stronger ($2.7\sigma$) when using the rms spectrum than when using the mean spectrum ($2\sigma$). This anti-correlation may suggest that the accretion rate plays a role in the BLR structure and kinematics, which in turn determines the line-profile shape. This is plausible if BLR geometry and kinematics depend on accretion rate. %Future studies may help shed more light on this finding and determine whether line-profile shape may serve as a tracer of the inner regions of AGNs and their accretion rate. 

Alternatively, it is also possible that the anti-correlation with Eddington ratio is merely a by-product of the apparent but not statistically significant ($1.3\sigma$ as defined by our confidence intervals) correlation between line profile shape and black hole mass, since $L_{Bol}/L_{Edd} \propto 1/M_{BH}$. Followup work \citep{v2022} will extend the analysis in this work and  examine correlations between scale factor and $M_{BH}$, $L_{Bol}/L_{Edd}$, as well as FWHM/$\sigma$. The additional investigation of correlations with scale factor will allow us to further explore the relationship between the H$\beta$-emitting BLR and black hole mass/Eddington ratio, and their possible interpretations.

\begin{deluxetable*}{c l c cccccc}
\tablecaption{Linear regression results for line profile shape}
\tablewidth{0pt}
\setlength{\tabcolsep}{8pt}
\tablehead{
\colhead{Line Profile Shape} &
\colhead{ } &
\colhead{$\log_{10}(M_{\rm bh}/M_{\odot})$ } &
\colhead{$\theta_i$ (${\rm deg.}$)} &
\colhead{$\theta_o$ (${\rm deg.}$)} &
\colhead{$\log_{10}(L_{\rm bol}/L_{\rm Edd})$ } 
}
\startdata
\multirow{ 3}{*}{\begin{tabular}{l} $\log_{10}\Big(\frac{\rm{FWHM}}{\sigma}\Big)_{\rm{mean}}$ \\ \end{tabular}} 
 & $\alpha$ &  0.08 $\pm$ $0.65$ &  0.31 $\pm$ $17.0$ &  0.24 $\pm$ $0.21$ &    0.14 $\pm$ $0.10$ \\
& $\beta$ & 0.03 $\pm$ $0.09$ &  $-0.001$ $\pm$ $0.78$ &  0.002 $\pm$ $0.01$ &  $-0.13$ $\pm0.07$ \\
& $\sigma_{\rm int}$  &  0.16 $\pm$ $0.01$ &   0.17 $^{+ 0.02} _{- 0.01}$ &  0.19 $\pm$ $0.02$ &   0.15 $\pm$ $0.01$ \\ \hline
 \multirow{ 3}{*}{\begin{tabular}{l} $\log_{10}\Big(\frac{\rm{FWHM}}{\sigma}\Big)_{\rm{rms}}$ \\ \end{tabular}}  
& $\alpha$   &  $-0.57$ $\pm$ $0.69$ &   0.40 $\pm$ $0.57$ &   0.30 $\pm$ $0.21$ &   0.13 $\pm$ $0.09$ \\
& $\beta$  &       0.12 $\pm$ $0.10$ &  $-0.005$ $\pm$ $0.02$ &  $-0.0004$ $\pm$ $0.01$ &  $-0.16$ $\pm$ $0.07$ \\
& $\sigma_{\rm int}$  & 0.15 $\pm$ $0.01$ &   0.15 $\pm$ $0.01$ &  0.18 $\pm$ $0.02$ &   0.11 $\pm$ $0.01$ \\
\enddata
\tablecomments{Linear regression results for line profile shape vs. BLR/AGN parameters using both the mean and rms spectrum. The parameter $\alpha$ represents the constant in the regression and $\beta$ represents the slope of the regression, while $\sigma_{\rm{int}}$ represents the standard deviation of the intrinsic scatter. The corresponding relationship is therefore given by $\log_{10}(\rm{FWHM/}\sigma)= \alpha + \beta\times \rm{parameter}+\mathcal{N}(0,\sigma_{\rm int})$.}
\label{tab:lineprofile_params_linear_regression}
\end{deluxetable*}

\section{Conclusions}
\label{sec: summary}
We have applied forward modeling techniques to a sample of nine AGNs from the LAMP 2016 reverberation mapping campaign, increasing the number of dynamically modeled sources by nearly 50$\%$. We constrained the geometry and dynamics of the H$\beta$-emitting BLR and combined our results with previous studies (\citetalias{pancoast14b}, \citetalias{Grier++17}, \citetalias{2018ApJ...866...75W}, \citetalias{2020ApJ...902...74W}, and \citetalias{bentz2021detailed}) to investigate the existence of any trends in BLR structure and kinematics. Our main results are as follows.
\begin{enumerate}[(i)]
    \item Overall, we find the H$\beta$-emitting BLR of the LAMP 2016 sources to be best described by a thick disk observed at low to moderate inclination angles.
    \item We find no luminosity-dependent trends in the H$\beta$-emitting BLR geometry and kinematics, and conclude that AGNs have diverse BLR structures and kinematics. 
    \item We find marginal evidence for an anti-correlation between the line-profile shape of the broad \Hb emission line and Eddington ratio. This may suggest that the accretion rate plays a role in BLR structure and kinematics. Alternatively, the anti-correlation could merely be a by-product of an correlation with black hole mass that we cannot detect given our uncertainties. Followup work will further examine these two possible interpretations.
\end{enumerate}
With our simple phenomenological model we are able to gain insight on the BLR structure and kinematics in a more quantitative manner than the traditional interpretation of velocity-delay plots used in many reverberation mapping studies. Although much still remains unknown about the BLR, our findings suggest diversity that is consistent with transient AGN/BLR conditions over timescales of order months to years, rather than systematic trends. We note, however, that our combined sample is still small and may not be representative of the AGN population as a whole. Future reverberation mapping campaigns with sufficient data quality and variability will allow us to increase our sample size and thus improve the statistical significance of our findings. 

\begin{figure}[ht]
    \centering
    \includegraphics[height=14.5cm, keepaspectratio]{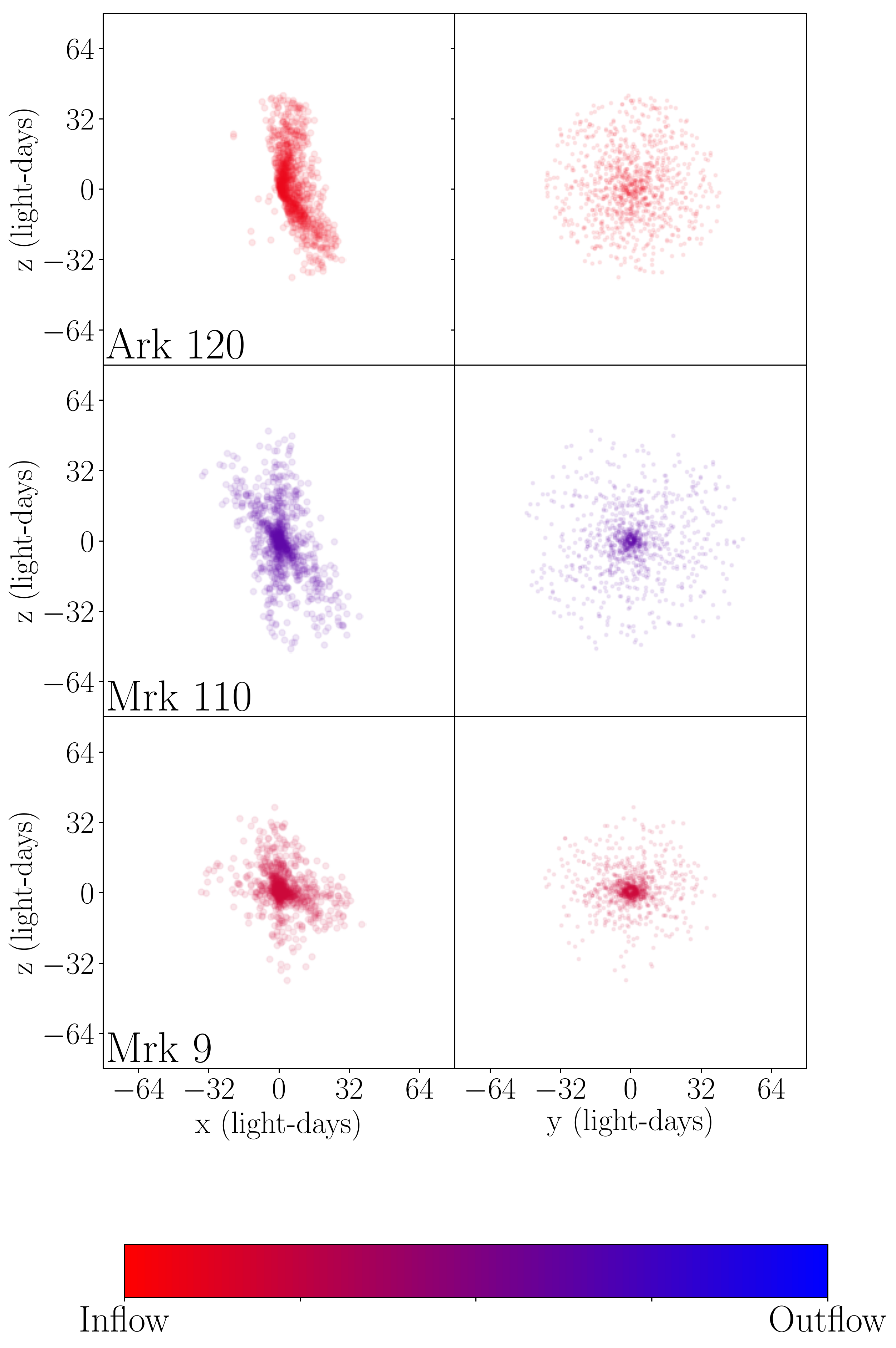}
    \caption{Geometric interpretation of BLR emission for the three LAMP 2016 sources excluded from our analysis (owing to moderate quality model fits) using median parameter estimates. For each source, the left panel shows an edge-on view while the right panel shows a face-on view. Each circle corresponds to one point particle in the model. The geometries are color-coded to indicate whether the BLR dynamics exhibit inflow (red) or outflow (blue).}
    \label{fig:geo_excluded}
\end{figure}

\begin{figure}[h!]
    \centering
    \includegraphics[height=9cm, keepaspectratio]{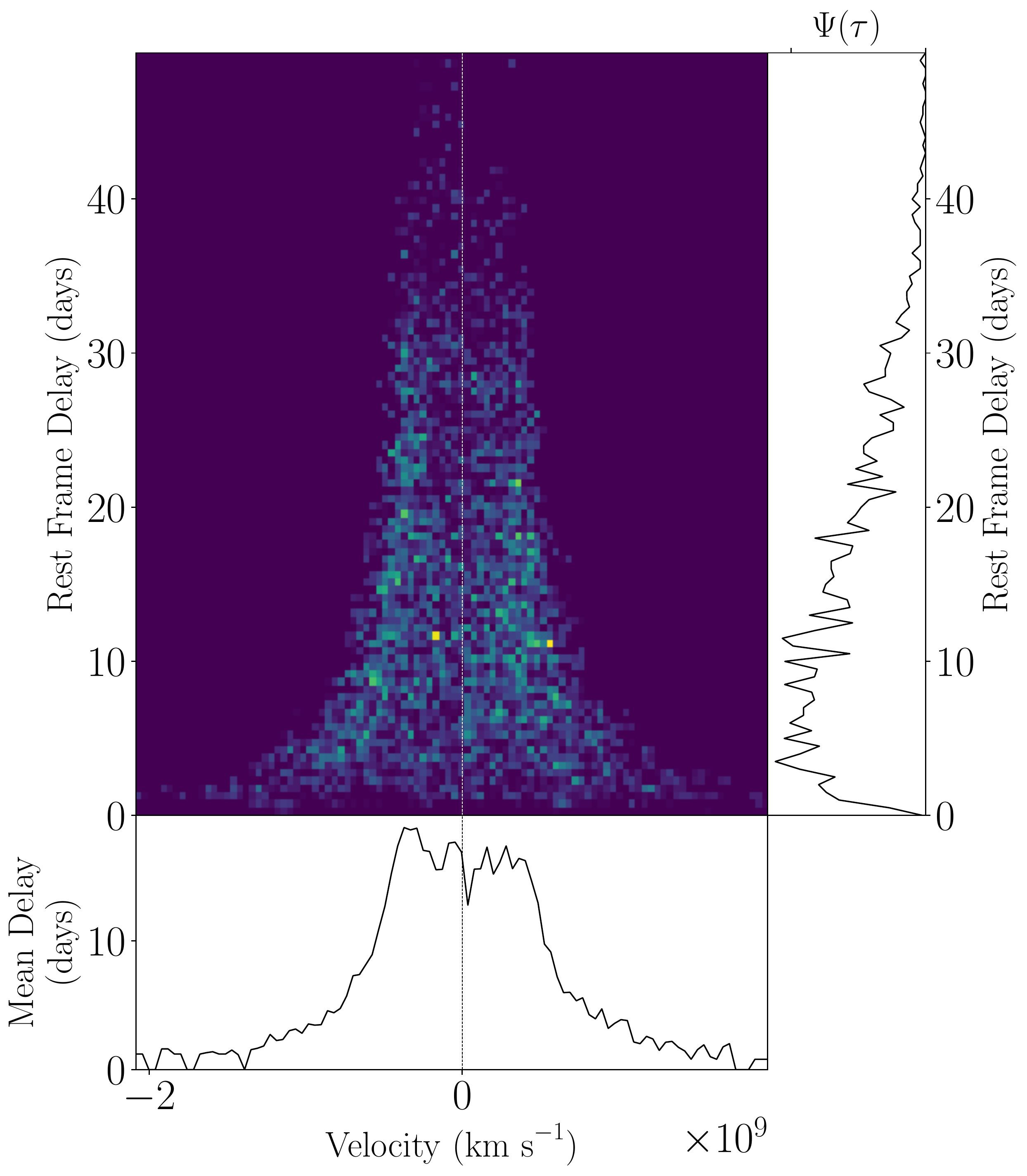}
    \caption{\ark\ transfer function produced using median model parameter estimates. The right-
    hand panel shows the velocity-integrated transfer function and the bottom panel shows the average time lag for each velocity pixel.}
    \label{fig:ark120_tf}
\end{figure}

\begin{figure}[h]
    \centering
    \includegraphics[height=9cm, keepaspectratio]{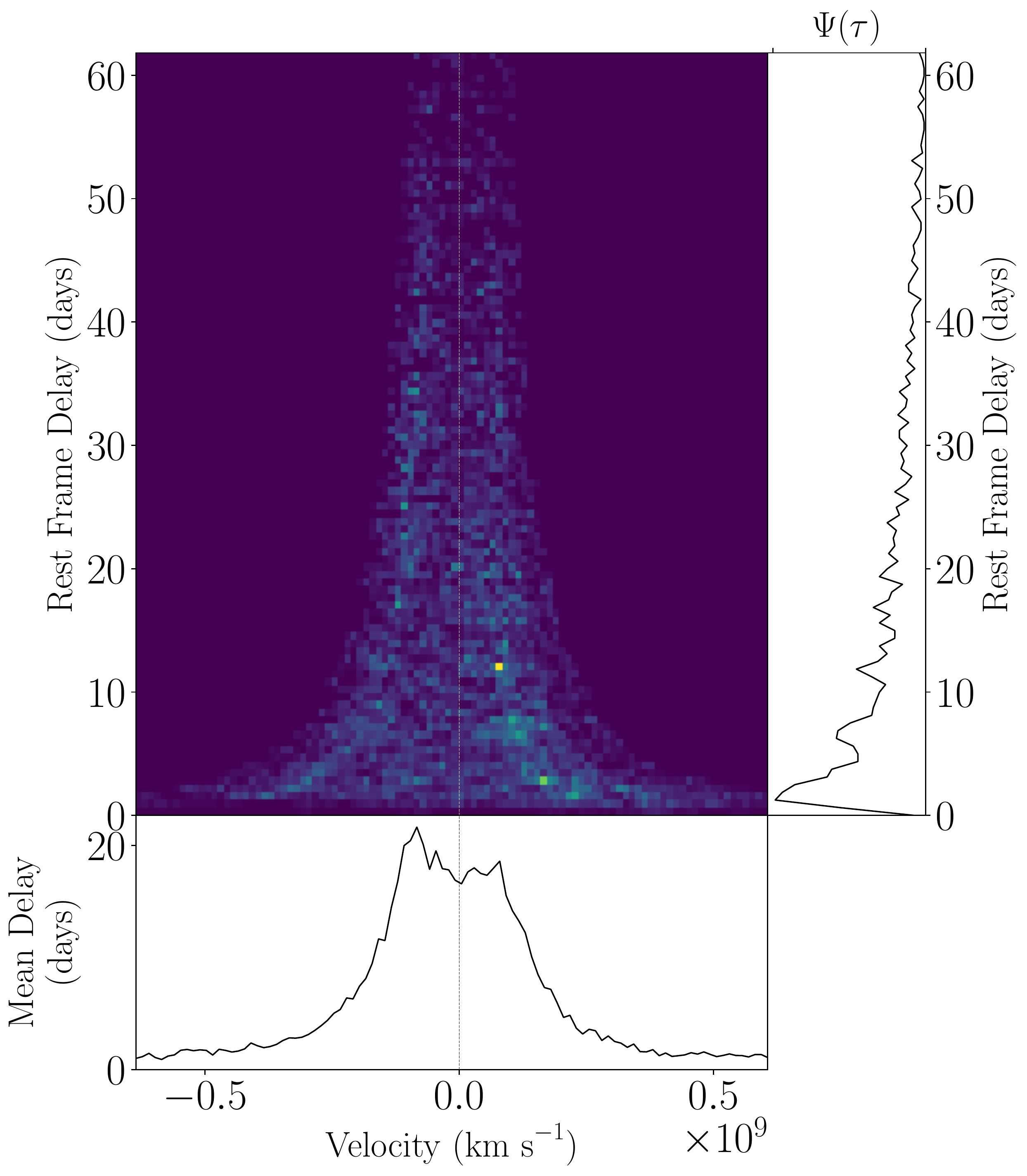}
    \caption{\mrkb\ transfer function produced using median model parameter estimates. See Figure \ref{fig:ark120_tf} caption for panel descriptions. }
    \label{fig:mrk110_tf}
\end{figure}

\begin{figure}[h]
    \centering
    \includegraphics[height=9cm, keepaspectratio]{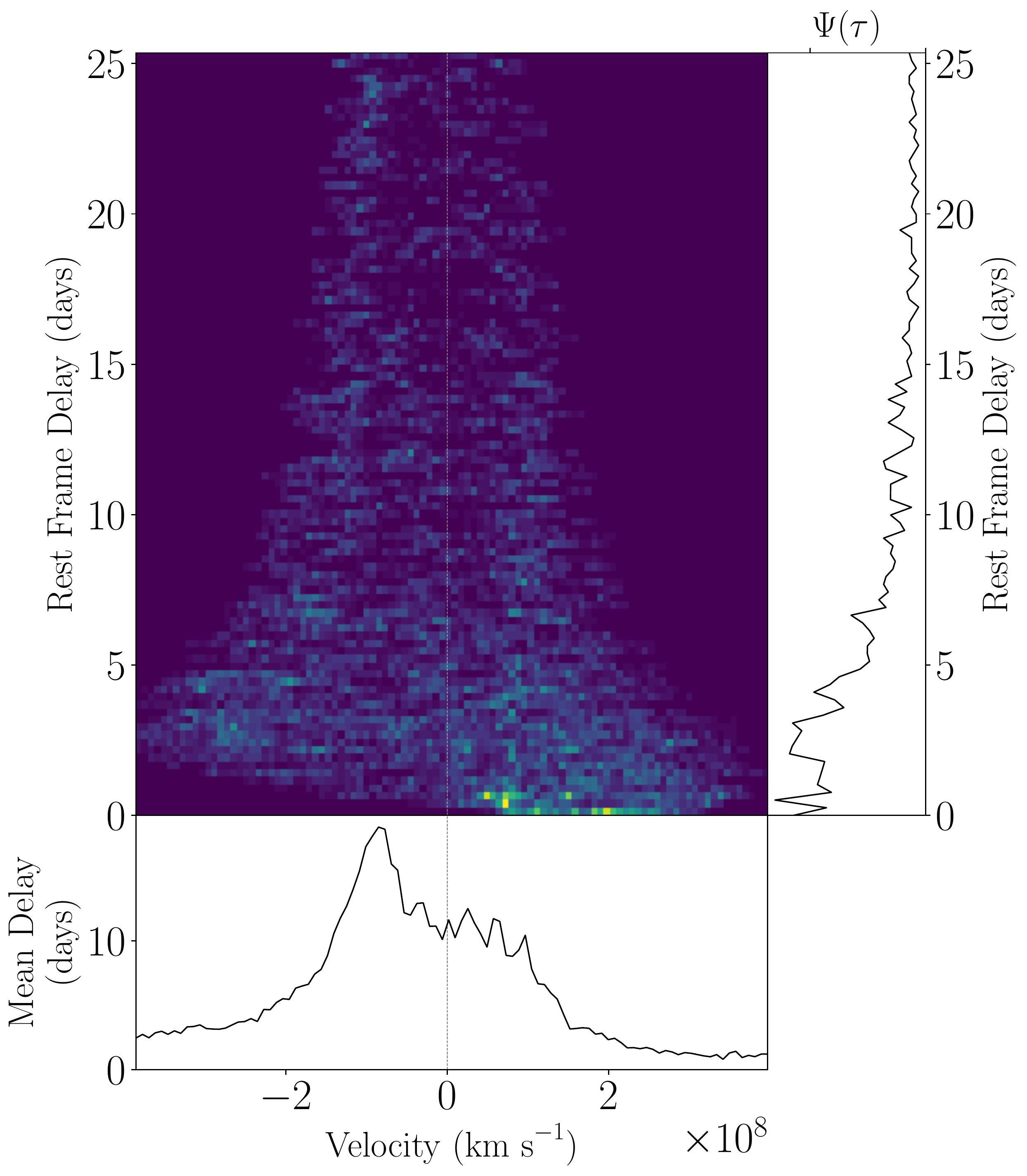}
    \caption{\mrkc\ transfer function produced using median model parameter estimates. See Figure \ref{fig:ark120_tf} caption for panel descriptions.}
    \label{fig:mrk9_tf}
\end{figure}

\begin{acknowledgements}

We are grateful to the following individuals for their time and effort contributed to the Lick observing campaign: (Shane) Zachary Parsons, Estefania Padilla Gonzalez, Noah Rivera, Cristilyn Gardner, Jake Haslemann, Sean Lewis, and Ellen Glad; (Nickel) Nick Choksi, Sameen Yunus, Jeffrey Molloy, Andrew Rikhter, and Haynes Stephens. 

Photometric data collection at MLO was supported by National Science Foundation (NSF) grant AST-1210311; we thank Robert Quimby, Emma Lee, Joseph Tinglof, Eric McLaughlin, Amy Igarashi, and Tariq Johnson for assistance with these observations. We are deeply grateful to the UCO/Lick staff for help with scheduling and supporting the observations. Research at Lick Observatory is partially supported by a generous gift from Google. 

The Kast red CCD detector upgrade, led by B. Holden, was made possible by the Heising-Simons Foundation, William and Marina Kast, and the University of California Observatories. KAIT and its ongoing operation were made possible by donations from Sun Microsystems, Inc., the Hewlett-Packard Company, AutoScope Corporation, Lick Observatory, the NSF, the University of California, the Sylvia \& Jim Katzman Foundation, and the TABASGO Foundation.

L.V., P.R.W, and T.T. acknowledge support from the NSF through grant AST-1907208 ``Collaborative Research: Establishing the Foundations of Black Hole Mass Measurements of AGN across Cosmic Time."
Data presented herein were obtained using the UCI Remote Observing Facility, made possible by a generous gift from John and Ruth Ann Evans. Research at UC Irvine has been supported by NSF grants AST-1412693 and AST-1907208. V.U acknowledges funding support from the University of California Riverside's Chancellor's Postdoctoral Fellowship and National Aeronautics and Space Administration (NASA) Astrophysics Data Analysis Program Grant \#80NSSC20K0450. Her work was conducted in part at the Aspen Center for Physics, which is supported by NSF grant PHY-1607611; she thanks the Center for its hospitality during the ``Astrophysics of Massive Black Holes Merger" workshop in June and July 2018.  T.T. acknowledges support by the Packard Foundation through a Packard research fellowship. V.N.B. and I.S. gratefully acknowledge assistance from NSF Research at Undergraduate Institutions (RUI) grants AST-1312296 and AST-1909297. Note that findings and conclusions do not necessarily represent views of the NSF. 

M.C.B. gratefully acknowledges support from the NSF through grant AST-2009230. G.C. acknowledges NSF support under grant AST-1817233. J.H.W. acknowledges funding from the Basic Science Research Program through the National Research Foundation of the Korean Government (NRF-2021R1A2C3008486). A.V.F.'s group at U.C. Berkeley is grateful for support from the TABASGO Foundation, the Christopher R. Redlich Fund, the Miller Institute for Basic Research in Science (in which he is a Miller Senior Fellow), and many individual donors. K.H. acknowledges support from STFC grant ST/R000824/1. We thank Marc J. Staley for a fellowship that partly funded B.E.S. whilst contributing to the work presented herein as a graduate student. I.S. acknowledges support from the Deutsche Forschungsgemeinschaft (DFG, German Research Foundation) under Germany’s Excellence Strategy --- EXC 2121 ``Quantum Universe" --- 390833306. Research by S.V. is supported by NSF grants AST–1813176 and AST-2008108.

This work makes use of observations from the LCOGT network. The Liverpool Telescope is operated on the island of La Palma by Liverpool John Moores University in the Spanish Observatorio del Roque de los Muchachos of the Instituto de Astrofisica de Canarias with financial support from the UK Science and Technology Facilities Council. Based on observations acquired at the Observatorio Astron\'{o}mico Nacional in the Sierra San Pedro M\'{a}rtir (OAN-SPM), Baja California, M\'{e}xico, we thank the daytime and night support staff at the OAN-SPM for facilitating and helping obtain our observations. Some of the data used in this paper were acquired with the RATIR instrument, funded by the University of California and NASA Goddard Space Flight Center, and the 1.5\,m Harold L.\ Johnson telescope at the Observatorio Astron\'omico Nacional on the Sierra de San Pedro M\'artir, operated and maintained by the Observatorio Astron\'omico Nacional and the Instituto de Astronom{\'\i}a of the Universidad Nacional Aut\'onoma de M\'exico. Operations are partially funded by the Universidad Nacional Aut\'onoma de M\'exico (DGAPA/PAPIIT IG100414, IT102715, AG100317, IN109418, IG100820, and IN105921). We acknowledge the contribution of Leonid Georgiev and Neil Gehrels to the development of RATIR.

This research was made possible through the use of the AAVSO Photometric All-Sky Survey (APASS), funded by the Robert Martin Ayers Sciences Fund and NSF grant AST-1412587 and contributed by observers worldwide. We acknowledge the use of The AGN Black Hole Mass Database as a compilation of some of the reverberation mapped black hole masses prior to 2015~\cite[][]{Bentz15}. This research has made use of the
NASA/IPAC Extragalactic Database (NED), which is operated by the Jet Propulsion Laboratory, California Institute of Technology, under contract with NASA. 
\end{acknowledgements}
\begin{appendix} \label{appendix}
We include a summary of our model parameter estimates for the three sources excluded from this work (owing to moderate quality model fits) in Table \ref{table_results_excluded} and their corresponding geometric interpretations in Figure \ref{fig:geo_excluded}. The model fits and a full detailed description of the BLR structure and kinematics for each source are found in the sections below. Transfer functions produced using the median model parameter estimates for each source are shown in Figures \ref{fig:ark120_tf}, \ref{fig:mrk110_tf}, and \ref{fig:mrk9_tf}.
\begin{deluxetable*}{lcccc}
\tablecaption{BLR Model Parameter Values}
\addtolength{\tabcolsep}{+4.5pt}
\tablehead{ 
\colhead{Parameter} & \colhead{${\rm Ark~ 120}$} & 
\colhead{${\rm Mrk ~ 110}$} &
\colhead{${\rm  Mrk ~ 9}$} 
}
\startdata
$\log_{10}(M_{\rm bh}/M_{\odot})$  & 
\arklogmbh & \mrkblogmbh & \mrkclogmbh \\
% $8.26^{+0.12}_{-0.17}$ & $7.17^{+0.67}_{-0.26}$ & $7.09^{+0.22}_{-0.23}$ \\
$r_{\rm mean}$ (light-days)  & \arkrmean & \mrkbrmean & \mrkcrmean \\
%$19.2^{+2.6}_{-2.2}$ & $17.6^{+1.6}_{-1.5}$ & $11.8^{+3.6}_{-2.7}$ \\
$r_{\rm median}$ (light-days)  & \arkrmedian & \mrkbrmedian & \mrkcrmedian \\
%$17.9^{+2.1}_{-2.4}$& $13.9^{+2.0}_{-1.8}$ & $8.0^{+2.8}_{-1.9}$\\
$r_{\rm min}$ (light-days)  & \arkrmin & \mrkbrmin & \mrkcrmin \\
%$1.16^{+1.40}_{-0.89}$ & $1.22^{+0.40}_{-0.44}$ & $2.21^{+0.81}_{-0.67}$ \\
$\sigma_r$ (light-days)  & \arksigmar & \mrkbsigmar & \mrkcsigmar \\
%$36^{+47}_{-21}$& $47^{+12}_{-14}$ & $36^{+150}_{-18}$ & \\
$\tau_{\rm mean}$ (days)  & \arktaumean & \mrkbtaumean & \mrkctaumean \\
%$12.8^{+1.4}_{-1.3}$& $19^{+2.0}_{-1.9}$ & $10.1^{+2.3}_{-2.2}$ \\
$\tau_{\rm median}$ (days)  & \arktaumedian & \mrkbtaumedian & \mrkctaumedian \\
%$11.0^{+1.5}_{-1.9}$ & $13.7^{+1.9}_{-1.8}$ & $5.6^{+1.5}_{-1.5}$ \\
$\beta$  & \arkbeta & \mrkbbeta & \mrkcbeta \\
%$0.89^{+0.09}_{-0.10}$& $1.20^{+0.09}_{-0.09}$ & $1.44^{+0.12}_{-0.15}$ \\
$\theta_o$ (degrees) & \arkthetao & \mrkbthetao & \mrkcthetao \\
%$31.9^{+7.0}_{-8.1}$& $27^{+16}_{-13}$ & $45^{+17}_{-17}$ \\
$\theta_i$  (degrees)  & \arkthetai & \mrkbthetai & \mrkcthetai \\
%$13.6^{+3.4}_{-3.2}$ & $19.9^{+9.6}_{-12}$ & $42^{+12}_{-15}$ \\
$\kappa$  & \arkkappa & \mrkbkappa & \mrkckappa \\
%$0.26^{+0.18}_{-0.22}$ & $-0.41^{+0.42}_{-0.06}$ & $0.02^{+0.11}_{-0.11}$ \\
$\gamma$  & \arkgamma & \mrkbgamma & \mrkcgamma \\
%$1.7^{+0.2}_{-0.6}$ & $1.6^{+0.3}_{-0.4}$ & $1.6^{+0.3}_{-0.3}$ \\
$\xi$  & \arkxi & \mrkbxi & \mrkcxi \\
%$0.02^{+0.04}_{-0.01}$ & $0.88^{+0.09}_{-0.19}$ & $0.52^{+0.21}_{-0.20}$ \\
$f_{\rm ellip}$  & \arkellip & \mrkbellip & \mrkcellip \\
%$0.14^{+0.02}_{-0.03}$ & $0.60^{+0.15}_{-0.20}$ & $0.12^{+0.17}_{-0.08}$ \\
$f_{\rm flow}$  &  \arkflow & \mrkbflow & \mrkcflow \\
%$0.25^{+0.17}_{-0.17}$ & $0.66^{+0.22}_{-0.39}$ & $0.27^{+0.20}_{-0.18}$ \\
$\theta_e$ (degrees)  & \arkthetae & \mrkbthetae & \mrkcthetae \\
%$7.2^{+6.5}_{-4.8}$ & $13.7^{+16}_{-9.9}$ & $45^{+15}_{-28}$ \\
${\rm In. - Out.}$  & \arkinout & \mrkbinout & \mrkcinout \\
%$-0.85^{+0.02}_{-0.03}$ & $0.30^{+0.21}_{-0.63}$ & $-0.59^{+0.23}_{-0.20}$ \\
$\sigma_{\rm turb}$  & \arksigmaturb & \mrkbsigmaturb & \mrkcsigmaturb \\
%$0.006^{+0.016}_{-0.004}$ & $0.011^{+0.038}_{-0.008}$ & $0.010^{+0.040}_{-0.008}$ \\
%$r_{\rm out}$ $({\rm light~days})$  & $56$ & $195$ & $240$\\
%$T$  & $500$ & $135$ & $600$\\
\enddata
\tablecomments{Median values and 68\% confidence intervals for BLR model parameters for three sources modeled but excluded from this work owing to moderate model fits.
\label{table_results_excluded}}
\end{deluxetable*}
\section{\ark}
\begin{figure}[h!]
    \centering
    \includegraphics[height=15cm, keepaspectratio]{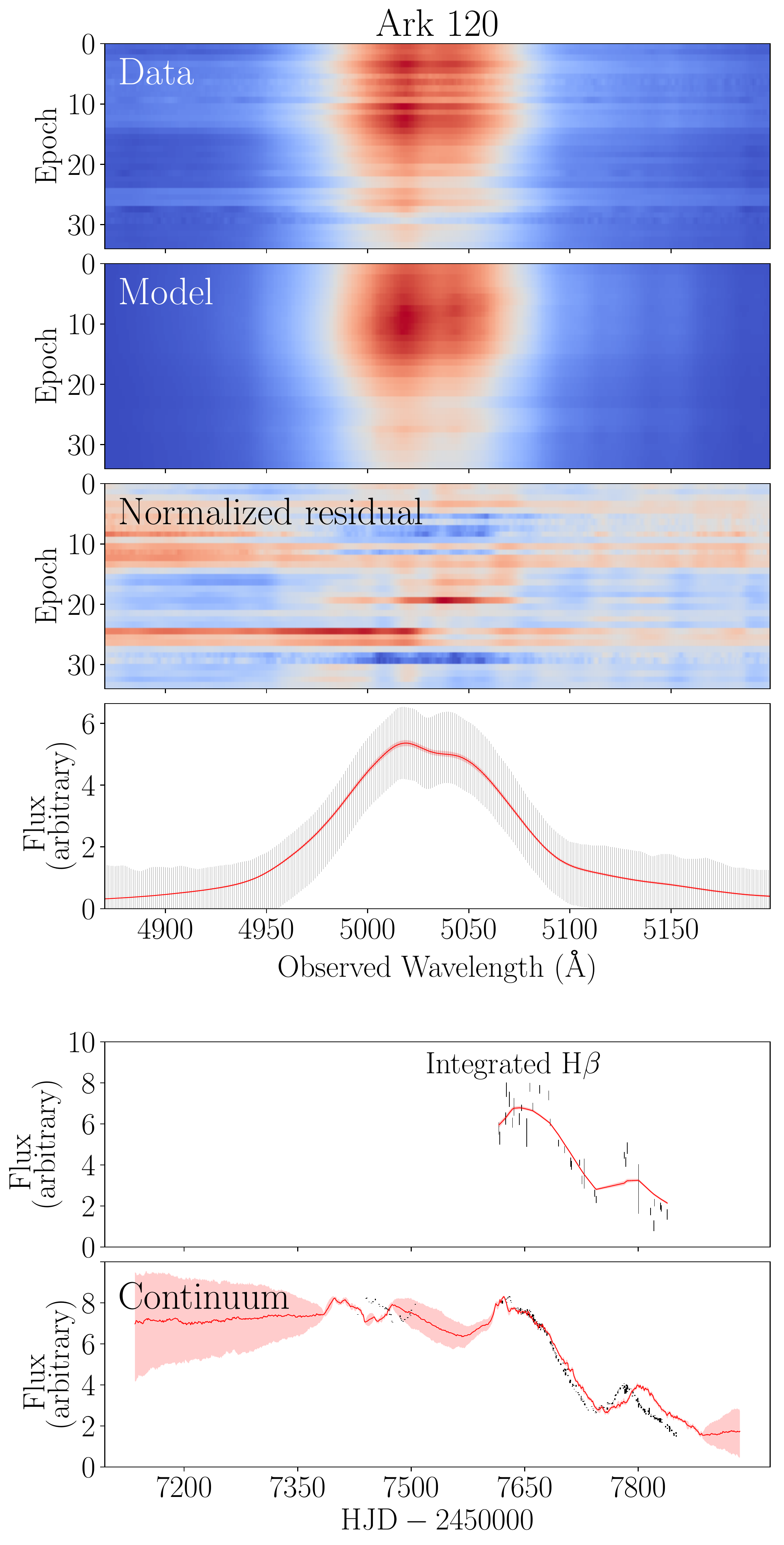}
    \caption{Model fits to the \Hb line profile, integrated \Hb flux, and AGN continuum flux for \ark. Labeling panels 1-6 from top to bottom, panels 1 and 2 show the observed intensity of the \Hb emission-line profile by observation epoch and the profile produced by one sample of the \textsc{caramel} BLR and continuum model. Panel 3 shows the resulting normalized residual. Panel 4 shows the observed \Hb profile of one randomly chosen epoch in black and the corresponding profile produced by the model in panel 2, in red. The corresponding error bars of the observed epoch have been multiplied by $\sqrt{T}$, where $T$ is the \textsc{dnest4} statistical ``temperature" that is used as a likelihood softening parameter post analysis.  Panels 5 \& 6 show the time series of the observed integrated \Hb and continuum flux in black and the corresponding model fits (of the model shown in panel 2) of the light curves in red.}
    \label{fig:ark120_modelfit}
\end{figure}

Our model was able to fit the large-scale variations in the integrated \Hb emission line and shape of the line profile relatively well, only missing some of the finer fluctuations of the integrated emission line toward later epochs and some finer variations in intensity toward the start of the campaign (see panels 5 and 2, respectively, in Figure \ref{fig:ark120_modelfit}). Ultimately, we decided to exclude this source from our analysis due to our model's inability to fit the continuum light curve (see panel 6) toward the end of the observational campaign. Considering that the model \Hb emission line long-scale variations fit the data pretty well, the structure and kinematics of \ark\ can be described by the our model description below.

Geometrically, the BLR is modeled as a thick disk ($\theta_o =$ \arkthetao\ degrees) inclined $\theta_i =$ \arkthetai\ degrees toward an observer with a median BLR radius of $r_{\rm{median}} =$ \arkrmedian\ light-days. The data best fit a mostly opaque BLR midplane with $\xi =$ \arkxi, slight preferrential emission from the near side of the BLR ($\kappa =$ \arkkappa), and slightly concentrated emission at the edges ($\gamma =$ \arkgamma). Dynamically, our model suggests that $\sim 14\%$ of particles have nearly circular orbits with ($f_{\rm{ellip}}) = $ \arkellip, with the remaining particles having velocities drawn from a Gaussian $v_r - v_{\phi}$ distribution rotated $\theta_e =$ \arkthetae\ degrees from radially inflowing ($f_{\rm flow}$ = \arkflow) escape velocity to circular velocity. The contribution from macroturbulent velocities is small, with $\sigma_{\rm turb} =$ \arksigmaturb. Finally, we estimate a black hole mass of $\log_{10}(M_{\rm bh}/M_{\odot}) =$ \arklogmbh\ that is consistent within $\sim 1.3 \sigma$ with the estimate $\log_{10}(M_{\rm bh}/M_{\odot}) = 7.86^{+0.14}_{-0.14}$ found by \citet{u2021lick}, with their standard assumption of virial coefficient \logfrmssigma$=0.65$. 

\section{\mrkb}
\begin{figure}[h!]
    \centering
    \includegraphics[height=15cm, keepaspectratio]{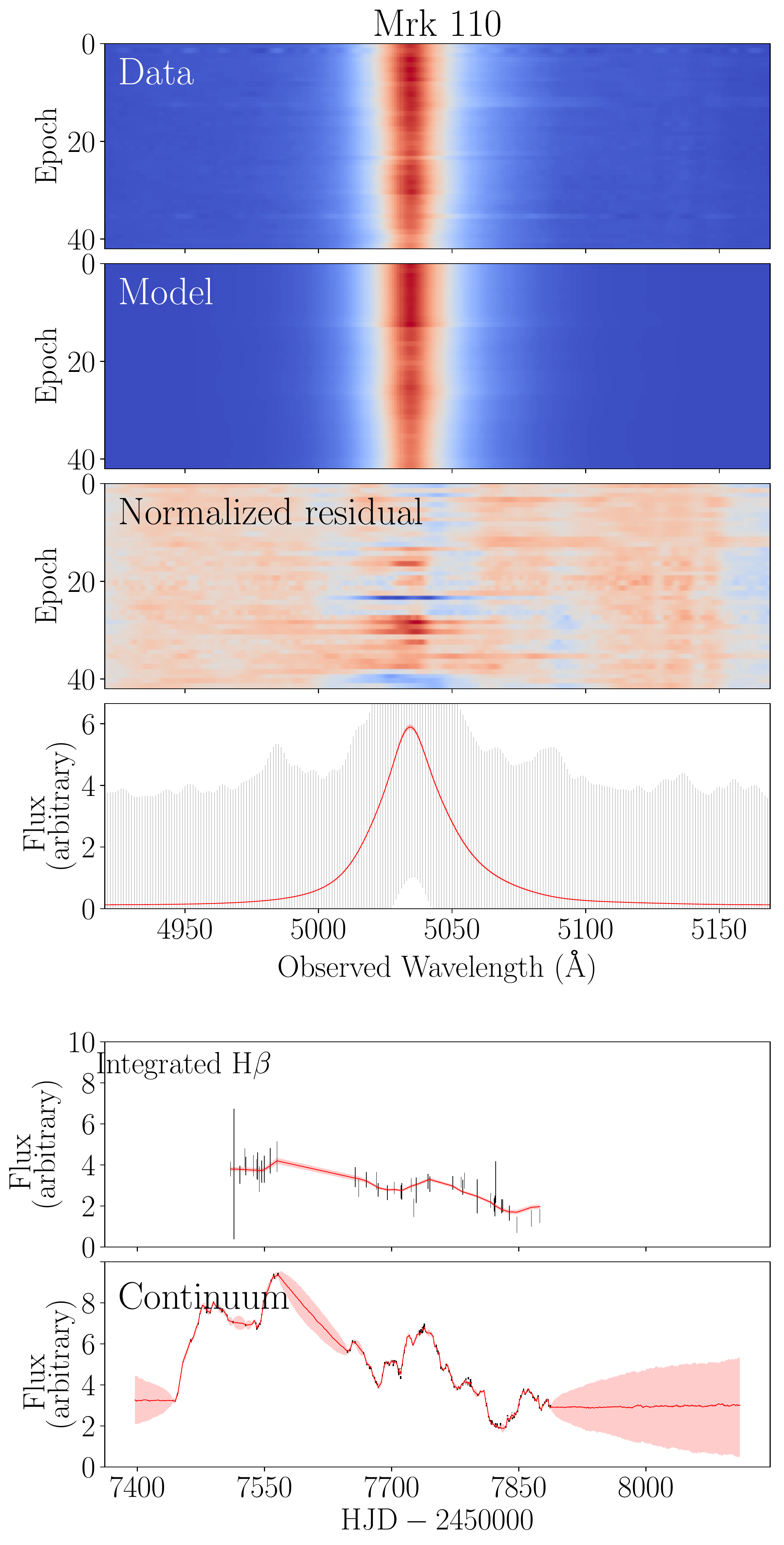}
    \caption{Model fits to the \Hb line profile, integrated \Hb flux, and AGN continuum flux for Mrk 110. See Figure \ref{fig:ark120_modelfit} caption for panel descriptions.}
    \label{fig:mrk110_modelfit}
\end{figure}

Our model was able to fit the large-scale variations in the integrated \Hb emission line and shape of the line profile very well, missing only some of the finer features of the \Hb emission line core toward later epochs (see panel 2 in Figure \ref{fig:mrk110_modelfit}). We now draw attention to panel 4, which depicts the implementation of a large statistical temperature in order to avoid overfitting, but results in very low S/N of the \Hb emission-line profile. Given the large uncertainty in the data (see panel 4) and thus increased (systematic) uncertainty in our model estimates, we decided to exclude the source from our analysis. This increased uncertainty, however, is taken into account in our model estimates which we describe below. 

Our model finds that the BLR is best described by a thick disk ($\theta_o =$ \mrkbthetao\ degrees) inclined $\theta_i = $ \mrkbthetai degrees toward the observer with a median radius of $r_{\rm median} = $ \mrkbrmedian\ light-days. The data favor a transparent BLR midplane ($\xi =$ \mrkbxi) and preferential emission from the far side of the BLR ($\kappa =$ \mrkbkappa). Our model is unable to constrain, however, whether emission is isotropic/concentrated at the edges ($\gamma =$ \mrkbgamma). Dynamically, our model suggests that over half of the particles have nearly circular orbits ($f_{\rm ellip} =$ \mrkbellip), with the remaining particles having velocities drawn from a Gaussian distribution in the $v_r - v_{\phi}$ distribution rotated $\theta_e =$ \mrkbthetae\ degrees from the radially outflowing  ($f_{\rm flow} =$ \mrkbflow) escape velocity toward circular velocity. The contribution from macroturbulent velocities is small, with $\sigma_{\rm turb} = $ \mrkbsigmaturb. Finally, we find a black hole mass of $\log_{10}(M_{\rm bh}/M_{\odot}) =$ \mrkblogmbh, which is consistent with the estimate of $\log_{10}(M_{\rm bh}/M_{\odot}) = 7.54^{+0.08}_{-0.13}$, found by \citet{u2021lick}. 
\section{\mrkc}
\begin{figure}[h]
    \centering
    \includegraphics[height=15cm, keepaspectratio]{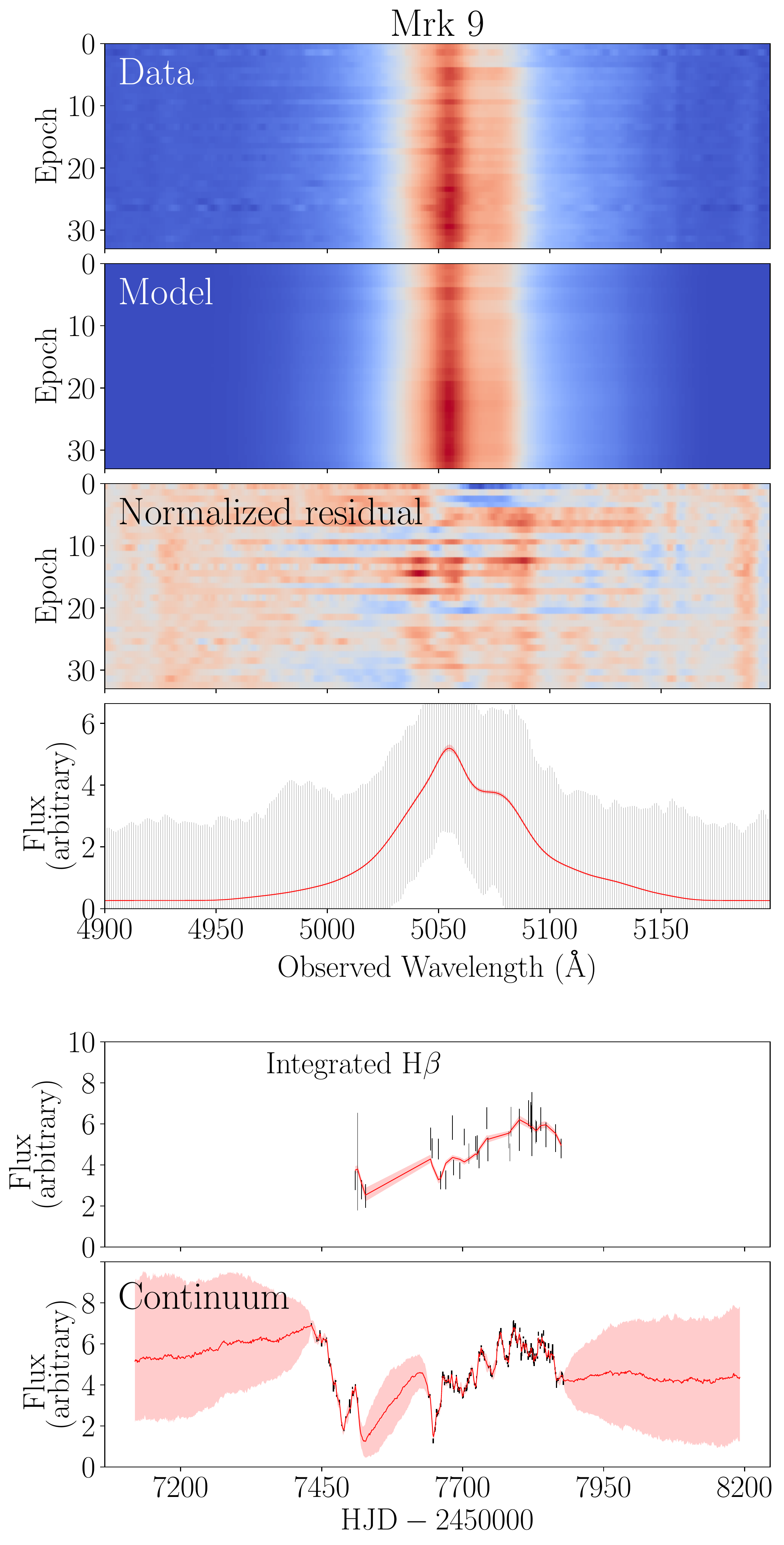}
    \caption{Model fits to the \Hb line profile, integrated \Hb flux, and AGN continuum flux for \mrkc. See Figure \ref{fig:ark120_modelfit} caption for panel descriptions.}
    \label{fig:mrk9_modelfit}
\end{figure}
Similar to the case of \mrkb, our model was able to fit the large-scale variations in the integrated \Hb emission line and shape of the line profile very well for \mrkc. As seen in Figure \ref{fig:mrk9_modelfit}, panel 2, our model only misses some of the finer features of the \Hb emission line core toward earlier epochs. Our model is also able to capture the long-scale variations in the integrated \Hb emission line (panel 5) and AGN continuum (panel 6). However, as seen in panel 4, the model for this source required implementing a large statistical temperature in order to avoid overfitting, which resulted in low S/N of the \Hb emission-line profile. Given the large uncertainty in the data (see panel 4) and thus increased uncertainty in our model estimates, we excluded the source from our analysis. This increased uncertainty, however, is taken into account in our model estimates which we describe below. 

The data best fit a thick disk ($\theta_o =$ \mrkcthetao) H$\beta$-emitting BLR, viewed at an inclination of $\theta_i =$ \mrkcthetai\ degrees with a median radius of $r_{\rm{median}} =$ \mrkcrmedian\ light-days. Our model finds a slight preference for an opaque BLR midplane with $\xi =$ \mrkcxi\ and a mostly isotropic BLR with $\kappa =$ \mrkckappa. The model is unable to constrain whether emission is uniformly emitted or concentrated at the edges ($\gamma =$ \mrkcgamma), however. Dynamically, our model finds $\sim 12\%$ of the particles have circular orbits ($f_{\rm ellip} =$ \mrkcellip). The remaining particles having velocities drawn from a Gaussian $v_r - v_{\phi}$ distribution rotated $\theta_e =$ \mrkcthetae\ degrees from radially inflowing ($f_{\rm flow}$ = \mrkcflow) escape velocity to circular velocity. The contribution from macroturbulent velocities is small, with $\sigma_{\rm turb} =$ \mrkcsigmaturb. Finally, we estimate a black hole mass of $\log_{10}(M_{\rm bh}/M_{\odot}) =$ \mrkclogmbh, which is consistent within $1\sigma$ uncertainties of the estimate $\log_{10}(M_{\rm bh}/M_{\odot}) = 7.61^{+0.12}_{-0.31}$ found by \citet{u2021lick}. 
\end{appendix}
\clearpage
\bibliographystyle{aasjournal}
\bibliography{references}

\begin{thebibliography}{}
\expandafter\ifx\csname natexlab\endcsname\relax\def\natexlab#1{#1}\fi
\providecommand{\url}[1]{\href{#1}{#1}}
\providecommand{\dodoi}[1]{doi:~\href{http://doi.org/#1}{\nolinkurl{#1}}}
\providecommand{\doeprint}[1]{\href{http://ascl.net/#1}{\nolinkurl{http://ascl.net/#1}}}
\providecommand{\doarXiv}[1]{\href{https://arxiv.org/abs/#1}{\nolinkurl{https://arxiv.org/abs/#1}}}

\bibitem[{Agudo {et~al.}(2012)Agudo, G{\'{o}}mez, Casadio, Cawthorne, \&
  Roca-Sogorb}]{Agudo_2012}
Agudo, I., G{\'{o}}mez, J.~L., Casadio, C., Cawthorne, T.~V., \& Roca-Sogorb,
  M. 2012, The Astrophysical Journal, 752, 92,
  \dodoi{10.1088/0004-637x/752/2/92}

\bibitem[{{Barth} {et~al.}(2011{\natexlab{a}}){Barth}, {Nguyen}, {Malkan},
  {Filippenko}, {Li}, {Gorjian}, {Joner}, {Bennert}, {Botyanszki}, {Cenko},
  {Childress}, {Choi}, {Comerford}, {Cucciara}, {da Silva}, {Duch{\^e}ne},
  {Fumagalli}, {Ganeshalingam}, {Gates}, {Gerke}, {Griffith}, {Harris},
  {Hintz}, {Hsiao}, {Kandrashoff}, {Keel}, {Kirkman}, {Kleiser}, {Laney},
  {Lee}, {Lopez}, {Lowe}, {Moody}, {Morton}, {Nierenberg}, {Nugent},
  {Pancoast}, {Rex}, {Rich}, {Silverman}, {Smith}, {Sonnenfeld}, {Suzuki},
  {Tytler}, {Walsh}, {Woo}, {Yang}, \& {Zeisse}}]{barth11}
{Barth}, A.~J., {Nguyen}, M.~L., {Malkan}, M.~A., {et~al.} 2011{\natexlab{a}},
  \apj, 732, 121, \dodoi{10.1088/0004-637X/732/2/121}

\bibitem[{{Barth} {et~al.}(2011{\natexlab{b}}){Barth}, {Pancoast}, {Thorman},
  {Bennert}, {Sand}, {Li}, {Canalizo}, {Filippenko}, {Gates}, {Greene},
  {Malkan}, {Stern}, {Treu}, {Woo}, {Assef}, {Bae}, {Brewer}, {Buehler},
  {Cenko}, {Clubb}, {Cooper}, {Diamond-Stanic}, {Hiner}, {H{\"o}nig}, {Joner},
  {Kandrashoff}, {Laney}, {Lazarova}, {Nierenberg}, {Park}, {Silverman}, {Son},
  {Sonnenfeld}, {Tollerud}, {Walsh}, {Walters}, {da Silva}, {Fumagalli},
  {Gregg}, {Harris}, {Hsiao}, {Lee}, {Lopez}, {Rex}, {Suzuki}, {Trump},
  {Tytler}, {Worseck}, \& {Yesuf}}]{barth11b}
{Barth}, A.~J., {Pancoast}, A., {Thorman}, S.~J., {et~al.} 2011{\natexlab{b}},
  \apjl, 743, L4, \dodoi{10.1088/2041-8205/743/1/L4}

\bibitem[{{Barth} {et~al.}(2015){Barth}, {Bennert}, {Canalizo}, {Filippenko},
  {Gates}, {Greene}, {Li}, {Malkan}, {Pancoast}, {Sand}, {Stern}, {Treu},
  {Woo}, {Assef}, {Bae}, {Brewer}, {Cenko}, {Clubb}, {Cooper},
  {Diamond-Stanic}, {Hiner}, {H{\"o}nig}, {Hsiao}, {Kandrashoff}, {Lazarova},
  {Nierenberg}, {Rex}, {Silverman}, {Tollerud}, \& {Walsh}}]{barth15}
{Barth}, A.~J., {Bennert}, V.~N., {Canalizo}, G., {et~al.} 2015, \apjs, 217,
  26, \dodoi{10.1088/0067-0049/217/2/26}

\bibitem[{{Bentz} \& {Katz}(2015{\natexlab{a}})}]{2015PASP..127...67B}
{Bentz}, M.~C., \& {Katz}, S. 2015{\natexlab{a}}, \pasp, 127, 67,
  \dodoi{10.1086/679601}

\bibitem[{{Bentz} \& {Katz}(2015{\natexlab{b}})}]{Bentz15}
---. 2015{\natexlab{b}}, \pasp, 127, 67, \dodoi{10.1086/679601}

\bibitem[{Bentz {et~al.}(2021)Bentz, Williams, Street, Onken, Valluri, \&
  Treu}]{bentz2021detailed}
Bentz, M.~C., Williams, P.~R., Street, R., {et~al.} 2021, A Detailed View of
  the Broad Line Region in NGC 3783 from Velocity-Resolved Reverberation
  Mapping.
\newblock \doarXiv{2108.00482}

\bibitem[{{Bentz} {et~al.}(2009){Bentz}, {Walsh}, {Barth}, {Baliber},
  {Bennert}, {Canalizo}, {Filippenko}, {Ganeshalingam}, {Gates}, {Greene},
  {Hidas}, {Hiner}, {Lee}, {Li}, {Malkan}, {Minezaki}, {Sakata}, {Serduke},
  {Silverman}, {Steele}, {Stern}, {Street}, {Thornton}, {Treu}, {Wang}, {Woo},
  \& {Yoshii}}]{bentz09}
{Bentz}, M.~C., {Walsh}, J.~L., {Barth}, A.~J., {et~al.} 2009, \apj, 705, 199,
  \dodoi{10.1088/0004-637X/705/1/199}

\bibitem[{{Blandford} \& {McKee}(1982)}]{blandford82}
{Blandford}, R.~D., \& {McKee}, C.~F. 1982, \apj, 255, 419,
  \dodoi{10.1086/159843}

\bibitem[{{Boroson} {et~al.}(2014){Boroson}, {Brown}, {Hjelstrom}, {Howell},
  {Lister}, {Pickles}, {Rosing}, {Saunders}, {Street}, \&
  {Walker}}]{2014SPIE.9149E..1EB}
{Boroson}, T., {Brown}, T., {Hjelstrom}, A., {et~al.} 2014, in Society of
  Photo-Optical Instrumentation Engineers (SPIE) Conference Series, Vol. 9149,
  Observatory Operations: Strategies, Processes, and Systems V, ed. A.~B.
  {Peck}, C.~R. {Benn}, \& R.~L. {Seaman}, 91491E, \dodoi{10.1117/12.2054776}

\bibitem[{{Brewer} {et~al.}(2011){Brewer}, {Treu}, {Pancoast}, {Barth},
  {Bennert}, {Bentz}, {Filippenko}, {Greene}, {Malkan}, \& {Woo}}]{brewer11b}
{Brewer}, B.~J., {Treu}, T., {Pancoast}, A., {et~al.} 2011, \apjl, 733, L33,
  \dodoi{10.1088/2041-8205/733/2/L33}

\bibitem[{{Brown} {et~al.}(2013){Brown}, {Baliber}, {Bianco}, {Bowman},
  {Burleson}, {Conway}, {Crellin}, {Depagne}, {De Vera}, {Dilday}, {Dragomir},
  {Dubberley}, {Eastman}, {Elphick}, {Falarski}, {Foale}, {Ford}, {Fulton},
  {Garza}, {Gomez}, {Graham}, {Greene}, {Haldeman}, {Hawkins}, {Haworth},
  {Haynes}, {Hidas}, {Hjelstrom}, {Howell}, {Hygelund}, {Lister}, {Lobdill},
  {Martinez}, {Mullins}, {Norbury}, {Parrent}, {Paulson}, {Petry}, {Pickles},
  {Posner}, {Rosing}, {Ross}, {Sand}, {Saunders}, {Shobbrook}, {Shporer},
  {Street}, {Thomas}, {Tsapras}, {Tufts}, {Valenti}, {Vander Horst}, {Walker},
  {White}, \& {Willis}}]{Brown++13}
{Brown}, T.~M., {Baliber}, N., {Bianco}, F.~B., {et~al.} 2013, \pasp, 125,
  1031, \dodoi{10.1086/673168}

\bibitem[{{Collin} {et~al.}(2006){Collin}, {Kawaguchi}, {Peterson}, \&
  {Vestergaard}}]{collin06}
{Collin}, S., {Kawaguchi}, T., {Peterson}, B.~M., \& {Vestergaard}, M. 2006,
  \aap, 456, 75, \dodoi{10.1051/0004-6361:20064878}

\bibitem[{{De Rosa} {et~al.}(2015){De Rosa}, {Peterson}, {Ely}, {Kriss},
  {Crenshaw}, {Horne}, {Korista}, {Netzer}, {Pogge}, {Ar{\'e}valo}, {Barth},
  {Bentz}, {Brandt}, {Breeveld}, {Brewer}, {Dalla Bont{\`a}}, {De
  Lorenzo-C{\'a}ceres}, {Denney}, {Dietrich}, {Edelson}, {Evans}, {Fausnaugh},
  {Gehrels}, {Gelbord}, {Goad}, {Grier}, {Grupe}, {Hall}, {Kaastra}, {Kelly},
  {Kennea}, {Kochanek}, {Lira}, {Mathur}, {McHardy}, {Nousek}, {Pancoast},
  {Papadakis}, {Pei}, {Schimoia}, {Siegel}, {Starkey}, {Treu}, {Uttley},
  {Vaughan}, {Vestergaard}, {Villforth}, {Yan}, {Young}, \&
  {Zu}}]{2015ApJ...806..128D}
{De Rosa}, G., {Peterson}, B.~M., {Ely}, J., {et~al.} 2015, \apj, 806, 128,
  \dodoi{10.1088/0004-637X/806/1/128}

\bibitem[{{De Rosa} {et~al.}(2018){De Rosa}, {Fausnaugh}, {Grier}, {Peterson},
  {Denney}, {Horne}, {Bentz}, {Ciroi}, {Dalla Bont{\`a}}, {Joner}, {Kaspi},
  {Kochanek}, {Pogge}, {Sergeev}, {Vestergaard}, {Adams}, {Antognini}, {Araya
  Salvo}, {Armstrong}, {Bae}, {Barth}, {Beatty}, {Bhattacharjee}, {Borman},
  {Boroson}, {Bottorff}, {Brown}, {Brown}, {Brotherton}, {Coker}, {Clanton},
  {Cracco}, {Crawford}, {Croxall}, {Eftekharzadeh}, {Eracleous}, {Fiorenza},
  {Frassati}, {Hawkins}, {Henderson}, {Holoien}, {Hutchison}, {Kellar},
  {Kilerci-Eser}, {Kim}, {King}, {La Mura}, {Laney}, {Li}, {Lochhaas}, {Ma},
  {MacInnis}, {Manne-Nicholas}, {Mason}, {McGraw}, {Mogren}, {Montouri},
  {Moody}, {Mosquera}, {Mudd}, {Musso}, {Nazarov}, {Nguyen}, {Ochner},
  {Okhmat}, {Onken}, {Ou-Yang}, {Pancoast}, {Pei}, {Penny}, {Poleski},
  {Portaluri}, {Prieto}, {Price-Whelan}, {Pulatova}, {Rafter}, {Roettenbacher},
  {Romero-Colmenero}, {Runnoe}, {Schimoia}, {Shappee}, {Sherf}, {Simonian},
  {Siviero}, {Skowron}, {Skowron}, {Somers}, {Spencer}, {Starkey}, {Stevens},
  {Stoll}, {Tamajo}, {Tayar}, {van Saders}, {Valenti}, {Villanueva},
  {Villforth}, {Weiss}, {Winkler}, {Zastrow}, {Zhu}, \&
  {Zu}}]{2018ApJ...866..133D}
{De Rosa}, G., {Fausnaugh}, M.~M., {Grier}, C.~J., {et~al.} 2018, \apj, 866,
  133, \dodoi{10.3847/1538-4357/aadd11}

\bibitem[{{Denney} {et~al.}(2009){Denney}, {Peterson}, {Pogge}, {Adair},
  {Atlee}, {Au-Yong}, {Bentz}, {Bird}, {Brokofsky}, {Chisholm}, {Comins},
  {Dietrich}, {Doroshenko}, {Eastman}, {Efimov}, {Ewald}, {Ferbey}, {Gaskell},
  {Hedrick}, {Jackson}, {Klimanov}, {Klimek}, {Kruse}, {Lad{\'e}route}, {Lamb},
  {Leighly}, {Minezaki}, {Nazarov}, {Onken}, {Petersen}, {Peterson},
  {Poindexter}, {Sakata}, {Schlesinger}, {Sergeev}, {Skolski}, {Stieglitz},
  {Tobin}, {Unterborn}, {Vestergaard}, {Watkins}, {Watson}, \&
  {Yoshii}}]{Denney++09}
{Denney}, K.~D., {Peterson}, B.~M., {Pogge}, R.~W., {et~al.} 2009, \apjl, 704,
  L80, \dodoi{10.1088/0004-637X/704/2/L80}

\bibitem[{{Denney} {et~al.}(2010){Denney}, {Peterson}, {Pogge}, {Adair},
  {Atlee}, {Au-Yong}, {Bentz}, {Bird}, {Brokofsky}, {Chisholm}, {Comins},
  {Dietrich}, {Doroshenko}, {Eastman}, {Efimov}, {Ewald}, {Ferbey}, {Gaskell},
  {Hedrick}, {Jackson}, {Klimanov}, {Klimek}, {Kruse}, {Lad{\'e}route}, {Lamb},
  {Leighly}, {Minezaki}, {Nazarov}, {Onken}, {Petersen}, {Peterson},
  {Poindexter}, {Sakata}, {Schlesinger}, {Sergeev}, {Skolski}, {Stieglitz},
  {Tobin}, {Unterborn}, {Vestergaard}, {Watkins}, {Watson}, \&
  {Yoshii}}]{denney10}
---. 2010, \apj, 721, 715, \dodoi{10.1088/0004-637X/721/1/715}

\bibitem[{{Du} {et~al.}(2015){Du}, {Hu}, {Lu}, {Huang}, {Cheng}, {Qiu}, {Li},
  {Zhang}, {Fan}, {Bai}, {Bian}, {Yuan}, {Kaspi}, {Ho}, {Netzer}, {Wang}, \&
  {SEAMBH Collaboration}}]{Du++15}
{Du}, P., {Hu}, C., {Lu}, K.-X., {et~al.} 2015, \apj, 806, 22,
  \dodoi{10.1088/0004-637X/806/1/22}

\bibitem[{{Du} {et~al.}(2016){Du}, {Lu}, {Hu}, {Qiu}, {Li}, {Huang}, {Wang},
  {Bai}, {Bian}, {Yuan}, {Ho}, {Wang}, \& {SEAMBH Collaboration}}]{Du++16}
{Du}, P., {Lu}, K.-X., {Hu}, C., {et~al.} 2016, \apj, 820, 27,
  \dodoi{10.3847/0004-637X/820/1/27}

\bibitem[{{Du} {et~al.}(2018){Du}, {Brotherton}, {Wang}, {Huang}, {Hu},
  {Kasper}, {Chick}, {Nguyen}, {Maithil}, {Hand}, {Li}, {Ho}, {Bai}, {Bian},
  {Wang}, \& {MAHA Collaboration}}]{2018ApJ...869..142D}
{Du}, P., {Brotherton}, M.~S., {Wang}, K., {et~al.} 2018, \apj, 869, 142,
  \dodoi{10.3847/1538-4357/aaed2c}

\bibitem[{{Feng} {et~al.}(2021){Feng}, {Liu}, {Bai}, {Yang}, {Hu}, {Li},
  {Yang}, {Lu}, \& {Xiao}}]{2021ApJ...912...92F}
{Feng}, H.-C., {Liu}, H.~T., {Bai}, J.~M., {et~al.} 2021, \apj, 912, 92,
  \dodoi{10.3847/1538-4357/abefe0}

\bibitem[{{Filippenko} {et~al.}(2001){Filippenko}, {Li}, {Treffers}, \&
  {Modjaz}}]{filippenko01}
{Filippenko}, A.~V., {Li}, W.~D., {Treffers}, R.~R., \& {Modjaz}, M. 2001, in
  Astronomical Society of the Pacific Conference Series, Vol. 246, IAU Colloq.
  183: Small Telescope Astronomy on Global Scales, ed. B.~{Paczynski}, W.-P.
  {Chen}, \& C.~{Lemme}, 121

\bibitem[{{Goad} {et~al.}(2012){Goad}, {Korista}, \& {Ruff}}]{goad12}
{Goad}, M.~R., {Korista}, K.~T., \& {Ruff}, A.~J. 2012, \mnras, 426, 3086,
  \dodoi{10.1111/j.1365-2966.2012.21808.x}

\bibitem[{{Grier} {et~al.}(2017){Grier}, {Pancoast}, {Barth}, {Fausnaugh},
  {Brewer}, {Treu}, \& {Peterson}}]{Grier++17}
{Grier}, C.~J., {Pancoast}, A., {Barth}, A.~J., {et~al.} 2017, \apj, 849, 146,
  \dodoi{10.3847/1538-4357/aa901b}

\bibitem[{{Grier} {et~al.}(2013){Grier}, {Martini}, {Watson}, {Peterson},
  {Bentz}, {Dasyra}, {Dietrich}, {Ferrarese}, {Pogge}, \& {Zu}}]{grier13b}
{Grier}, C.~J., {Martini}, P., {Watson}, L.~C., {et~al.} 2013, \apj, 773, 90,
  \dodoi{10.1088/0004-637X/773/2/90}

\bibitem[{{Horne}(1994)}]{horne94}
{Horne}, K. 1994, in Astronomical Society of the Pacific Conference Series,
  Vol.~69, Reverberation Mapping of the Broad-Line Region in Active Galactic
  Nuclei, ed. P.~M. {Gondhalekar}, K.~{Horne}, \& B.~M. {Peterson}, 23--25

\bibitem[{{Horne} {et~al.}(2004){Horne}, {Peterson}, {Collier}, \&
  {Netzer}}]{2004PASP..116..465H}
{Horne}, K., {Peterson}, B.~M., {Collier}, S.~J., \& {Netzer}, H. 2004, \pasp,
  116, 465, \dodoi{10.1086/420755}

\bibitem[{{Jorstad} {et~al.}(2005){Jorstad}, {Marscher}, {Lister}, {Stirling},
  {Cawthorne}, {Gear}, {G{\'o}mez}, {Stevens}, {Smith}, {Forster}, \&
  {Robson}}]{2005AJ....130.1418J}
{Jorstad}, S.~G., {Marscher}, A.~P., {Lister}, M.~L., {et~al.} 2005, \aj, 130,
  1418, \dodoi{10.1086/444593}

\bibitem[{{Kara} {et~al.}(2021){Kara}, {Mehdipour}, {Kriss}, {Cackett}, {Arav},
  {Barth}, {Byun}, {Brotherton}, {De Rosa}, {Gelbord}, {Hern{\'a}ndez
  Santisteban}, {Hu}, {Kaastra}, {Landt}, {Li}, {Miller}, {Montano},
  {Partington}, {Aceituno}, {Bai}, {Bao}, {Bentz}, {Brink}, {Chelouche},
  {Chen}, {Colmenero}, {Bont{\`a}}, {Dehghanian}, {Du}, {Edelson}, {Ferland},
  {Ferrarese}, {Fian}, {Filippenko}, {Fischer}, {Goad}, {Gonz{\'a}lez
  Buitrago}, {Gorjian}, {Grier}, {Guo}, {Hall}, {Ho}, {Homayouni}, {Horne},
  {Ili{\'c}}, {Jiang}, {Joner}, {Kaspi}, {Kochanek}, {Korista}, {Kynoch}, {Li},
  {Liu}, {McHardy}, {McLane}, {Mitchell}, {Netzer}, {Olson}, {Pogge},
  {Popovi{\'c}}, {Proga}, {Storchi-Bergmann}, {Strasburger}, {Treu},
  {Vestergaard}, {Wang}, {Ward}, {Waters}, {Williams}, {Yang}, {Yao},
  {Zastrocky}, {Zhai}, \& {Zu}}]{2021ApJ...922..151K}
{Kara}, E., {Mehdipour}, M., {Kriss}, G.~A., {et~al.} 2021, \apj, 922, 151,
  \dodoi{10.3847/1538-4357/ac2159}

\bibitem[{{Kelly}(2007)}]{Kelly07}
{Kelly}, B.~C. 2007, \apj, 665, 1489, \dodoi{10.1086/519947}

\bibitem[{{Li} {et~al.}(2021){Li}, {Yang}, {Yang}, {Chen}, {Songsheng}, {Liu},
  {Du}, {Luo}, {Yu}, {Hu}, {Jiang}, {Bao}, {Guo}, {Zhang}, {Li}, {Xiao}, {Lu},
  {Ho}, {Bai}, {Bian}, {Aceituno}, {Minezaki}, {Horne}, {Kokubo}, \&
  {Wang}}]{2021ApJ...920....9L}
{Li}, S.-S., {Yang}, S., {Yang}, Z.-X., {et~al.} 2021, \apj, 920, 9,
  \dodoi{10.3847/1538-4357/ac116e}

\bibitem[{{Pancoast} {et~al.}(2011){Pancoast}, {Brewer}, \&
  {Treu}}]{pancoast11}
{Pancoast}, A., {Brewer}, B.~J., \& {Treu}, T. 2011, \apj, 730, 139,
  \dodoi{10.1088/0004-637X/730/2/139}

\bibitem[{{Pancoast} {et~al.}(2014){Pancoast}, {Brewer}, {Treu}, {Park},
  {Barth}, {Bentz}, \& {Woo}}]{pancoast14b}
{Pancoast}, A., {Brewer}, B.~J., {Treu}, T., {et~al.} 2014, \mnras, 445, 3073,
  \dodoi{10.1093/mnras/stu1419}

\bibitem[{{Pancoast} {et~al.}(2018){Pancoast}, {Barth}, {Horne}, {Treu},
  {Brewer}, {Bennert}, {Canalizo}, {Gates}, {Li}, {Malkan}, {Sand}, {Schmidt},
  {Valenti}, {Woo}, {Clubb}, {Cooper}, {Crawford}, {H{\"o}nig}, {Joner},
  {Kandrashoff}, {Lazarova}, {Nierenberg}, {Romero-Colmenero}, {Son},
  {Tollerud}, {Walsh}, \& {Winkler}}]{2018ApJ...856..108P}
{Pancoast}, A., {Barth}, A.~J., {Horne}, K., {et~al.} 2018, \apj, 856, 108,
  \dodoi{10.3847/1538-4357/aab3c6}

\bibitem[{{Pei} {et~al.}(2017){Pei}, {Fausnaugh}, {Barth}, {Peterson}, {Bentz},
  {De Rosa}, {Denney}, {Goad}, {Kochanek}, {Korista}, {Kriss}, {Pogge},
  {Bennert}, {Brotherton}, {Clubb}, {Dalla Bont{\`a}}, {Filippenko}, {Greene},
  {Grier}, {Vestergaard}, {Zheng}, {Adams}, {Beatty}, {Bigley}, {Brown},
  {Brown}, {Canalizo}, {Comerford}, {Coker}, {Corsini}, {Croft}, {Croxall},
  {Deason}, {Eracleous}, {Fox}, {Gates}, {Henderson}, {Holmbeck}, {Holoien},
  {Jensen}, {Johnson}, {Kelly}, {Kim}, {King}, {Lau}, {Li}, {Lochhaas}, {Ma},
  {Manne-Nicholas}, {Mauerhan}, {Malkan}, {McGurk}, {Morelli}, {Mosquera},
  {Mudd}, {Muller Sanchez}, {Nguyen}, {Ochner}, {Ou-Yang}, {Pancoast}, {Penny},
  {Pizzella}, {Poleski}, {Runnoe}, {Scott}, {Schimoia}, {Shappee}, {Shivvers},
  {Simonian}, {Siviero}, {Somers}, {Stevens}, {Strauss}, {Tayar}, {Tejos},
  {Treu}, {Van Saders}, {Vican}, {Villanueva}, {Yuk}, {Zakamska}, {Zhu},
  {Anderson}, {Ar{\'e}valo}, {Bazhaw}, {Bisogni}, {Borman}, {Bottorff},
  {Brandt}, {Breeveld}, {Cackett}, {Carini}, {Crenshaw}, {De
  Lorenzo-C{\'a}ceres}, {Dietrich}, {Edelson}, {Efimova}, {Ely}, {Evans},
  {Ferland}, {Flatland}, {Gehrels}, {Geier}, {Gelbord}, {Grupe}, {Gupta},
  {Hall}, {Hicks}, {Horenstein}, {Horne}, {Hutchison}, {Im}, {Joner}, {Jones},
  {Kaastra}, {Kaspi}, {Kelly}, {Kennea}, {Kim}, {Kim}, {Klimanov}, {Lee},
  {Leonard}, {Lira}, {MacInnis}, {Mathur}, {McHardy}, {Montouri}, {Musso},
  {Nazarov}, {Netzer}, {Norris}, {Nousek}, {Okhmat}, {Papadakis}, {Parks},
  {Pott}, {Rafter}, {Rix}, {Saylor}, {Schn{\"u}lle}, {Sergeev}, {Siegel},
  {Skielboe}, {Spencer}, {Starkey}, {Sung}, {Teems}, {Turner}, {Uttley},
  {Villforth}, {Weiss}, {Woo}, {Yan}, {Young}, \& {Zu}}]{Pei++17}
{Pei}, L., {Fausnaugh}, M.~M., {Barth}, A.~J., {et~al.} 2017, \apj, 837, 131,
  \dodoi{10.3847/1538-4357/aa5eb1}

\bibitem[{{Peterson}(1993)}]{peterson93}
{Peterson}, B.~M. 1993, \pasp, 105, 247, \dodoi{10.1086/133140}

\bibitem[{{Planck Collaboration} {et~al.}(2016){Planck Collaboration}, {Ade},
  {Aghanim}, {Arnaud}, {Ashdown}, {Aumont}, {Baccigalupi}, {Banday},
  {Barreiro}, {Bartlett}, \& et~al.}]{Planck16}
{Planck Collaboration}, {Ade}, P.~A.~R., {Aghanim}, N., {et~al.} 2016, \aap,
  594, A13, \dodoi{10.1051/0004-6361/201525830}

\bibitem[{{Raimundo} {et~al.}(2020){Raimundo}, {Vestergaard}, {Goad}, {Grier},
  {Williams}, {Peterson}, \& {Treu}}]{2020MNRAS.493.1227R}
{Raimundo}, S.~I., {Vestergaard}, M., {Goad}, M.~R., {et~al.} 2020, \mnras,
  493, 1227, \dodoi{10.1093/mnras/staa285}

\bibitem[{{Skielboe} {et~al.}(2015){Skielboe}, {Pancoast}, {Treu}, {Park},
  {Barth}, \& {Bentz}}]{skielboe15}
{Skielboe}, A., {Pancoast}, A., {Treu}, T., {et~al.} 2015, \mnras, 454, 144,
  \dodoi{10.1093/mnras/stv1917}

\bibitem[{{Steele} {et~al.}(2004){Steele}, {Smith}, {Rees}, {Baker}, {Bates},
  {Bode}, {Bowman}, {Carter}, {Etherton}, {Ford}, {Fraser}, {Gomboc}, {Lett},
  {Mansfield}, {Marchant}, {Medrano-Cerda}, {Mottram}, {Raback}, {Scott},
  {Tomlinson}, \& {Zamanov}}]{2004SPIE.5489..679S}
{Steele}, I.~A., {Smith}, R.~J., {Rees}, P.~C., {et~al.} 2004, in Society of
  Photo-Optical Instrumentation Engineers (SPIE) Conference Series, Vol. 5489,
  Ground-based Telescopes, ed. J.~{Oschmann}, Jacobus~M., 679--692,
  \dodoi{10.1117/12.551456}

\bibitem[{U {et~al.}(2022)U, Barth, Vogler, Guo, Treu, Bennert, Canalizo,
  Filippenko, Gates, Hamann, Joner, Malkan, Pancoast, Williams, Woo, Abolfathi,
  Abramson, Armen, Bae, Bohn, Boizelle, Bostroem, Brandel, Brink, Channa,
  Cooper, Cosens, Donohue, Fillingham, Gonz{\'{a}}lez-Buitrago, Halevi, Halle,
  Hood, Horne, Horst, de~Kouchkovsky, Kuhn, Kumar, Leonard, Loveland,
  Manzano-King, McHardy, Michel, Olaes, Park, Park, Pei, Ross, Runco, Samuel,
  S{\'{a}}nchez, Scott, Sexton, Shin, Shivvers, Spencer, Stahl, Stegman,
  Stomberg, Valenti, Villafa{\~{n}}a, Walsh, Yuk, \& Zheng}]{u2021lick}
U, V., Barth, A.~J., Vogler, H.~A., {et~al.} 2022, The Lick {AGN} Monitoring
  Project 2016: Velocity-resolved H$\beta$ Lags in Luminous Seyfert Galaxies,
  American Astronomical Society, \dodoi{10.3847/1538-4357/ac3d26}

\bibitem[{Villafaña {et~al.}(In Preparation)Villafaña, Williams, Treu, U,
  Barth, Bennert, Bentz, Vogler, Guo, Canalizo, Filippenko, Gates, Hamann,
  Joner, Malkan, Pancoast, Williams, Woo, Abolfathi, Abramson, Armen, Bae,
  Bohn, Boizelle, Bostroem, Brandel, Brink, Channa, Cooper, Cosens, Donohue,
  Fillingham, González-Buitrago, Halevi, Halle, Hood, Horne, Horst,
  de~Kouchkovsky, Kuhn, Kumar, Leonard, Loveland, Manzano-King, McHardy,
  Michel, Olaes, Park, Park, Pei, Ross, Runco, Samuel, Sánchez, Scott, Sexton,
  Shin, Shivvers, Spencer, Stahl, Stegman, Stomberg, Valenti, , Walsh, Yuk, \&
  Zheng}]{v2022}
Villafaña, L., Williams, P., Treu, T., {et~al.} In Preparation, Establishing
  the Foundations of BH Mass Estimators I: Searching for Observational Proxies
  to Determine Individual Scale Factors for Future Reverberation Mapping
  Campaigns

\bibitem[{{Whittle}(1992)}]{whittle92}
{Whittle}, M. 1992, \apjs, 79, 49, \dodoi{10.1086/191644}

\bibitem[{{Williams} {et~al.}(2018){Williams}, {Pancoast}, {Treu}, {Brewer},
  {Barth}, {Bennert}, {Buehler}, {Canalizo}, {Cenko}, {Clubb}, {Cooper},
  {Filippenko}, {Gates}, {Hoenig}, {Joner}, {Kandrashoff}, {Laney}, {Lazarova},
  {Li}, {Malkan}, {Rex}, {Silverman}, {Tollerud}, {Walsh}, \&
  {Woo}}]{2018ApJ...866...75W}
{Williams}, P.~R., {Pancoast}, A., {Treu}, T., {et~al.} 2018, \apj, 866, 75,
  \dodoi{10.3847/1538-4357/aae086}

\bibitem[{{Williams} {et~al.}(2020){Williams}, {Pancoast}, {Treu}, {Brewer},
  {Peterson}, {Barth}, {Malkan}, {De Rosa}, {Horne}, {Kriss}, {Arav}, {Bentz},
  {Cackett}, {Dalla Bont{\`a}}, {Dehghanian}, {Done}, {Ferland}, {Grier},
  {Kaastra}, {Kara}, {Kochanek}, {Mathur}, {Mehdipour}, {Pogge}, {Proga},
  {Vestergaard}, {Waters}, {Adams}, {Anderson}, {Ar{\'e}valo}, {Beatty},
  {Bennert}, {Bigley}, {Bisogni}, {Borman}, {Boroson}, {Bottorff}, {Brandt},
  {Breeveld}, {Brotherton}, {Brown}, {Brown}, {Canalizo}, {Carini}, {Clubb},
  {Comerford}, {Corsini}, {Crenshaw}, {Croft}, {Croxall}, {Deason}, {De
  Lorenzo-C{\'a}ceres}, {Denney}, {Dietrich}, {Edelson}, {Efimova}, {Ely},
  {Evans}, {Fausnaugh}, {Filippenko}, {Flatland}, {Fox}, {Gardner}, {Gates},
  {Gehrels}, {Geier}, {Gelbord}, {Gonzalez}, {Gorjian}, {Greene}, {Grupe},
  {Gupta}, {Hall}, {Henderson}, {Hicks}, {Holmbeck}, {Holoien}, {Hutchison},
  {Im}, {Jensen}, {Johnson}, {Joner}, {Jones}, {Kaspi}, {Kelly}, {Kennea},
  {Kim}, {Kim}, {Kim}, {King}, {Klimanov}, {Knigge}, {Krongold}, {Lau}, {Lee},
  {Leonard}, {Li}, {Lira}, {Lochhaas}, {Ma}, {MacInnis}, {Manne-Nicholas},
  {Mauerhan}, {McGurk}, {McHardy}, {Montuori}, {Morelli}, {Mosquera}, {Mudd},
  {M{\"u}ller-S{\'a}nchez}, {Nazarov}, {Norris}, {Nousek}, {Nguyen}, {Ochner},
  {Okhmat}, {Papadakis}, {Parks}, {Pei}, {Penny}, {Pizzella}, {Poleski},
  {Pott}, {Rafter}, {Rix}, {Runnoe}, {Saylor}, {Schimoia}, {Scott}, {Sergeev},
  {Shappee}, {Shivvers}, {Siegel}, {Simonian}, {Siviero}, {Skielboe}, {Somers},
  {Spencer}, {Starkey}, {Stevens}, {Sung}, {Tayar}, {Tejos}, {Turner},
  {Uttley}, {Van Saders}, {Vaughan}, {Vican}, {Villanueva}, {Villforth},
  {Weiss}, {Woo}, {Yan}, {Young}, {Yuk}, {Zheng}, {Zhu}, \&
  {Zu}}]{2020ApJ...902...74W}
---. 2020, \apj, 902, 74, \dodoi{10.3847/1538-4357/abbad7}

\end{thebibliography}
\end{document}